\documentclass[twocolumn,astrosymb]{aastex631}
\newcommand\kms{\ensuremath{{\rm km}~{\rm s}^{-1}}}

\newcommand\nevi{[\ion{Ne}{6}]}
\newcommand\neiiwave{[\ion{Ne}{2}] $\lambda$12.813$\mu$m}
\newcommand\neii{[\ion{Ne}{2}]}

\newcommand\nev{[\ion{Ne}{5}]}
\newcommand\oviwave{\ion{O}{6} $\lambda\lambda$1032,1038\angst}

\newcommand\ergs{\ensuremath{{\rm erg}~{\rm s}^{-1}}}
\newcommand\ergscm{\ensuremath{{\rm erg}~{\rm s}^{-1}~{\rm cm}^{-2}}}
\newcommand\msun{\ensuremath{M_\odot}}
\newcommand\msunyr{\ensuremath{M_\odot~{\rm yr}^{-1}}}

\newcommand\angst{\ensuremath{\mathring{\rm A}}}
\newcommand\kmsMpc{\ensuremath{{\rm km}~{\rm s}^{-1}~{\rm Mpc}^{-1}}}

\newcommand\rev{}
\newcommand\rrev{}

\usepackage{threeparttable}
\usepackage{amsmath}
\usepackage{amssymb}
\usepackage{amsfonts}
\usepackage{mathbbol}
\usepackage{physics}
\usepackage{gensymb}

\let\tablenum\relax
\usepackage{siunitx}

\shorttitle{Cold gas in the Phoenix Cluster}
\shortauthors{Reefe et al.}

\graphicspath{{./}{figures/}}

\begin{document}

\title{Cold Gas and Star Formation in the Phoenix Cluster with JWST}

\correspondingauthor{Michael Reefe}
\email{mreefe@mit.edu}

\author[0000-0003-4701-8497]{Michael Reefe}
\altaffiliation{National Science Foundation, Graduate Research Fellow}
\affiliation{Kavli Institute for Astrophysics \& Space Research, Massachusetts Institute of Technology, Cambridge, MA 02139, USA}

\author[0000-0001-5226-8349]{Michael McDonald}
\affiliation{Kavli Institute for Astrophysics \& Space Research, Massachusetts Institute of Technology, Cambridge, MA 02139, USA}

\author[0000-0002-8823-0606]{Marios Chatzikos}
\affiliation{Department of Physics \& Astronomy, University of Kentucky, Lexington, KY 40506, USA}

\author{Jerome Seebeck}
\affiliation{Department of Astronomy \& Joint Space-Science Institute, 
University of Maryland, College Park, College Park, MD 20740, USA}

\author[0000-0002-7962-5446]{Richard Mushotzky}
\affiliation{Department of Astronomy \& Joint Space-Science Institute, 
University of Maryland, College Park, College Park, MD 20740, USA}

\author[0000-0002-3158-6820]{Sylvain Veilleux}
\affiliation{Department of Astronomy \& Joint Space-Science Institute, 
University of Maryland, College Park, College Park, MD 20740, USA}

%%% alphabetical

\author[0000-0003-0667-5941]{Steven W. Allen}
\affiliation{Kavli Institute for Particle Astrophysics and Cosmology, Stanford University, 452 Lomita Mall, Stanford, CA 94305, USA}
\affiliation{Department of Physics, Stanford University, 382 Via Pueblo Mall, Stanford, CA 94305, USA}
\affiliation{SLAC National Accelerator Laboratory, 2575 Sand Hill Road, Menlo Park, CA 94025, USA}

\author[0000-0003-1074-4807]{Matthew Bayliss}
\affiliation{Department of Physics, University of Cincinnati, Cincinnati, OH 45221, USA}

\author[0000-0002-2238-2105]{Michael Calzadilla}
\affiliation{Kavli Institute for Astrophysics \& Space Research, Massachusetts Institute of Technology, Cambridge, MA 02139, USA}

\author[0000-0003-1398-5542]{Rebecca Canning}
\affiliation{Institute of Cosmology \& Gravitation, University of Portsmouth, Dennis Sciama Building, Portsmouth, PO1 3FX, UK}

\author[0000-0002-2808-0853]{Megan Donahue}
\affiliation{Department of Physics and Astronomy, Michigan State University, East Lansing, MI 48824, USA}

\author[0000-0003-4175-571X]{Benjamin Floyd}
\affiliation{Institute of Cosmology \& Gravitation, University of Portsmouth, Dennis Sciama Building, Portsmouth, PO1 3FX, UK}
\affiliation{Department of Physics and Astronomy, University of Missouri-Kansas City, Flarsheim Hall, 5110 Rockhill Road, Kansas City, MO 64110, USA}

\author[0000-0003-2754-9258]{Massimo Gaspari}
\affiliation{Department of Physics, Informatics and Mathematics, University of Modena and Reggio Emilia, 41125 Modena, Italy}

\author[0000-0001-7271-7340]{Julie Hlavacek-Larrondo}
\affiliation{Department of Physics, Universit\'e de Montr\'eal, Montreal, QC H3T 1J4, Canada}

\author[0000-0002-2622-2627]{Brian McNamara}
\affiliation{Department of Physics and Astronomy, University of Waterloo, Waterloo, ON N2L 3G1, Canada}

\author[0000-0001-5208-649X]{Helen Russell}
\affiliation{School of Physics \& Astronomy, University of Nottingham, University Park, Nottingham NG7 2RD, UK}

\author[0000-0002-5222-1337]{Arnab Sarkar}
\affiliation{Kavli Institute for Astrophysics \& Space Research, Massachusetts Institute of Technology, Cambridge, MA 02139, USA}

\author[0000-0002-7559-0864]{Keren Sharon}
\affiliation{Department of Astronomy, University of Michigan, 1085 S. University Ave, Ann Arbor, MI 48109, USA}

\author[0000-0003-3521-3631]{Taweewat Somboonpanyakul}
\affiliation{Department of Physics, Faculty of Science, Chulalongkorn University, 254 Phayathai Road, Pathumwan, Bangkok 10330, Thailand}

\begin{abstract}
We present integral field unit observations of the Phoenix Cluster with the \textit{JWST} Mid-infrared Instrument's Medium Resolution Spectrometer (MIRI/MRS). We focus this study on the molecular gas, dust, and star formation in the brightest cluster galaxy (BCG).  We use precise spectral modeling to produce maps of the silicate dust, molecular gas, and polycyclic aromatic hydrocarbons (PAHs) in the inner $\sim$50 kpc of the cluster.  \rev{We measure} the optical depth from silicates by comparing the observed H$_2$ line ratios to those predicted by excitation models.  We provide updated measurements of the total molecular gas mass of \rrev{$1.9^{+0.5}_{-0.4} \times 10^{10}$ \msun\,,} which agrees with CO-based estimates, providing an estimate of the CO-to-H$_2$ conversion factor of \rrev{$\alpha_{\rm CO} = 0.8 \pm 0.2\,\msun\,{\rm pc}^{-2}\,({\rm K}\,\kms)^{-1}$}; an updated stellar mass of $M_* = 2.6 \pm 0.5 \times 10^{10}$ \msun\,; and star formation rates averaged over 10 and 100 Myr of $\langle{\rm SFR}\rangle_{\rm 10} = 1340 \pm 100$ \msunyr\, and $\langle{\rm SFR}\rangle_{\rm 100} = 740 \pm 80$ \msunyr, respectively. The H$_2$ emission seems to be powered predominantly by \rev{shocks and star formation} within the central $\sim 20$ kpc\rev{, induced by stellar feedback and radio jets from the active galactic nucleus}.  Additionally, we find nearly an order of magnitude drop in the star formation rates estimated by PAH fluxes in cool core BCGs compared to field galaxies, suggesting that hot particles from the intracluster medium are destroying PAH grains even in the centralmost 10s of kpc.
\end{abstract}


\section{Introduction} \label{sec:intro}

The hot intracluster medium (ICM) that permeates clusters of galaxies exists at temperatures of order $10^7$ K and emits thermal bremsstrahlung radiation in the X-ray.  This radiation causes it to lose energy and cool down over time.  Once the gas cools to $\sim 10^{6.5}$ K, it \rev{will} start to radiate through other mechanisms, primarily metal line emission, increasing the rate of cooling.  \rev{Individual ionization states of metal species will contribute to the cooling at different temperatures, determined by the temperatures at which their collisional ionization equilibrium (CIE) abundances are maximized}. The many complex physical processes that contribute to the cooling rate of \rev{the} gas \rev{at different temperatures} are commonly conglomerated into a single entity called the \textit{cooling function} $\Lambda(T)$ \citep[i.e.][]{1993ApJS...88..253S}.  This, in combination with the initial thermal energy (per unit volume) of the gas $\mathcal{E}$, can give an idea of the overall \textit{cooling time}, $t_{\rm cool} \sim \mathcal{E}/n^2\Lambda \propto T^{1/2}n^{-1}$. If the initial temperature of the gas is not too high, and its density is not too low, then a \textit{cooling flow} may develop within cooling times less than the age of the universe, meaning they should be observable \citep{1984Natur.310..733F}.  In the classical (isobaric) picture, a cooling flow unfolds as follows \citep[see, e.g.,][]{2010gfe..book.....M}.  Consider a parcel of ICM gas that cools by an incremental amount---its temperature drops, and by the ideal gas law so does its pressure.  The weight of the surrounding gas now pushes on the parcel, causing it to move inwards and increase in density.  The change in density causes a corresponding change in pressure until it can again reach a state of equilibrium with the surrounding gas.

\rev{These cooling flow models provide compelling explanations for the presence and origin of $10^{5}$-$10^{6}$ K ionized gas in the interstellar, circumgalactic, intragroup, and intracluster media \citep{2017ApJ...848..122B}. However}, ideal cooling flows like described above cannot exist in nature because cooling can never be 100\% efficient. Heating, turbulence, and magnetic fields are just a few examples of physical processes that reduce cooling efficiency and prevent unabated cooling flows from developing. Indeed, if the gas were able to cool unabated from the hot $\sim$10$^7$ K atmosphere to the cold 10 K molecular phase and rapidly form stars, we should observe star formation rates (SFRs) in brightest cluster galaxies (BCGs) of a similar magnitude to the classical cooling rates measured from the ICM. In reality, the observed SFRs are typically suppressed by 2 orders of magnitude relative to the cooling rates \citep{1989AJ.....98.2018M, 1995MNRAS.276..947A, 1999MNRAS.306..857C, 2005ApJ...635L...9H, 2007MNRAS.379..100E, 2007MNRAS.380...33H, 2008ApJ...681.1035O, 2010ApJ...721.1262M, 2018ApJ...858...45M, 2012ApJS..199...23H, 2012ApJ...747...29R, 2015ApJ...805..177D, 2015MNRAS.450.2564M, 2016A&A...595A.123M, 2023ApJ...947...44C}. \rev{Studying the cold gas content and stellar populations in galaxy clusters is, therefore, crucial for understanding the end stages of these cooling flows, which illuminates how they affect the evolution of BCGs.}

\rev{Cooling} flows are expected to last for a significant fraction of a cluster's lifetime based on the incidence rate of cool core clusters in the general galaxy cluster population.  For timescales of Gyr, the amount of molecular gas formed should be on the order $10^{11}$--$10^{12}$ \msun.  Nevertheless, observed molecular gas masses consistently fall short of this prediction, typically ranging from $10^8$--$10^{10}$ \msun~ \citep{2001MNRAS.328..762E, 2003A&A...412..657S, 2008A&A...483..793S}. This, in conjunction with the low SFRs, is evidence that a large fraction of the gas is not able to fully cool and is being reheated, likely by feedback from an active galactic nucleus (AGN) in the core \citep[see reviews by][]{2007ARA&A..45..117M, 2012ARA&A..50..455F, 2020NatAs...4...10G}. Thus, the classical steady isobaric picture of cooling flows is most likely not an accurate representation of how cooling actually takes place in these clusters. Simulations and theories now suggest that cooling proceeds as a more chaotic, turbulent cascade of condensing cold clouds that periodically precipitate onto and fuel accretion (feeding) and feedback episodes of the central supermassive black hole \citep[SMBH;][]{2011MNRAS.411..349G, 2015ApJ...811..108P, 2017ApJ...847..106L, 2018ApJ...854..167G}.

Dust content is also important to consider in the context of cooling flows---dust grains are closely linked with the cold molecular gas, acting as ``seeds'' for the gas to clump into star forming regions and filaments, boosting condensation and the cooling cascade and enriching the stellar populations. The presence of dust can inform the presence of cold gas and stars, and vice versa \citep{2008A&A...479..669C, 2012MNRAS.423...49K}.  It is especially important when considering attenuation effects, which are dependent on both wavelength and geometry and can complicate measurements of the cooling rates and inferences made from these measurements.  Cooling rates measured from emission lines in the UV, such as \oviwave\,, can be significantly underestimated if an extinction correction is not taken into account.  Absorption effects may also partially explain why many lines in the soft X-ray regime below $10^{6.5}$ K seem to be absent or much weaker than expected in many systems \citep{1988ASIC..229...63C, 2001ApJ...557..546D, 2003ApJ...590..207P}, implying much of the cooling gas is ``hidden,'' but may still be fueling massive cooling flows down to the molecular gas regime \citep{2022MNRAS.515.3336F, 2023MNRAS.521.1794F}.

The Phoenix Cluster \citep[SPT-CLJ2344-4243;][]{2011ApJ...738..139W, 2012Natur.488..349M} is an interesting case study in the field of cooling flows. Its extreme 500--800 \msunyr~ starburst represents $\sim$10--20\% of the 3,800 \msunyr~ classical cooling rate \citep[compared to the typical 1\%;][]{2012Natur.488..349M}, and it contains a massive $2.1 \pm 0.3 \times 10^{10}$ \msun~ reservoir of cold molecular gas \citep{2017ApJ...836..130R}, both signs which point to it being one of the most rapid and uninhibited cooling flows in the known universe.  This makes it an ideal playground to study the effects of large-scale cooling on host galaxies.  However, due to the complex nature of this system, which hosts an extremely luminous type II QSO on top of a massive cooling flow, many of these measurements are marred by systematic uncertainties relating to how much of the underlying emission can be attributed to the QSO, how much dust extinction there is, and the metallicity of the ICM.  The molecular gas mass estimate in particular relies on an uncertain conversion factor between CO and H$_2$, $\alpha_{\rm CO}$, which may vary by a factor of a few depending on the density, temperature, and metallicity of the molecular clouds \citep{2013ARA&A..51..207B}. 

Our recent work \rev{\citep[][hereafter R25]{Reefe_2025}} looked at this unique cluster through an infrared lens with observations using \textit{JWST}'s Mid-infrared Instrument in Medium Resolution Spectroscopy mode (MIRI/MRS), which has the advantage of being in a wavelength regime with essentially no extinction.  R25 focused on the intermediate-temperature ($10^{5.5}$ K) coronal gas, using the high-ionization \nevi\,, \nev\,, [\ion{Fe}{8}], [\ion{Fe}{7}], and [\ion{Mg}{7}] lines to reveal the morphology, kinematics, and cooling rate of the gas in between the hot and cold phases. In this work, we wish to examine this same dataset through a new lens, now with a focus on the molecular gas, dust, and star formation.  The mid-infrared spectral range covered by MIRI/MRS is rich in spectral features from warm molecular gas (the \rev{rotational} H$_2$ emission lines), dust (the thermal dust continuum and the 9.7 $\mu$m silicate absorption feature), and very small dust grains known as Polycyclic Aromatic Hydrocarbons (PAHs, which exhibit broad emission features). We have 4 primary goals with this analysis, which we tackle in sequence throughout this paper: (1) Understand the morphology of the warm molecular gas, dust, and PAHs and how they relate to heating and feedback sources (i.e. radio jets) and the other gas phases, most notably the cold molecular gas seen in CO; (2) obtain an independent measurement of the total molecular gas mass directly from the H$_2$ emission lines, without relying on the uncertain $\alpha_{\rm CO}$ conversion factor; (3) use this new data to get better constraints on the star formation history, and a more precise measurement of the current star formation rate; and (4) use correlations and known scaling relations between IR spectral features to obtain constraints on the importance of different heating mechanisms in exciting the molecular gas and dust, and compare these with the typical cool core BCG population.

To aid with our fourth goal, we build upon the work of \cite{2011ApJ...732...40D}, hereafter D11, who studied a sample of cool core (CC) BCGs using infrared spectra and photometry from the \textit{Spitzer Space Telescope}.  Correlations between the emission features from dust, PAHs, molecular gas, and warm ionized gas are often seen in galaxies of different types and can give insights into the dominant physical mechanisms that power the emission.  Gas and dust can be heated through star formation, cosmic rays, suprathermal ICM electrons, photoelectrons ejected from dust grains, shocks, and/or AGN feedback in the form of mechanical radio jets, radiative winds, and turbulence \citep{2015ApJ...805...35H, 2011MNRAS.411..349G, 2015ApJ...811..108P}.  D11 found a number of such correlations in the CC BCG population using spectral features in the mid-infrared, and the Infrared Spectrograph on \textit{Spitzer} covers a very similar spectral range as MIRI/MRS, so we find these correlations useful as a comparison point for the Phoenix Cluster.

This paper is organized as follows.  The observations and data reduction methods are presented in $\S$\ref{sec:obs}. \rev{Then, $\S$\ref{sec:results} goes over our analysis of the gas and dust morphology ($\S$\ref{sec:gas_dust}) and the stellar populations ($\S$\ref{sec:stellarpop}). $\S$\ref{sec:discussion} then goes into the correlation analysis, discussing and interpreting our findings in the context of the global picture of the cooling flow and the theory that Phoenix may contain an undermassive SMBH.}  We then summarize our findings in $\S$\ref{sec:conc}.  Throughout this work, we assume a flat $\Lambda$CDM cosmology with $\Omega_m = 0.27$, $\Omega_\Lambda = 0.73$, and $H_0 = 70$ \kmsMpc.

% Then, $\S$\ref{sec:results} goes over the analysis, subdivided by our individual goals as outlined above: $\S$\ref{sec:gas_dust} covers the gas and dust morphology and the gas mass, $\S$\ref{sec:stellarpop} covers the stellar populations and star formation, and $\S$\ref{sec:scaling} covers the correlation analysis. $\S$\ref{sec:discuss} discusses our findings in the context of the theory that Phoenix contains an undermassive SMBH, and goes over the expected evolution of the cluster in this scenario. 

% the presence of multiphase gas in their central brightest cluster galaxies (BCGs) spanning 10--10$^7$ K is ubiquitous.  Observations have confirmed the presence of emission nebulae in the millimeter \citep{2001MNRAS.328..762E, 2003A&A...412..657S, 2008A&A...483..793S}, tracing 10 K molecular gas; the optical/IR \citep{1985ApJS...59..447H, 1987MNRAS.224...75J, 1989ApJ...338...48H, 1999MNRAS.306..857C, 2002MNRAS.337...49E, 2007MNRAS.379..100E, 2007MNRAS.380...33H, 2010ApJ...721.1262M}, tracing 10$^4$ K ionized gas; and the UV \citep{2001ApJ...553L.125B, 2006ApJ...642..746B, 2001ApJ...560..187O, 2014ApJ...791L..30M}, tracing 10$^{5.5}$ K coronal gas. 
% % Nevertheless, the quantity of cool gas observed in these clusters is often far less than what would be expected for an unabated cooling flow running for the cluster's lifetime (ref).

\section{Observations} \label{sec:obs}

\subsection{Data Collection} \label{sec:obs_collect}
We obtained \textit{JWST} MIRI/MRS observations of the Phoenix Cluster on UTC 2023 July 27--28 with program ID 2439.  Exposures were taken in all channel/band combinations.  The SHORT (A) band exposures were the longest, at 6.95 hours (since the primary science goal of this program was measuring the [\ion{Ne}{6}] emission, i.e. R25), while the MEDIUM (B) and LONG (C) band exposures were 51.0 and 85.8 minutes, respectively.  Dedicated background exposures were also taken, with exposure times of 51.0, 16.8, and 25.2 minutes. On-source exposures used a 4-point dither pattern while background exposures used a 2-point dither pattern.  In the rest-frame of Phoenix A ($z = 0.597$), MIRI/MRS covers a wavelength range of 3.1--17.5 $\mu$m.

\subsection{Data Reduction} \label{sec:obs_reduce}
Here, we provide a brief overview of the general steps taken in the data reduction, cleaning, and correction processes. For full details on these procedures, see R25. We use the STScI pipeline version 1.12.3 and CRDS context \texttt{jwst\_1140.pmap} to reduce the MIRI/MRS data. We use a few non-standard settings, including more aggressive cosmic ray flagging (lowering the jump rejection threshold to 3.5$\sigma$ and enabling cosmic ray shower flagging\footnote{\href{https://jwst-pipeline.readthedocs.io/en/stable/jwst/jump/description.html\#large-events-snowballs-and-showers}{JWST pipeline cosmic ray snowball/shower information}}), 2D residual fringe correction (removing fixed-frequency modulations in the spectrum caused by standing waves\footnote{\href{https://jwst-pipeline.readthedocs.io/en/latest/jwst/residual\_fringe/main.html\#fringe-background-information}{JWST Pipeline fringing information}}), and 2D pixel-by-pixel background subtraction\footnote{\href{https://jwst-pipeline.readthedocs.io/en/latest/jwst/background\_subtraction/main.html\#spectroscopic-modes}{JWST pipeline background subtraction information}}. We also perform a few additional data cleaning procedures. We remove hot/warm pixels by flagging outliers in the background exposures and masking them in both the background and science frames, following \citet{2023Natur.618..708S}. The thresholds for flagging outliers were chosen by hand for each channel and tried to keep a balance between keeping as much data as possible and making sure to exclude obviously bad pixels. The (low, high) threshold pixel values for channels 1 \& 2, 3, and 4, after subtracting the median, were $(-0.14,+0.13)$, $(-0.16,+0.10)$, and $(-0.16,+0.27)$, respectively. We also remove stripe artifacts caused by cosmic ray hits, also following \citet{2023Natur.618..708S}. Additionally, we rescale the data to account for the loss in sensitivity over time of MIRI. \rev{This effect has been accounted for in the 1.12.3 version of the STSci pipeline, but we have found that the corrected fluxes still underpredict independently measured infrared fluxes from WISE. We therefore take the wavlenegth-dependent correction vector from the STSci pipeline and rescale it until the data match the observed WISE fluxes within 1$\sigma$.} We also replace the pipeline-produced errors with our own estimates of the pixel-by-pixel scatter by measuring the RMS variation between the data and a subic spline fit with a spacing of 7 pixels between knots (masking emission lines). 

\begin{figure*}
    \centering
    \quad
    \includegraphics[width=0.9\textwidth]{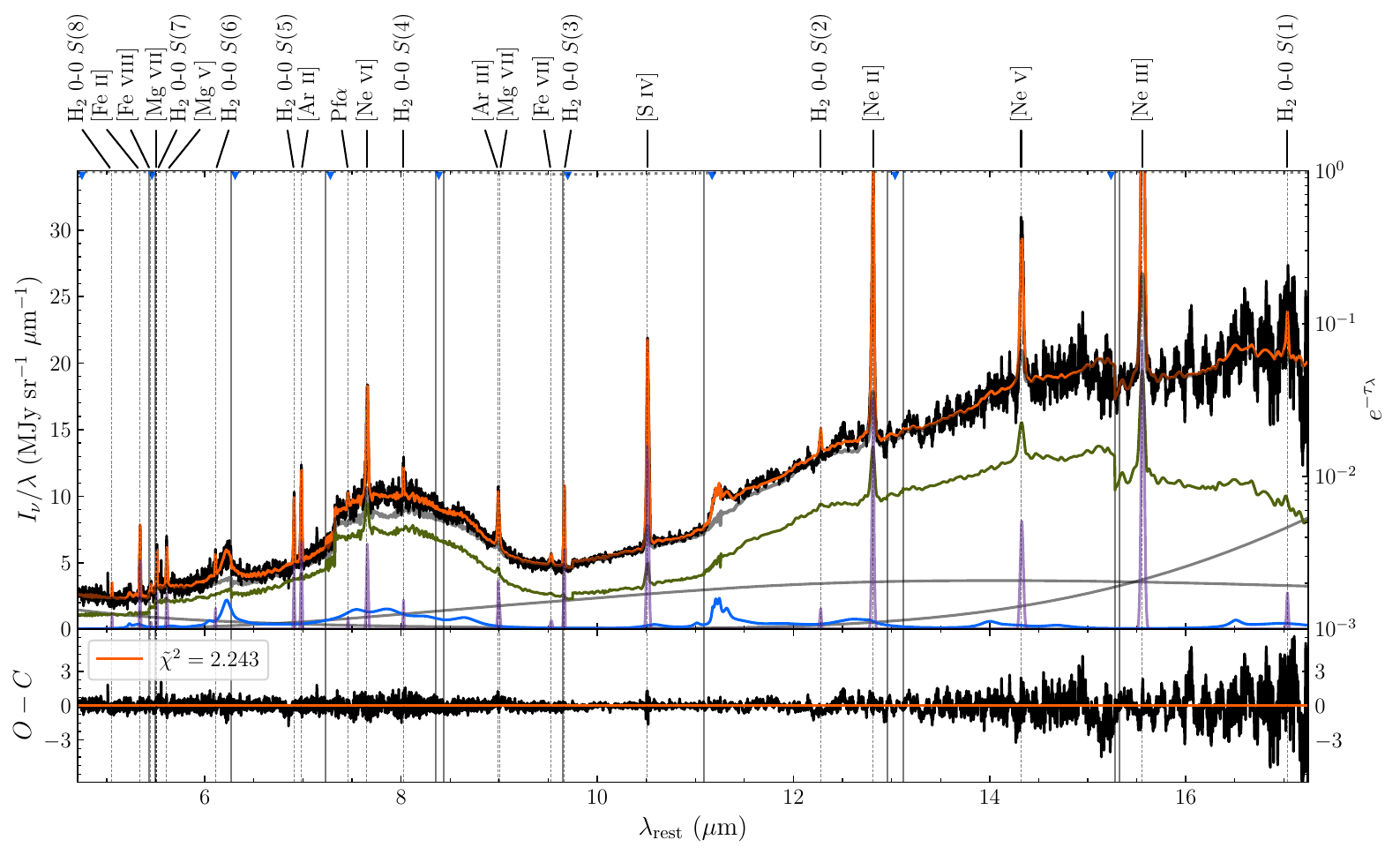}
    \includegraphics[width=0.9\textwidth]{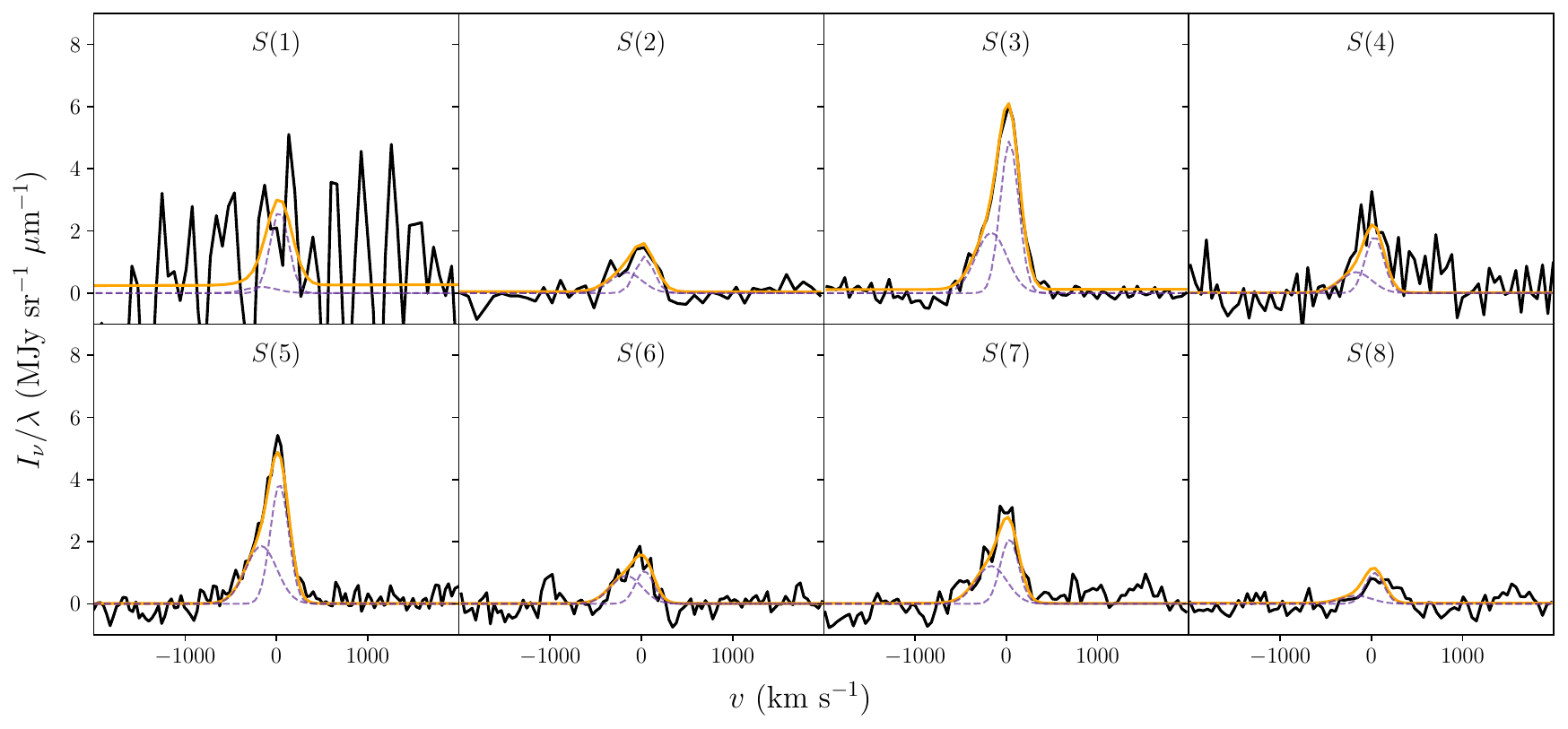}
    \caption{The MIRI/MRS spectrum of a bright off-nuclear spaxel.  \rev{The spaxel has coordinates $(x=23,y=16)$, corresponding to $(\alpha=\ang[angle-symbol-over-decimal, angle-symbol-degree=^{\rm h}, angle-symbol-minute=^{\rm m}, angle-symbol-second=^{\rm s}]{23;44;44.02},\delta=\ang[angle-symbol-over-decimal]{-42;43;12.19})$ and a side length of $\ang[angle-symbol-over-decimal]{;;0.17}$}. \rev{\textit{Top:}} The upper panel shows the data, model, and individual model components, and the lower panel shows the residuals. The data is presented in black, the full model in orange, thermal dust continua in gray, PAH emission in blue, emission lines in purple, and the QSO PSF model in green.  The extinction profile is shown by the gray dotted line across the top of the plots and is read from the right axis.  Emission lines are labeled with vertical dashed lines, and boundaries between MIRI/MRS channels are labeled with blue triangles at the top of the plots. \rev{This plot} shows the full \rev{mid-infrared} spectrum, incorporating data from channels 2--4.  The vertical jumps that occur in the QSO PSF component are due to differences in the size and shape of the PSF between MIRI/MRS channels. \rev{\textit{Bottom}:} A zoom-in of the same spectrum around \rev{each of the} H$_2$ emission line\rev{s}, \rev{showing their velocity profiles. As in the top plot, the data is in black and the model is in orange, with individual velocity components shown with thin purple dashed lines.  The continuum model has been subtracted in these zoom-in plots to highlight the line profiles.}}
    \label{fig:1dspectrum}
\end{figure*}

Our data has a bright point source due to the IR-bright QSO at the center of Phoenix A. We are, therefore, interested in modeling the PSF of MIRI/MRS. To do so, we use data of the bright star 16 Cygni B from program ID 1538. We shift the centroid of the star to match the position of our Phoenix A data, normalize such that it integrates to 1 at each wavelength slice, and fit a cubic spline along the wavelength axis with 100 pixels between each knot. The normalization is done to remove the spectral shape of the star, and the spline fitting is done to reduce the noise in the extracted PSF profile, which should only vary gradually with wavelength. Again, for full details on these procedures, refer to R25. \rev{Figure \ref{fig:1dspectrum} shows an example of a bright off-nuclear spectrum and model}.

\section{Results} \label{sec:results}

\subsection{Gas \& Dust Content} \label{sec:gas_dust}

We use the Likelihood Optimization of gas Kinematics in IFUs \rev{\citep[LOKI;][]{reefe_2025_15069243}\footnote{\href{https://github.com/Michael-Reefe/Loki.jl}{https://github.com/Michael-Reefe/Loki.jl}}} code to perform least-squares fits of the spectra in each spaxel. During the fitting procedure, the light from each spaxel is separated into a host galaxy component and a QSO component. The former traces the underlying emission from the gas, dust, and stars within and along the line of sight of the BCG at the location of the spaxel. The latter traces light from the central QSO that has been dispersed according to the PSF and contaminates the observed spectrum of the spaxel. For these fits, we follow the methods of R25. 

\subsubsection{Silicate Dust Obscuration} \label{sec:dust_obs}

\begin{figure}
    \centering
    \includegraphics[width=\columnwidth]{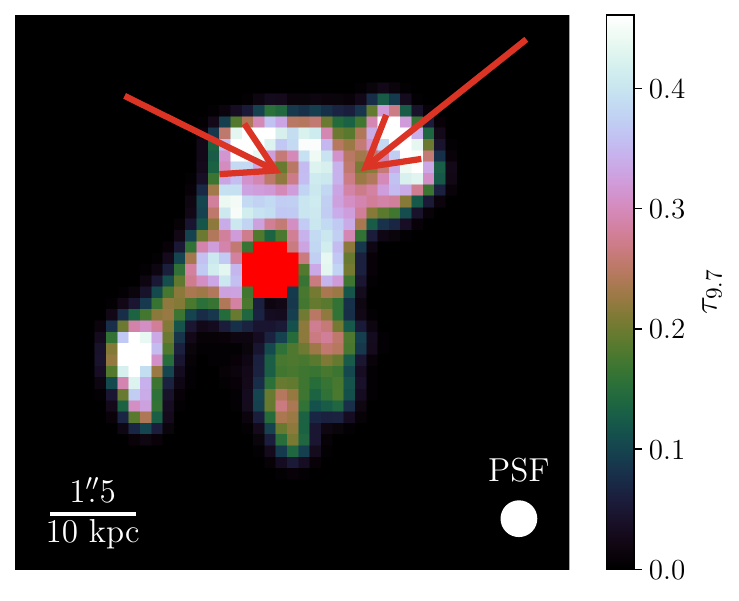}
    \caption{The optical depth of the Phoenix cluster at 9.7 $\mu$m from the silicate absorption feature, estimated using the H$_2$ line ratios. Note that the central spaxels are masked out (denoted by red) since the H$_2$ lines are not significantly detected in these spaxels. \rev{The red arrows highlight two loop structures to the north of the nucleus that have not been seen (or resolved) in dust maps traced by the Balmer decrement or UV continuum.}}
    \label{fig:opticaldepth}
\end{figure}

We first perform a set of initial fits for each individual channel. In these initial fits, the recovered optical depth $\tau$ of the host galaxy is unreliable because the QSO spectrum outshines the host galaxy within $\sim$15 kpc of the center (due to the PSF). The faint continuum of the host galaxy becomes dominated by instrumental systematics that make it difficult to measure $\tau$ from the shape of the silicate absorption feature. As such, we measure the optical depth of the host galaxy using an independent method that only relies on the flux ratios of the rotational H$_2$ lines obtained from these initial fits. In contrast to the continuum and higher ionization lines, the H$_2$ lines have a much smaller equivalent width in the QSO spectrum compared to the host galaxy, so their fluxes are dominated by their host galaxy fluxes away from the nucleus. Most of these lines are unaffected by extinction, but there is one---the H$_2$ 0-0 $S(3)$ line---that fortuitously lands very near the peak of the 9.7 $\mu$m silicate absorption feature (see Figure \ref{fig:1dspectrum}). We therefore use the $S(4)/S(3)$ line ratio as a proxy for the extinction. We estimate what this ratio should \rev{intrinsically} be in the absence of extinction by fitting excitation models with the $S(3)$ line masked out (see $\S$\ref{sec:gas_mass}). We then compare this to our observed $S(4)/S(3)$ ratio to obtain an estimate of the optical depth at the peak of the silicate absorption feature:
\begin{equation}
    \tau_{9.7} \approx \ln\frac{[S(4)/S(3)]_{\rm observed}}{[S(4)/S(3)]_{\rm intrinsic}}
\end{equation}
\rev{Similar techniques have been used in previous studies \citep[i.e.][]{2000A&A...356..705R}, including recently in observations of the Orion Bar and young stellar objects \citep{2024A&A...687A..86V, 2024A&A...687A..36T}}. This results in the optical depth map shown in Figure \ref{fig:opticaldepth}. Note once again that this is the optical depth just in the host galaxy, and does not accurately represent the obscuration of the QSO at the center (marked by the red spaxels). The estimated uncertainty in these measurements is $\simeq 5$\% in the brightest spaxels, rising up to a median of $\simeq 20$\% in all spaxels, due to a combination of uncertainties in the measured and modeled line ratios.

% In reality, the QSO has a much larger optical depth---this can be much more easily measured since the central spaxels are dominated by the QSO and we no longer have to worry about the decomposition between QSO and host galaxy, allowing us to measure the optical depth directly from the shape of the continuum.

The optical depth at 9.7 $\mu$m is a direct tracer of the column density of silicate dust grains, which create broad absorption features in the MIR through bending and stretching modes. We can see that the silicates are concentrated north of the nuclear region where they form two loop structures (marked with red arrows), with two filaments extending to the south and southeast. The overall structure is reminiscent of the $E(B-V)$ maps obtained from the UV continuum \citep{2013ApJ...765L..37M} and Balmer lines \citep{2014ApJ...784...18M}, showing that the silicate dust grains generally follow the rest of the dust, but the substructure in the silicates is distinct, particularly in the loops.  The structure is statistically significant given our relatively low uncertainties in the brighter spaxels.

\begin{figure*}
    \centering
    \includegraphics[width=\textwidth]{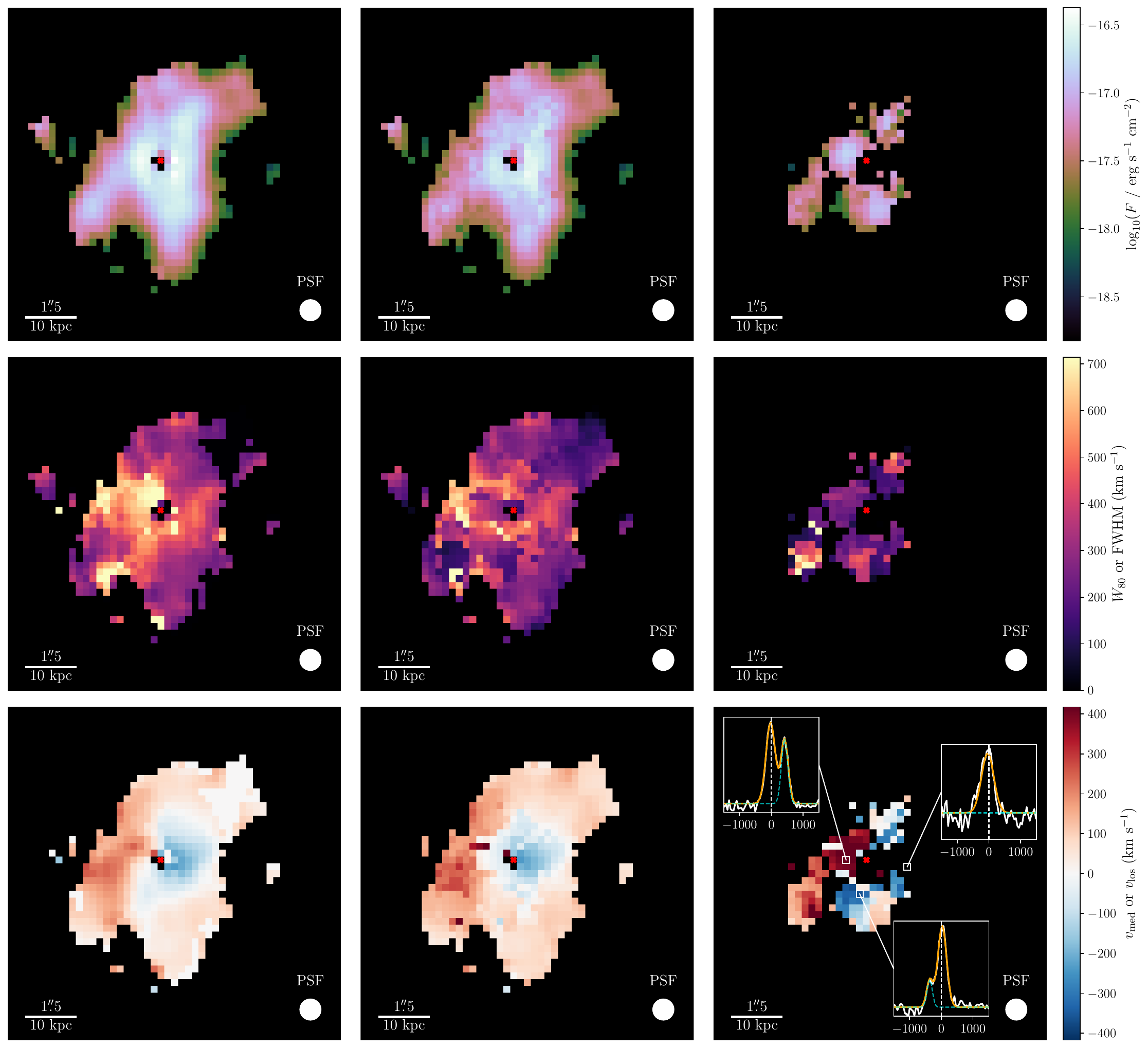}
    \caption{Maps of the H$_2$ 0-0 $S(3)$ line over the channel 3 FOV. The top row shows the flux with a logarithmic colormap. The left panel is the total flux, and the right 2 panels show the flux of the individual components. The middle row shows the line width in \kms. The left panel is $W_{80}$, the width containing 80\% of the flux, and the right 2 panels are the FWHMs of the individual components. The bottom row shows the LOS velocity in \kms. The left panel is the median velocity, and the right 2 panels are the LOS velocities of the individual components. In all of the middle and right column, the components are sorted by flux, with the middle column having the larger flux and the right column having the smaller flux. Only spaxels with an $S/N \geqslant 3$ are shown. \rev{In the bottom-right panel, the inset panels show the velocity profiles of the line at the indicated spaxel locations. In each inset, the white vertical dashed line shows the velocity zero point, and the cyan dashed line shows the individual velocity component which corresponds to the value shown in this map.}}
    \label{fig:h2s3}
\end{figure*}

\begin{figure}
    \centering
    \includegraphics[width=\columnwidth]{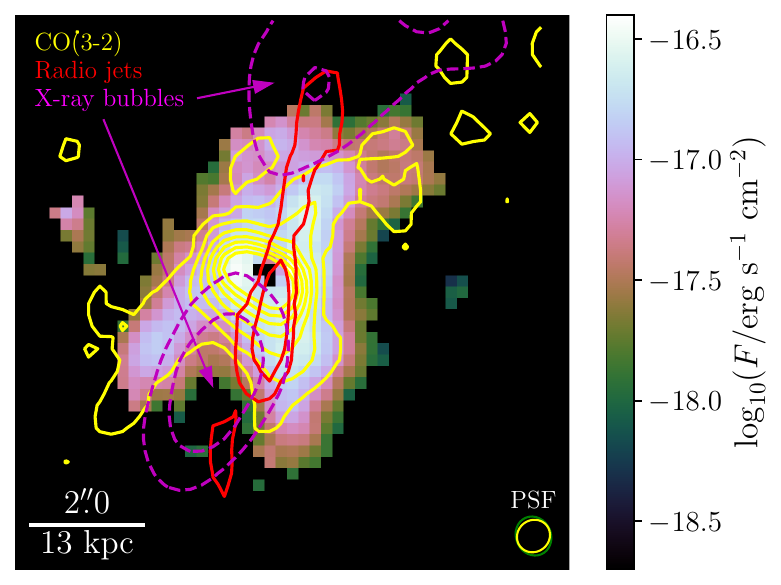}
    \caption{\rev{A map of the H$_2$ 0-0 $S(3)$ flux (identical to the top-left panel of Figure \ref{fig:h2s3}). CO emission from ALMA observations \citep{2017ApJ...836..130R} is shown with the yellow contours (ranging from 0--7$\sigma$ above the mean).  Radio jets from VLA observations \citep{2021A&A...646A..38T} are shown with the red contours (at 2 and 4$\sigma$ above the background).  X-ray cavities from \textit{Chandra} observations \citep{2019ApJ...885...63M} are shown in the magneta dashed contours (at 2 and 4$\sigma$ below the mean).  The PSF shape of MIRI/MRS and ALMA are both shown in the bottom-right corner, with MIRI in green and ALMA in yellow.  Both have FWHMs of $\sim\ang[angle-symbol-over-decimal]{;;0.6}$.}}
    \label{fig:h2_co_compare}
\end{figure}

\subsubsection{The Molecular Gas Phase} \label{sec:mol_gas}

After obtaining the optical depth map from the excitation models, we run a second iteration of our individual channel fits while locking the optical depth of the host galaxy to these values, with the primary purpose of obtaining more accurate molecular H$_2$ line fluxes. Note that this analysis inherently assumes that the extinction of the line-emitting H$_2$ gas is the same as the extinction of the thermal continuum-emitting dust. Previous studies from \citet{2009ApJS..182..628V} have suggested that there may not be a simple 1:1 correlation between these two optical depths, with the emission lines and PAHs being systematically extincted at lower optical depths than the continuum. We find that due to the low continuum level of the host galaxy relative to the QSO, we require some external constraint on the optical depth parameter to keep it within a parameter space that makes physical sense, and this assumption does not significantly hinder our fitting results. Throughout the rest of our analysis, we are primarily concerned with using optical depths to correct the flux values of emission lines, so the values derived using the $S(4)/S(3)$ line ratios are the most relevant and accurate ones to use.

The flux, velocity width, and velocity shift of the $S(3)$ line are shown in Figure \ref{fig:h2s3}. The line is decomposed into 2 Gaussian velocity components with distinct centers and widths---the leftmost column shows the combined values of both components, the middle column shows the brighter component, and the rightmost column shows the dimmer component.  \rev{The FWHM of the MIRI/MRS line spread function, which ranges from $\sim$80--200 \kms\, \citep{2021A&A...656A..57L}, has been subtracted in quadrature from all line width measurements.} The morphology of the H$_2$ lines \rev{in Figure \ref{fig:h2s3}} strongly resembles the CO emission from \citet{2017ApJ...836..130R}\rev{, but the H$_2$ appears slightly more extended than the CO. We compare these morphologies by showing CO contours on top of the H$_2$ emission in Figure \ref{fig:h2_co_compare}.  The size of the MIRI/MRS PSF at the wavelength of the $S(3)$ line is roughly the same as the ALMA PSF at the wavelength of CO(3-2) ($\sim \ang[angle-symbol-over-decimal]{;;0.6}$). However, the H$_2$ appears more extended due to the difference in sensitivity of the two instruments.  The ALMA observations report an uncertainty $\sigma = 0.067$ Jy/beam \kms $= 5.76 \times 10^{-20}$ \ergscm$\,{\rm spaxel}^{-1}$. Using the integrated H$_2$-to-CO ratio over the whole FOV, this translates to a 3$\sigma$ lower limit of $10^{-16.83}$ \ergscm$\,{\rm spaxel}^{-1}$, which roughly aligns with the outer CO contour in Figure \ref{fig:h2_co_compare}. Therefore, we can conclude that the difference in extent is likely a sensitivity effect and not physical}. The H$_2$ and CO both share three filamentary structures extending to the southeast, south, and northwest, and a bright extended nuclear region. \rev{In the core region immediately surrounding the nucleus,} the brighter H$_2$ component has broader line widths of up to 600 \kms\, and smaller velocity shifts within $|v| \lesssim 200$ \kms. \rev{It follows that this} component \rev{primarily} traces the motions of unresolved clouds drifting slowly \rev{within the turbulent atmosphere of the cluster. Their motions are dominated by large-scale turbulence (e.g. on the scale of a star-forming region) induced by stellar and AGN feedback, with sub-dominated bulk motions.} The dimmer component\rev{, which is only significantly detected within this core region,} is \rev{narrower}, with line widths of $\sim 200$--$300$ \kms\,, and with faster shifts up to $\pm 400$ \kms. This \rev{indicates that this} component \rev{is capturing} the bulk motion of high-velocity gas clouds, \rev{which may originate} in the filaments that feed into or out of the nucleus, fueling episodes of chaotic cold accretion \citep[see i.e.][]{2018ApJ...854..167G}. \rev{Moving away from the core region, along these three filaments, the velocity profile becomes dominated by a single component with generally low FWHM ($\sim 100$--$300$ \kms) and low shifts ($|v| \lesssim 200$ \kms). This may in fact be the natural extension of the high-velocity component in the core, if our interpretation that this component traces gas feeding in from the filaments is correct.} The other H$_2$ lines generally follow the same morphology as the $S(3)$ line, with varying levels of brightness. We see \rev{overall} that the warm molecular gas \rev{shows a similar morphology to UV and optical continuum emission \citep{2013ApJ...765L..37M, 2015ApJ...811..111M} and warm ionized nebular emisison \citep[i.e. \lbrack\ion{O}{2}\rbrack, H$\beta$ in][]{2014ApJ...784...18M}, tracing regions of rapid and recent star formation.} \rev{Figure \ref{fig:h2_co_compare} also shows that the molecular gas wraps around the cavities in the X-ray emitting ICM (magenta dashed contours), inflated by the quasar's radio jets (red contours). The H$_2$ gas is possibly being uplifted in the bubble's turbulent wakes or forming in-situ through turbulent mixing at the boundaries of the bubbles \citep[][R25]{2015ApJ...811..111M}.}

\subsubsection{Molecular Gas Mass} \label{sec:gas_mass}

\begin{figure*}
    \centering
    \quad
    \includegraphics[width=0.5\textwidth]{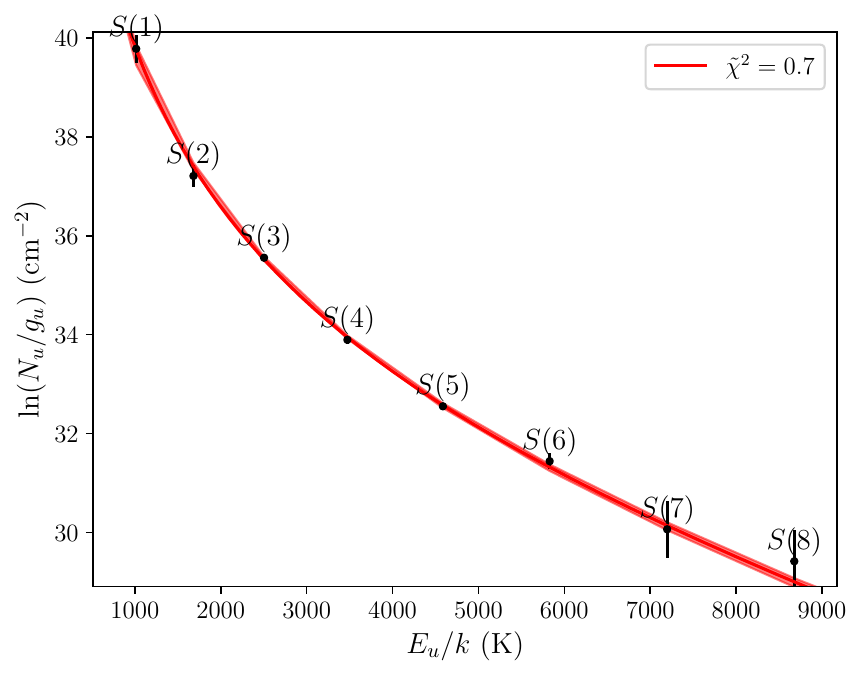}
    \includegraphics[width=0.45\textwidth]{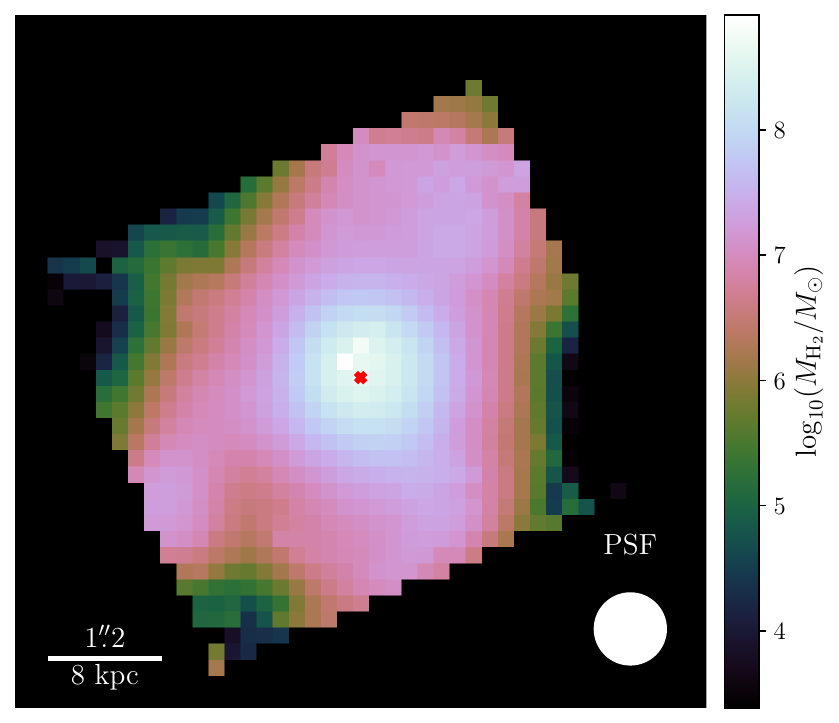}
    \caption{Left: An excitation diagram of the H$_2$ rotational lines integrated over the full channel 2 FOV. The abscissa is the energy of the upper level of the transition in Kelvin, $E_u/k$, and the ordinate is the column density of the upper level of the transition in cm$^{-2}$, $N_u$, normalized by the degeneracy $g_u$.  The black points show the observed values, with each H$_2$ line labeled accordingly, and the red line shows the best-fit continuous temperature model, with a \rrev{$\tilde{\chi}^2 = 0.7$}. Right: A map of the total molecular gas mass over the channel 2 FOV, obtained by fitting a continuous temperature model to the H$_2$ excitation diagram, extrapolating to lower temperatures, and integrating over temperature. Note that the color scale is logarithmic (base 10).}
    \label{fig:excitation}
\end{figure*}

\begin{figure*}
    \centering
    \includegraphics[width=\textwidth]{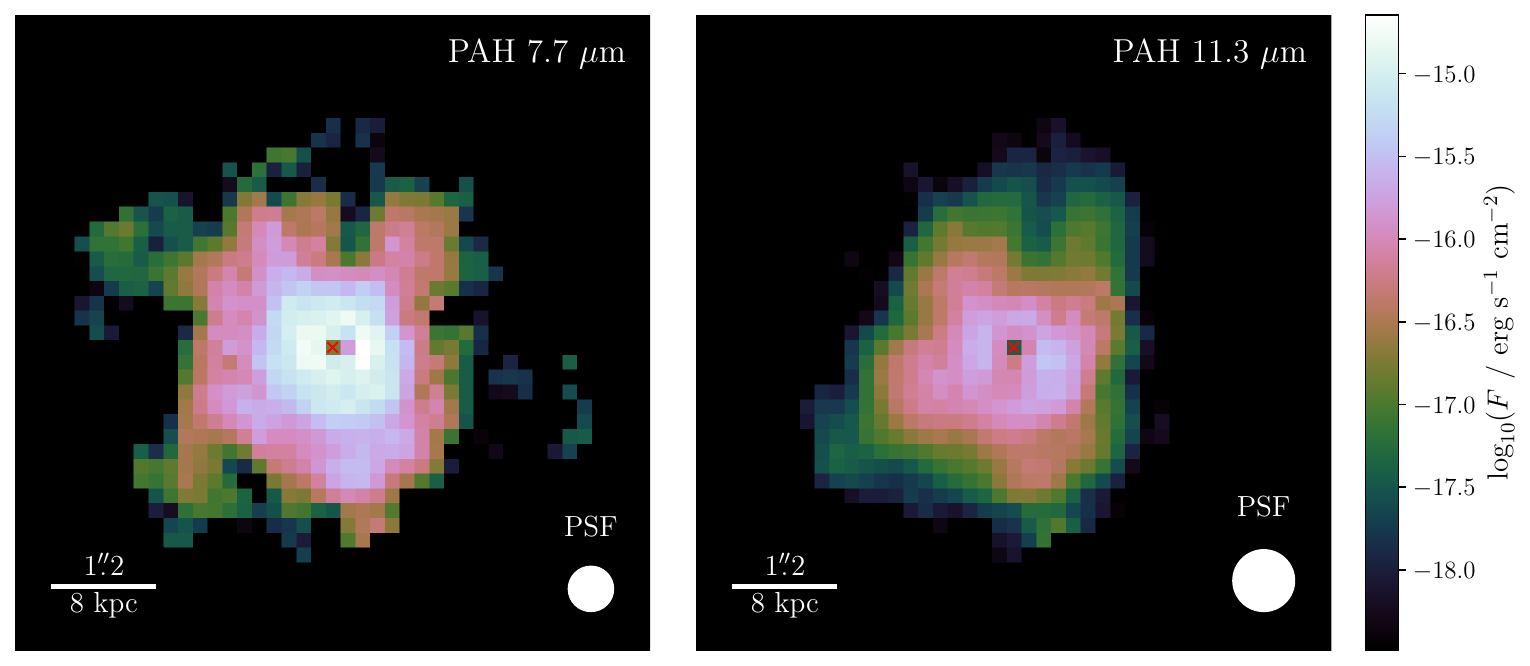}
    \caption{Maps of the 7.7 $\mu$m and 11.3 $\mu$m PAH features over the channel 2 FOV.}
    \label{fig:PAHs}
\end{figure*}

Using the fluxes for the set of pure rotational H$_2$ lines that fall within our wavelength range, \rrev{$S(1)$} to $S(8)$, we construct excitation diagrams of the upper level column densities of these transitions and fit a continuous temperature model from \citet{2016ApJ...830...18T}:
\begin{equation}
    \frac{N_u}{g_u} = \frac{N_{\rm tot}(n-1)}{T_\ell^{1-n}-T_u^{1-n}}\int_{T_\ell}^{T_u}\frac{e^{-E_u/kT}}{Z(T)}T^{-n}{\rm d}T
\end{equation}
where $N_u$, $E_u$, and $g_u$ are the column density, energy, and degeneracy of the upper energy level, $T_\ell$ and $T_u$ are the limits of the temperature distribution, $N_{\rm tot}$ is the total column density of H$_2$, $Z(T)$ is the partition function of the H$_2$, and $n$ is a generic power-law index that sets the temperature distribution. We set $T_u = 2000$ K following \citet{2016ApJ...830...18T} and allow $N_{\rm tot}$, $n$, and $T_\ell$ to vary. 

We have 2 goals with this analysis: 1) Obtain ``intrinsic'' $S(4)/S(3)$ line ratios so that we may reconstruct the optical depth of the emission lines in the host galaxy; and 2) Obtain an estimate for the total molecular gas mass and how this gas is distributed spatially. For the first goal, we take the H$_2$ line fluxes from our first iteration fits and mask the $S(3)$ line. Once we obtain a modeled value for the $S(3)$ line, we can compute the inferred line ratio as
\begin{equation}
    \frac{F_i}{F_j} = \frac{N_i}{N_j}\frac{\lambda_j}{\lambda_i}\frac{A_i}{A_j}
\end{equation}
where $F_i$, $N_i$, $\lambda_i$, and $A_i$ are the fluxes, column densities, wavelengths, and Einstein $A$ coefficients of each line. We then compare this with the observed line ratio to compute an optical depth as described in $\S$\ref{sec:dust_obs}. After obtaining the optical depth map, we use line fluxes from our second iteration fits, including the $S(3)$ line, to obtain estimates for the molecular gas mass. Note that we use the \textit{total} flux of the H$_2$ lines in both of these procedures, recombining the components from the QSO PSF and host galaxy decomposition. The decomposition into these two components is still important, however, since they may be extincted by different amounts (this is only relevant for the $S(3)$ line, so it is not an important distinction for the first goal). Note that including the rotational lines up to $S(8)$ limits us to the channel 2 FOV. 

We show the excitation diagram for the H$_2$ rotational lines, integrated over the whole FOV of channel 2, in the left panel of Figure \ref{fig:excitation}. The errors here have been obtained by bootstrapping the spectral fits 100 times, so they are purely statistical. The data are well fit by a temperature distribution with a power law index of \rrev{$n = 4.7 \pm 0.1$} and a lower temperature limit of \rrev{$T_\ell = 44^{+143}_{-5}$ K}. \rrev{The large upper uncertainty on $T_\ell$ indicates that the model loses sensitivity to H$_2$} at temperatures \rrev{$\lesssim$ 200 K \citep[see][]{2016ApJ...830...18T}}. However, the continuous temperature model allows us to extrapolate to lower temperatures and get an estimate of the total molecular gas mass.  We use a lower temperature \rrev{limit} appropriate for ULIRGs and LIRGs of $T_{\rm \ell}' = 80$ K \citep{2016ApJ...830...18T}, which yields a total molecular gas mass of \rrev{$M_{\rm H_2} = 1.9^{+0.5}_{-0.4} \times 10^{10}$ \msun}. This is in excellent agreement with previous measurements from \citet{2014ApJ...784...18M} and \citet{2017ApJ...836..130R}, who obtained an H$_2$ mass of $2.1 \pm 0.3 \times 10^{10}\,\msun$ using a CO-to-H$_2$ conversion factor $X_{\rm CO} = 0.4 \times 10^{20}\,{\rm cm}^{-2}\,({\rm K}\,\kms)^{-1}$, equivalent to $\alpha_{\rm CO} = 0.86\,\msun\,{\rm pc}^{-2}\,({\rm K}\,\kms)^{-1}$. \rev{Spaxel-by-spaxel fits of these models, with the molecular gas mass in each spaxel, are shown in the right panel of Figure \ref{fig:excitation}. The mass distribution generally follows the same morphology as the surface brightness.}

\rev{By comparing the H$_2$ mass measurements of \citet{2017ApJ...836..130R} to ours}, assuming the same underlying CO mass, we can obtain an estimate of the CO-to-H$_2$ conversion factor of \rrev{$\alpha_{\rm CO} = 0.8 \pm 0.2\,\msun\,{\rm pc}^{-2}\,({\rm K}\,\kms)^{-1}$}. This is roughly a factor of 5 smaller than the standard Milky Way disk value of $\alpha_{\rm CO,Gal} = 4.3\,\msun\,{\rm pc}^{-2}\,({\rm K}\,\kms)^{-1}$ \citep{2013ARA&A..51..207B}.  Lower values of $\alpha_{\rm CO}$ are common in mergers, ULIRGs, and starburst galaxies because these systems tend to have more turbulence and gas inflows which lead to molecular gas at higher temperatures and densities.  This increases the molecular gas luminosity per unit mass, lowering $\alpha_{\rm CO}$ \citep{2016ApJ...830...18T, 1993ApJ...414L..13D, 1998ApJ...507..615D}.  The extreme cooling and star formation in Phoenix make it comparable to these systems and is the most likely explanation behind the lowered $\alpha_{\rm CO}$ value.  Variations in metallicity can also produce differences in the conversion between CO and H$_2$, and are likely the reason that our H$_2$ mass measurement does not match exactly with \citet{2017ApJ...836..130R} (aside from pure statistical variations). 

\rev{We also compare these measurements of the total (cold) molecular gas mass to the (warm) molecular gas mass traced by the MIR H$_2$ transitions by fitting a discrete 2-temperature model with no extrapolation. This results in a warm H$_2$ mass of \rrev{$3.4 \pm 1.4 \times 10^{8}$ \msun}\, at characteristic temperatures of \rrev{$350^{+40}_{-30}$ K} and \rrev{$1100 \pm 100$ K}. The fit quality is \rrev{marginally worse than} the continuous temperature model, with a \rrev{$\tilde{\chi}^2 = 1.63$}. This implies that the warm H$_2$ accounts for only a very small fraction \rrev{($\sim$2\%)} of the total H$_2$ mass, which is to be expected based on the mass distributions for the \citet{2016ApJ...830...18T} models ($\dv*{M}{T} \propto T^{-n}$). Indeed, the warm H$_2$ mass fraction estimated \rrev{from} these models, using a lower temperature limit \rrev{$T_\ell = 200$ K (from the sensitivity floor of the continuous model)} relative to a lower integration temperature $T_{\rm \ell}' = 80$ K, is of a similarly small magnitude:
\begin{equation}
    \frac{\int_{T_\ell}^{T_u}(\dv*{M}{T}){\rm d}T}{\int_{T_{\rm \ell}'}^{T_u}(\dv*{M}{T}){\rm d}T} \approx \bigg(\frac{T_\ell}{T_{\rm \ell}'}\bigg)^{1-n} \approx \rrev{3\%}~.
\end{equation}
}

\subsubsection{PAH Features} \label{sec:pah_morph}

Due to a rather unfortunate alignment in the wavelength boundaries between channels of the MIRI instrument, at the redshift of the Phoenix cluster, the prominent PAH features at 7.7 $\mu$m and 11.3 $\mu$m both land at the edges between channels 2/3 and 3/4, meaning they are partially cut off in each channel. This motivated us to, in addition to our single-channel fits, also run a multi-channel fit by combining data from channels 2--4, projecting everything onto the channel 2 grid. This way, we are able to measure the full integrated fluxes of these PAH features. 

Maps of the total flux of the 7.7 $\mu$m and 11.3 $\mu$m PAH complexes are shown in Figure \ref{fig:PAHs}. The PAH emission is more centrally concentrated than the molecular gas and does not have distinct filaments extending to the south and southeast, but it does have protrusions from the nucleus that generally align with the directions of the filaments. \rev{However, we are cautionary in drawing any conclusions from the morphology of these protrusions, as they may be partially contaminated by PSF residuals that have not been fully subtracted. This is likely due to the low equivalent widths of the PAH features in this system, making it difficult for our spectral modeling procedure to cleanly separate the PAH features from the continuum and PSF.}
\rev{Despite these shortcomings, we can argue that the PAH morphology} is clearly extended to both the north and the south, whereas the warm/hot ionized gas phases at $\sim$10$^{5.5}$ K are only extended to the north (see i.e. R25).  Qualitatively speaking, then, the PAHs more closely resemble the morphologies of the colder gas phases. The physical extent of the PAHs is likely constrained by the necessity for them to be shielded from hard ionizing UV and X-ray photons from the AGN, and photons and suprathermal electrons from the hot ICM. It is interesting, then, that we still see significant PAH emission even in the centralmost spaxels closest to the X-ray-luminous AGN. \rev{This seems to be a ubiquitous feature of AGN---they may have profound effects on the PAH grain size distribution and ionization fraction, preferentially destroying smaller and more ionized grains, but they seemingly do not cause the unilateral destruction of PAHs \citep[see, i.e.,][]{2024A&A...691A.162G}. The highly obscured nature of Phoenix's AGN may also play a role in the survivability of PAHs within such close distances. We explore this further in $\S$\ref{sec:pah_correlations}.}
% This emission likely comes from dusty gas clumps in and aroud the torus and in the surrounding regions that have lines of sight to the AGN that are shielded by the torus.

\subsection{Stellar Populations and a Robust SFR} \label{sec:stellarpop}

\begin{deluxetable*}{llll}
    \tabletypesize{\footnotesize}
    \tablecaption{CIGALE Parameters}
    \label{tab:cigale}
    \tablehead{\colhead{Parameter Description} & \colhead{Name} & \colhead{Searched Values} & \colhead{Bayesian fit value}}
    \decimals
    \startdata
    \hline
    Star Formation History & \texttt{sfhdelayed} \\
    \hline
    Main population e-folding time & \texttt{tau\_main} & 100, 500, 1000, 2000, 4000, 6000, 8000 Myr & $6000 \pm 2000$ Myr \\
    Main population age & \texttt{age\_main} & 8000 Myr \\
    Burst population e-folding time & \texttt{tau\_burst} & 1, 10, 100, 1000 Myr & $600 \pm 400$ Myr \\
    Burst population age & \texttt{age\_burst} & 1, 5, 10, 20, 50, 100 Myr & $26 \pm 13$ Myr \\
    Burst population mass fraction & \texttt{f\_burst} & 0, 0.0001, 0.0005, 0.001, 0.005, 0.01 & $0.006 \pm 0.002$ \\
    \hline
    Stellar Populations & \texttt{bc03} \\
    \hline
    Initial Mass Function & \texttt{imf} & Salpeter \\
    Stellar metallicity & \texttt{metallicity} & 0.0004, 0.004, 0.008 & $0.006 \pm 0.002$ \\
    Old \& young population separation age & \texttt{separation\_age} & 10 Myr \\
    \hline
    Nebular Emission & \texttt{nebular} \\
    \hline
    Ionization parameter & \texttt{logU} & $-3$ \\
    Gas metallicity & \texttt{zgas} & 0.005 \\
    Electron density & \texttt{ne} & 1000 cm$^{-3}$ \\
    % Lyman continuum photon escape fraction & \texttt{f\_esc} & 0 \\
    % Lyman continuum photon absorption fraction & \texttt{f\_dust} & 0 \\
    Lyman continuum dust absorption fraction & \texttt{f\_dust} & 0, 0.001, 0.01, 0.1 & $0.01 \pm 0.02$ \\
    Line width & \texttt{lines\_width} & 800 \kms \\
    \hline
    Dust Attenuation & \texttt{dustatt\_modified\_starburst} \\
    \hline
    Gas reddening & \texttt{E\_BV\_lines} & 0.1, 0.2, 0.3, 0.4, 0.5, 0.6, 0.7 & $0.500 \pm 0.002$ \\
    Stellar reddening reduction factor & \texttt{E\_BV\_factor} & 0.44 \\
    % UV bump wavelength & 2175 \angstrom \\
    % UV bump width & 35 \angstrom \\
    UV bump amplitude & \texttt{uv\_bump\_amplitude} & 0 \\
    Attenuation curve power law slope modifier & \texttt{powerlaw\_slope} & 0 \\
    % Emission line extinction law & CCM89 \\
    % Extinction law slope ($R_V = A_V/E(B-V)$) & 3.1 \\
    \hline
    Dust Emission & \texttt{dl2014} \\
    \hline
    PAH mass fraction & \texttt{q\_pah} & 0.47 \\
    Minimum radiation field & \texttt{umin} & 10 \\
    Power law slope & \texttt{alpha} & 2 \\
    Illumination fraction & \texttt{gamma} & 0.25 \\
    \hline
    AGN & \texttt{skirtor2016} \\
    \hline
    % Disk type & Skirtor \\
    Average edge-on optical depth at 9.7 $\mu$m & \texttt{t} & 9 \\
    Dust density radial slope & \texttt{pl} & 1 \\
    Dust density polar slope & \texttt{q} & 0 \\
    Torus opening angle & \texttt{oa} & \ang{50;;} \\
    Ratio of outer to inner radius & \texttt{R} & 30 \\
    % Fraction of total mass inside clumps ($M_{\rm cl}$) & 0.97 \\
    Inclination angle & \texttt{i} & 60, 70, 80, 90\degree & $80 \pm 1$\degree \\
    Optical disk power law slope modifier & \texttt{delta} & 1.0 \\
    AGN fraction of IR luminosity & \texttt{fracAGN} & 0, 0.1, 0.2, 0.3, 0.4, 0.5, 0.6, 0.7, 0.8, 0.9, 0.99 & $0.50 \pm 0.01$ \\
    % Polar dust extinction law & Calzetti 2000 \\
    Polar dust extinction & \texttt{EBV} & 0, 0.2, 0.4 & $0.21 \pm 0.05$ \\
    % Polar dust temperature & 100 K \\
    % Polar dust emissivity & 1.6 \\
    \hline
    % X-ray emission (\texttt{xray}) \\
    % \hline
    % Photon index ($\Gamma$) & 1.4 \\
    % Exponential cutoff energy of the AGN spectrum ($E_{\rm cut}$) & 300 keV \\
    % Power-law slope connecting $L_\nu$ at 2500 \angstrom\, and 2 keV ($\alpha_{\rm ox}$) & $-1.4$ \\
    % Maximum allowed deviation of $\alpha_{\rm ox}$ from empirical & 0.3 \\
    % Polynomial coefficients of X-ray dependence on AGN viewing angle & 0 \& 0, 0.5 \& 0, 1 \& 0, 0.33 \& 0.67 \\
    % % Deviation from expected LMXB luminosity & 0 \\
    % % Deviation from expected HMXB luminosity & 0 \\
    % \hline
    Radio emission & \texttt{radio} \\
    \hline
    FIR/radio correlation coefficient & \texttt{qir\_sf} & 2.7 \\
    Star formation synchrotron slope & \texttt{alpha\_sf} & 0 \\
    AGN radio loudness & \texttt{R\_agn} & 0.1, 1, 10, 100, 1000 & $10$ \\
    AGN radio emission slope & \texttt{alpha\_agn} & 1.35 \\
    \hline
    \enddata
    \begin{tablenotes}
        \item[1] \textbf{Note}: Any parameters not listed here take their default values.
    \end{tablenotes}
\end{deluxetable*}

With these new \textit{JWST} data in the MIR, we obtain valuable constraints on the shape of Phoenix's SED in a region that is dominated by the AGN and thermal dust emission. As such, we calculate the integrated flux within each subchannel (using a consistently sized aperture, which takes up the full FOV of channel 1) and create simulated broadband photometry measurements. We combine these measurements with archival broadband data spanning from the UV to the radio (GALEX, HST, 2MASS, WISE, Herschel/PACS, \& Herschel/SPIRE: \citet{2012Natur.488..349M}; ATCA \& SUMSS: \citet{2014ApJ...784...18M}) and fit SED models using the CIGALE software \citep{2019A&A...622A.103B, 2020MNRAS.491..740Y}\footnote{\href{https://cigale.lam.fr/}{CIGALE website}}. The modules used and parameter space searched are detailed in Table \ref{tab:cigale}. The full range of parameters we wished to search was large enough that doing a single run varying everything at the same time would have been computationally infeasible. As such, we took an iterative approach and varied parameters in groups. The values shown in Table \ref{tab:cigale} for the nebular, dust, AGN, and radio modules that are fixed have been selected from previous iterations. During these iterations, the other parameters are given values based on educated guesses of the conditions in Phoenix A---for example, given that it hosts a bright type II AGN, we assumed an inclination angle of $\ang{70;;}$ and an AGN fraction of 50\%. In most cases, we see that the final values of these parameters in our final fits are not too far from these initial guesses. The new \textit{JWST} points in the MIR provide stronger constraints on the shape of the AGN and dust emission, allowing us to adjust many of the AGN torus parameters from their default values. However, a major strength of CIGALE comes from its ability to constrain parameters in different wavelength bands using the principles of energy balancing and correlations between parameters from the underlying physics. This is why, in our final iteration, we still allowed certain parameters in these modules to vary from their previously found values, such as the AGN fraction and radio loudness.

A notable caveat to this analysis is the spatially agnostic nature of the CIGALE fit, which presents itself most prominently in the decomposition between the ``dust emission'' and ``AGN'' components of the model.  The templates CIGALE uses for dust emission only consider starlight-heated dust, whereas a powerful AGN such as the one in Phoenix may also contribute to heating the dust all throughout the host galaxy. In contrast, the SKIRTOR AGN templates only consider the radiative transfer process through the dust in the nucleus/torus region, not throughout the whole galaxy.  A study from \citet{2020A&A...638A.150V} looking at NGC 1068 showed, using their own radiative transfer analysis, that considering the effects of the AGN separately caused it to contribute up to 15\% of the total dust heating, depending on viewing angle.  In the case of Phoenix, with a much more luminous AGN, this fraction could be even higher.  Nevertheless, we expect the ramifications of this on our interpretation of the results to be minor: \citet{2020A&A...638A.150V} performed comparison CIGALE fits and found the SFRs to be in agreement with their models.  Without the inclusion of an AGN component, they found the young stellar populations to be overly attenuated in order to boost the MIR dust emission to the observed levels, but in our case the separate inclusion of an AGN component with a distinct shape in the MIR likely alleviates this concern.

% CIGALE SED fits and SFR
\begin{figure}
    \centering
    \includegraphics[width=\columnwidth]{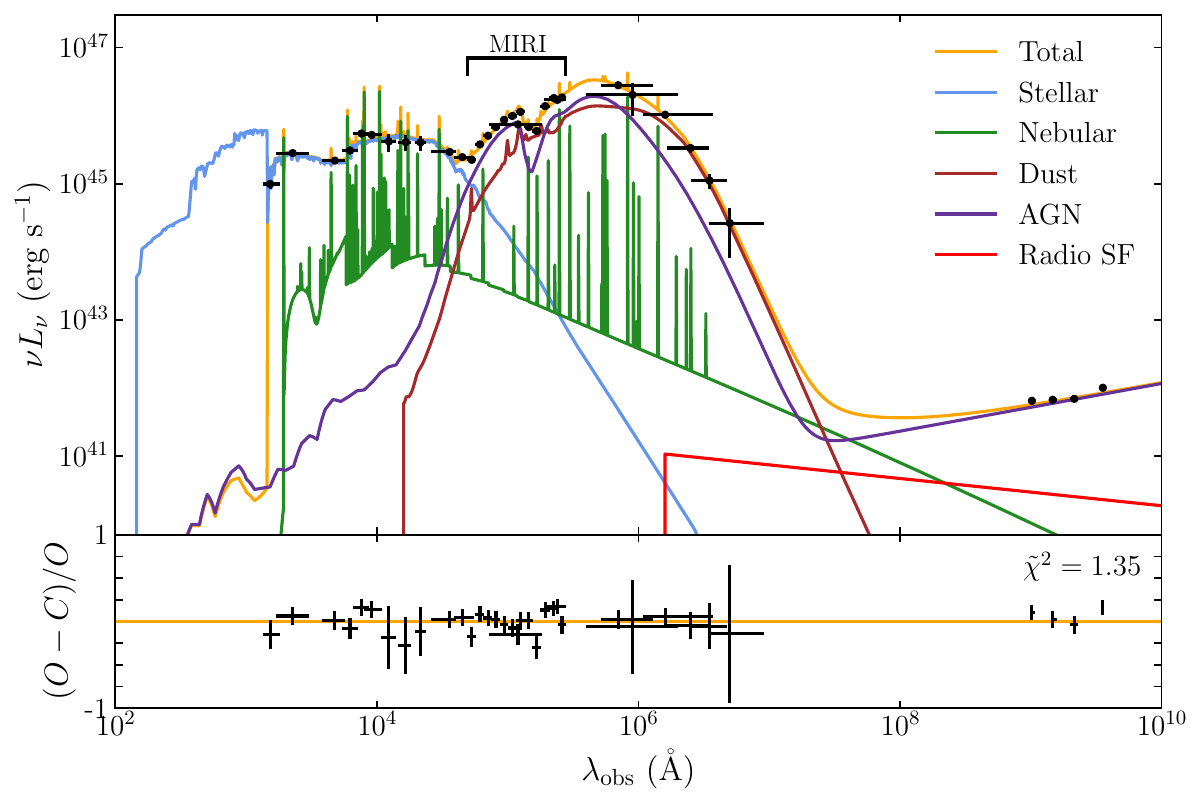}
    \caption{The best-fit SED model for Phoenix A. Photometry data points are shown in black with error bars. The full SED model spectrum is the amber line, and the individual components of the SED are labeled in the legend. The bottom panel shows the residuals of the fit.}
    \label{fig:sed}
\end{figure}

The final SED model is shown in Figure \ref{fig:sed}. The attenuated stellar populations (blue) dominate in the UV and optical due to the obscuration of the AGN, while the AGN (purple) and dust emission (brown) dominate in the MIR-FIR. The data from $\sim 5-10$ $\mu$m in the observed frame (MIRI channel 1 and 2) provide the strongest constraints on the relative power of the AGN and dust emission, since dust emission in this range would correspond to dust heated above $\gtrsim 300$ K, which is difficult to achieve without the presence of an AGN. Notice that the AGN component dominates the SED in this range, whereas the dust emission becomes the dominant component in the FIR above $\sim 100$ $\mu$m, corresponding to much colder ($\sim 30$ K) temperatures. Interpolating our model at a rest-frame wavelength of 24 $\mu$m gives a total luminosity of $\nu L_\nu(24\,{\rm \mu m}) \approx 3.0 \times 10^{46}$ \ergs\, and an AGN-subtracted (i.e. host galaxy) luminosity of $\approx 1.3 \times 10^{46}$ \ergs.  We will use these values in the next section for a more in-depth analysis of the heating mechanisms of the different gas and dust phases.

\begin{figure}
    \centering
    \includegraphics[width=\columnwidth]{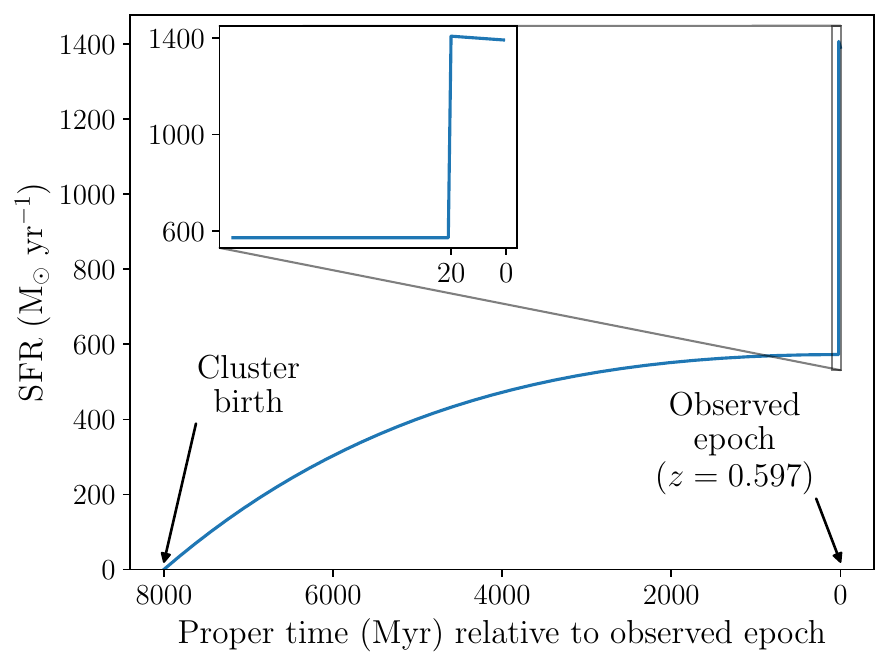}
    \caption{The star formation rate is shown as a function of time in Phoenix's rest frame, relative to the observed epoch. As the main ``old'' stellar population has an age fixed at 8 Gyr, the furthest back the plot goes is 8 Gyr relative to the observed epoch (corresponding roughly to the Big Bang). An inset is shown zooming in on the most recent 100 Myr of star formation, showing the rapid spike at 20 Myr. \rev{The shape of the star formation history is flexible aside from the late-time burst.}}
    \label{fig:sfh}
\end{figure}

The star formation history is shown in Figure \ref{fig:sfh}. The models settle on a fairly consistent star formation rate throughout the past 8 Gyr of the galaxy's history, with the average SFR over all time being around 400 \msunyr.  The recent burst of star formation started $26 \pm 13$ Myr ago and \rev{marginally} favors models with large $e$-folding times of $600 \pm 400$ Myr, bringing the current instantaneous value of the SFR well above 1000 \msunyr.  \rev{The exact shape of the star formation history is not well constrained by the photometry, which is evident in the large uncertainties of the $e$-folding times for both the main and burst populations. The data could be nearly equally well described by, for example, a model with a more heavy weighting towards stars that formed earlier (corresponding to shorter $e$-folding times). This may have an effect on the relative old and young population masses that we measure, but it will have a minimal effect on the long-time averaged star formation rates. These values are summarized in Table \ref{tab:cigale_star}, along with other parameters of the star formation history, which are derived from the parameters given in Table \ref{tab:cigale}.  The fact that the late-time burst occurs within $\lesssim 20$ Myr suggests that} even the youngest O-type stars with lifetimes $\leqslant 10$ Myr should be present in the galaxy.  The presence of extremely luminous O-type stars explains why, despite the young population contributing a considerable amount to the observed optical/UV luminosity, it only contributes $\sim$0.5\% of the total stellar mass. 

\begin{deluxetable}{lll}
    \tabletypesize{\footnotesize}
    \tablecaption{CIGALE Stellar Population Parameters}
    \label{tab:cigale_star}
    \tablehead{\colhead{Parameter} & \colhead{Symbol} & \colhead{Value}}
    \decimals
    \startdata
    \hline
    Old population age & $t_{\rm *,old}$ & 8 Gyr \\
    Young population age & $t_{\rm *,young}$ & $26 \pm 13$ Myr \\
    Old population mass$^\dagger$ & $M_{\rm *,old}$ & $2.6 \pm 0.5 \times 10^{12}$ \msun \\
    Young population mass$^\dagger$ & $M_{\rm *,young}$ & $1.3 \pm 0.1 \times 10^{10}$ \msun \\
    Stellar metallicity & $Z_*$ & $0.006 \pm 0.002$ \\
    Instantaneous SFR$^\ddag$ & SFR & $1330 \pm 130$ \msunyr \\
    10 Myr averaged SFR$^\ddag$ & $\langle {\rm SFR} \rangle_{10}$ & $1340 \pm 100$ \msunyr \\
    100 Myr averaged SFR$^\ddag$ & $\langle {\rm SFR} \rangle_{100}$ & $740 \pm 80$ \msunyr \\
    \hline
    \enddata
    \begin{tablenotes}
        \item[0] All measured stellar masses and SFRs assume a Salpeter IMF.
        
        \item[1] $^\dagger$To convert stellar masses to those that would be measured with other IMFs, multiply by 0.61 (Chabrier) or 0.66 (Kroupa) \citep{2014ARAA..52..415M}.
        
        \item[2] $^\ddag$To convert SFRs to those that would be measured with other IMFs, multiply by 0.63 (Chabrier) or 0.67 (Kroupa) \citep{2014ARAA..52..415M}.
    \end{tablenotes}
\end{deluxetable}

\section{Discussion}
\label{sec:discussion}

\subsection{\rev{Gas \& Dust Energetics}} \label{sec:scaling}

D11 studied a sample of cool core (CC) BCGs and found that the thermal dust continuum and PAH features were consistent with being heated primarily by star formation, whereas the warm molecular H$_2$ gas and ionized [\ion{Ne}{2}] gas \rev{required} an additional heating source, which they attributed to energetic particles from the ICM. It is useful to reexamine some of these correlations with the Phoenix cluster in mind, for a few reasons: (1) To understand the relative contributions of different heating sources in Phoenix itself, and compare them to the typical cool core population; (2) To isolate the effects of AGN photoionization heating, since none of the BCGs studied in the D11 sample exhibited evidence of AGN activity; and (3) To identify the relative importances of different heating sources in offsetting cooling flows.

\subsubsection{\rev{Rotational H$_2$ Emission}} \label{sec:h2_corr}

\begin{figure}
    \centering
    \quad
    \includegraphics[width=\columnwidth]{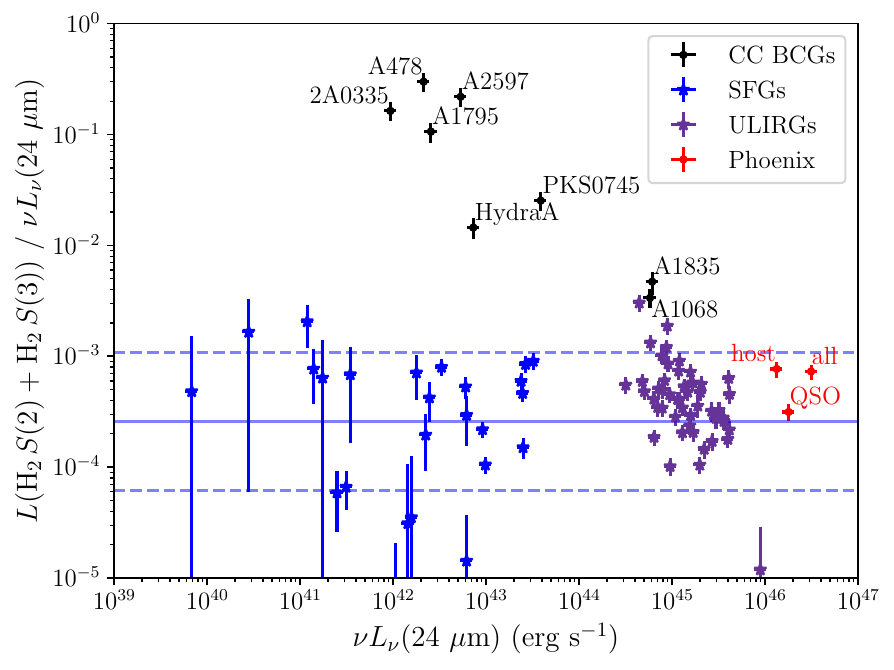}
    \caption{The ratio of the combined H$_2$ 0--0 $S(2)$ and $S(3)$ luminosity over the 24 $\mu$m continuum luminosity $\nu L_\nu(24\,{\rm \mu m})$ plotted as a function of $\nu L_\nu(24\,{\rm \mu m})$. The black points show the cool core (CC) BCGs from the D11 sample, and the red points show the Phoenix cluster: ``all'' corresponds to the integrated values over the whole field of view, ``host'' corresponds to the integrated values after the QSO PSF has been subtracted, and ``QSO'' corresponds to the integrated QSO PSF values. The blue stars show galaxies from the SINGS sample designated as ``\ion{H}{2}'' or star-forming galaxies (SFGs) by \citet{2007ApJ...656..770S}. The solid blue line shows the average value of these galaxies, and the dashed blue lines show the 1-$\sigma$ standard deviation. The purple stars show ULIRGs from \citet{2007ApJ...667..149F}.
    % \textit{Right}: A 2D map showing the $L_{{\rm H}_2}/L_{15}$ ratio in each spaxel, where $L_{15} \equiv \nu L_\nu(15\,{\rm \mu m})$, covering the channel 2 FOV. The color scale is logarithmic, and the data has been smoothed to the PSF at 15 $\mu$m.
    }
    \label{fig:L24_H2}
\end{figure}

\begin{figure}
    \centering
    \includegraphics[width=\columnwidth]{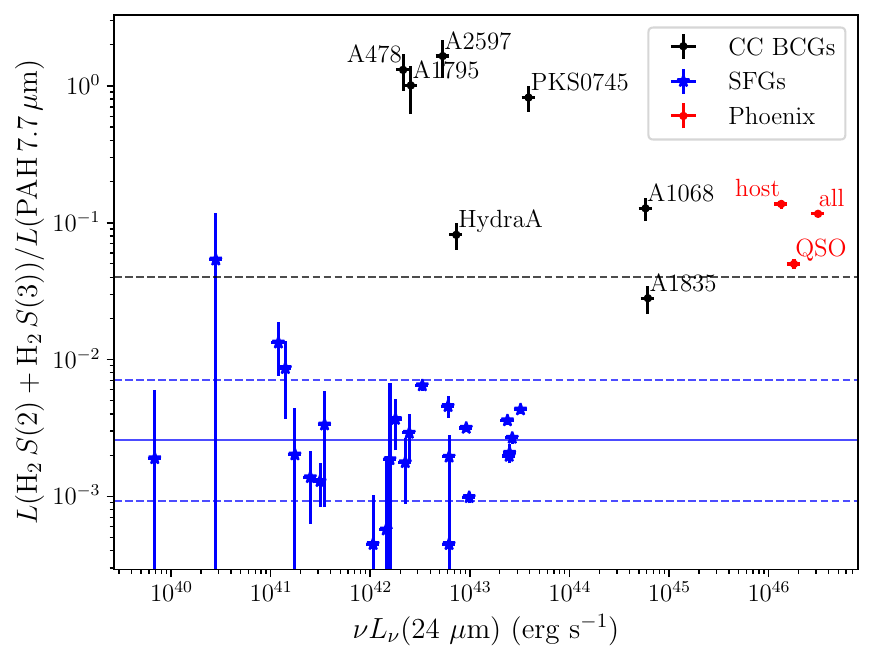}
    \caption{\rev{The ratio of the combined H$_2$ 0-0 $S(2)$ and $S(3)$ luminosity over the 7.7 $\mu$m PAH feature luminosity, plotted against $\nu L_\nu(24\,{\rm \mu m})$. The color-coding of the points is identical to Figure \ref{fig:L24_H2}.  The black dashed line shows the upper limit of the $L_{\rm H_2}/L_{\rm PAH}$ ratio in photodissociation region simulations using the Meudon code \citep{2006ApJS..164..506L}, showing that points above this line must have some non-radiative contribution to their H$_2$ emission.}}
    \label{fig:L24_H2_PAH}
\end{figure}

Being the gas phase most closely linked to star formation, cold molecular gas is an important tool in studying the characteristics of star-forming regions. However, it is often difficult to detect directly, as cold molecular H$_2$ does not emit strongly, and must be traced by the second most abundant molecule in the ISM (CO) instead. Warm molecular gas heated to a few 100 K \rev{-- a few 1000 K}, on the other hand, has rovibrational emission lines in the infrared that can be used to directly detect the warm H$_2$ gas, but no longer has the direct connection to star formation due to its increased excitation. Untangling how much of this emission is due to star formation, as opposed to other sources, can be an important clue in distinguishing Phoenix from other BCGs. In the D11 sample of BCGs, the ratio of the combined luminosity of the H$_2$ 0--0 $S(2)$ and $S(3)$ lines ($L_{{\rm H}_2}$) and the 24 $\mu$m continuum luminosity ($L_{24} \equiv \nu L_\nu$ at 24 $\mu$m) was elevated orders of magnitude above typical values seen in \rev{star-forming} galaxies \citep{2010ApJ...719.1191T} (from $\sim 4 \times 10^{-4}$ to as high as $0.3$), suggesting that star formation is not a significant heating mechanism, and other energy sources, such as cosmic rays, likely play an important role in the observed H$_2$ luminosities.  However, the massive starburst in the Phoenix cluster gives it the unique position of having emission components comparable to both the BCGs and the star-forming galaxies.

In order to make this comparison, we need a measurement of $L_{24}$ for Phoenix. For this, we take the values interpolated from our CIGALE fit in $\S$\ref{sec:stellarpop}.  Indeed, on average we see a much smaller $L_{{\rm H}_2}/L_{24}$ ratio of $\sim 8 \times 10^{-4}$.  It lies along the anticorrelation between $L_{{\rm H}_2}/L_{24}$ and $L_{24}$ noted by D11, which is primarily due to its increased $L_{24}$ relative to the other BCGs. We recreate Figure 10b from D11 in Figure \ref{fig:L24_H2}, which shows this anticorrelation. We compare the BCG sample to a sample of nearby infrared-luminous galaxies: the Spitzer Infrared Nearby Galaxies Survey (SINGS), with photometry measurements taken from \citet{2007ApJ...655..863D}, H$_2$ measurements taken from \citet{2007ApJ...669..959R}, and PAH and Neon line measurements taken from \citet{2007ApJ...656..770S}. We select only the star-forming galaxies (SFGs), designated by the type ``\ion{H}{2}'' by \citet{2007ApJ...656..770S}, and show this sample with blue stars in Figure \ref{fig:L24_H2}. These galaxies have $L_{\rm H_2}/L_{24}$ ratios much smaller than the BCGs and more similar to Phoenix, but with smaller overall $L_{24}$.  We also include a selection of Ultraluminous Infrared Galaxies (ULIRGs) from \citet{2007ApJ...667..149F}.  Powered by vigorous starbursts or AGNs, the ULIRGs bridge the gap between the normal star-forming galaxies and Phoenix.  This shows that Phoenix is an extension of this population into the cluster environment---a unique case where a galaxy in the highest mass regime hosts both a QSO and a starburst. This would seem to imply that the majority of the H$_2$ emission in Phoenix is produced by stellar heating, which would make it unique among the cool core BCGs.  \rev{However, we must n}ote that the presence of the AGN increases $L_{24}$ substantially above that produced by dust-reprocessed stellar light in the host galaxy. We have attempted to account for this by subtracting the portion of $L_{24}$ generated by the AGN and recovering the $L_{24}$ just from the host galaxy. But as discussed in $\S$\ref{sec:stellarpop}, if the AGN is a significant source of heating for the dust in the host galaxy and not just the nucleus, this may not be properly separated by our decomposition. \rev{If the true value of $L_{24}$ from just starlight is a few times smaller than our analysis shows, this could indicate a non-negligible elevation of the $L_{\rm H_2}/L_{24}$ ratio, which in turn implies a non-radiative contribution to the H$_2$ emission \citep[e.g.][]{2010ApJ...724.1193O}.}

\rev{To investigate this, we plot the ratio of H$_2$ to PAH luminosity, which should be more resistant to the effects of AGN contamination, in Figure \ref{fig:L24_H2_PAH}.  This Figure reveals that, indeed, the ratio of H$_2$ to PAH emission in Phoenix is noticeably similar to that of its nearest cool-core cluster neighbors, Abell 1068 and Abell 1835.  Like these other cool cores, Phoenix appears to have either enhanced H$_2$ emission or suppressed PAH emission (or both) relative to the pure star-forming galaxies of the SINGS sample, though the effect is not quite as prominent as in the most quiescent cool cores.  It is likely that the H$_2$ emission in Phoenix is boosted by shocks induced by both the mechanical influence of the radio jets from the AGN and from stellar feedback in the form of supernovae.  The PAH emission may be suppressed due to grain destruction and sputtering caused by the harsh ionizing radiation from the AGN and the ICM.  We investigate both of these possiblities in the next few subsections.}

\begin{figure}
    \centering
    \quad
    \includegraphics[width=\columnwidth]{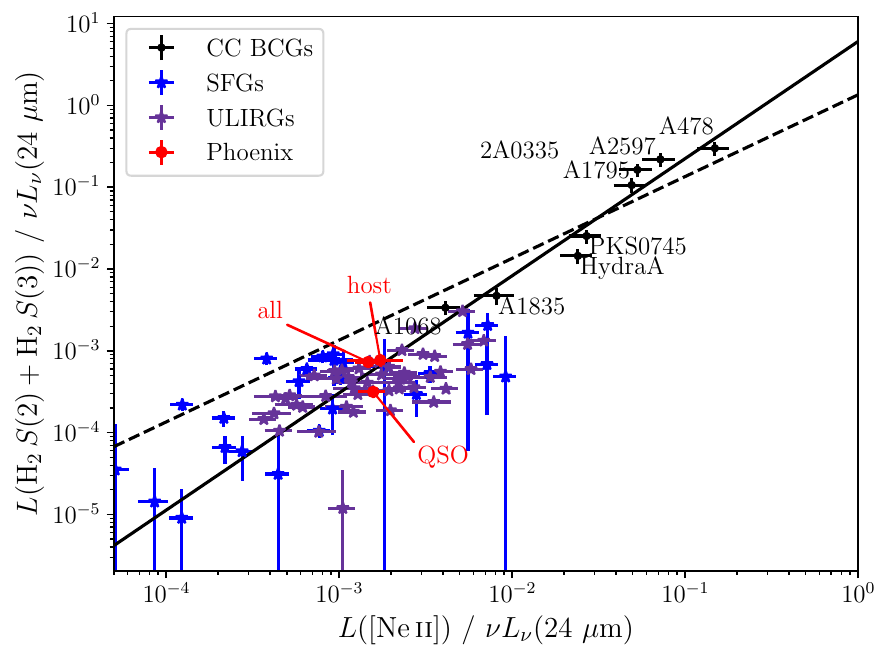}
    \caption{The ratio of the combined H$_2$ 0--0 $S(2)$ and $S(3)$ luminosity over the 24 $\mu$m continuum luminosity $\nu L_\nu(24\,{\rm \mu m})$ versus the ratio of the \neiiwave\, luminosity over $\nu L_\nu(24\,{\rm \mu m})$. Both luminosities are normalized to $\nu L_\nu(24\,{\rm \mu m})$ to remove any ``bigger is bigger'' effects \citep{1990ASSL..161..405K}. The data points are labeled identically to Figure \ref{fig:L24_H2}. The dashed line shows a 1:1 correlation, and the solid line shows the best-fit power law, which is marginally steeper.}
    \label{fig:NeII_H2}
\end{figure}

We also see a strong correlation between the H$_2$ luminosity and the \neiiwave\, luminosity ($L_{\rm [Ne\,\textsc{ii}]}$), corroborating that found in D11 (their Figure 11, our Figure \ref{fig:NeII_H2}). Phoenix lies right along the best-fit power law, $(L_{{\rm H}_2}/L_{24}) \propto (L_{\rm [Ne\,\textsc{ii}]}/L_{24})^{1.4 \pm 0.1}$, but lands in a region of parameter space similar to the SINGS star-forming galaxies and ULIRGs as opposed to the other BCGs. This correlation means that the heating sources for the molecular gas and the warm ionized gas must be tightly interlinked. In other words, if the \rev{observed} H$_2$ excitation is \rev{due to a combination of stellar and shock heating,} then the \neii\, excitation \rev{can likely also be attributed to these sources.}

\subsubsection{\rev{Shock-Driven H$_2$ Emission}}
\label{sec:shocks}

\begin{figure}
    \centering
    \includegraphics[width=\columnwidth]{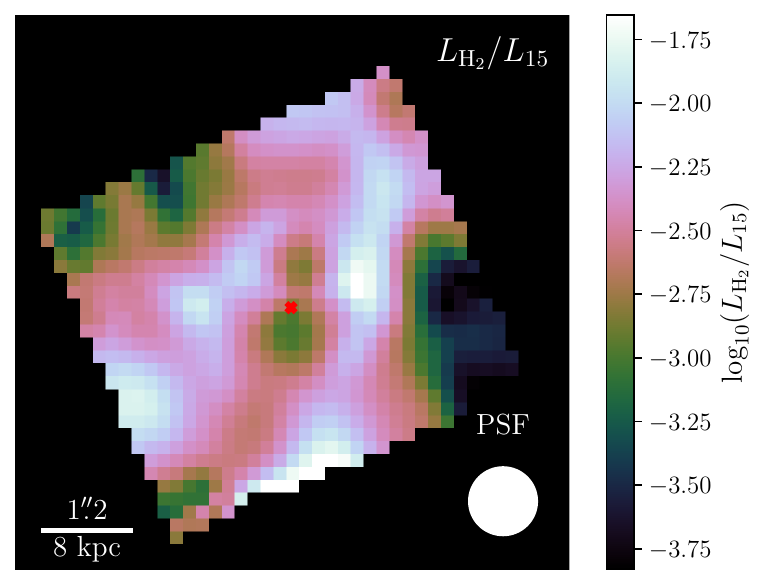}
    \caption{\rev{A map of the ratio of the H$_2$ $S(2) + S(3)$ luminosity over the 15 $\mu$m continuum luminosity in each spaxel, over the channel 2 FOV. This map has been spatially convolved with a Gaussian with a standard deviation of 1 pixel. Areas with a higher $L_{\rm H_2}/L_{15}$ ratio may have more contributions from shocks and other non-radiative emission mechanisms compared to regions with lower $L_{\rm H_2}/L_{15}$ ratios.}}
    \label{fig:L15_H2_map}
\end{figure}

\rev{Shocks may constitute a significant fraction of the H$_2$ emission observed in this cluster.  To better quantify this contribution, here we examine the H$_2$ emission in further detail both spatially and spectrally.  We show the spatial distribution of the H$_2$-to-continuum ratio in Figure \ref{fig:L15_H2_map}. Note that in this figure we use the 15 $\mu$m continuum luminosity, rather than the 24 $\mu$m continuum luminosity, since our MIRI/MRS spectral coverage for this system does not extend to 24 $\mu$m. 15 $\mu$m was chosen as a proxy due to it being relatively free from contamination from lines and PAH features, while not being too short in wavelength that the continuum becomes dominated by emission from the AGN and stellar populations. Focusing on the spatial distribution, rather than the absolute value of this ratio, we find that the H$_2$ emission is enhanced relative to the continuum to the east and west of the nucleus, forming a cocoon-like shape. This may be the result of bowed shock fronts expanding out from the north-south aligned radio jets \citep[see, for example, Figure 3 in][]{2000ApJ...540..678B}.}

\rev{To constrain how much of the H$_2$ emission can be attributed to shocks, we implement a 2-component toy model that describes the H$_2$ emission as a combination of shocks (with some distribution over velocities) and photodissociation regions (PDRs).  We follow the methods of \citet{2024A&A...688A..96V} in constructing this model. For the PDR component, we use pre-computed models from the Meudon PDR code\footnote{\href{https://ism.obspm.fr/pdr.html}{https://ism.obspm.fr/pdr.html}} with a density $n_{\rm H} = 10$ cm$^{-3}$, illumination factor $G_0=10$, and maximum extinction $A_{V,{\rm max}}=10$. For the shock component, we run a small set of models with the Paris-Durham shock code\footnote{\href{https://ism.obspm.fr/shock.html}{https://ism.obspm.fr/shock.html}}, with density $n_{\rm H} = 10$ cm$^{-3}$, illumination factor $G_0=10$, extinction $A_V=1$, magnetic field strength $b = 0.1$, and velocities from 1--40 \kms\, in steps of 5 \kms. The velocity distribution of the shocks is assumed to follow an exponential decay: $f_s{\rm d}v_s \propto (1/\sigma_s){\rm exp}(-v_s/\sigma_s){\rm d}v_s$. All of the PDRs are assumed to have the same $G_0 = 10$.}

\begin{figure}
    \centering
    \includegraphics[width=\columnwidth]{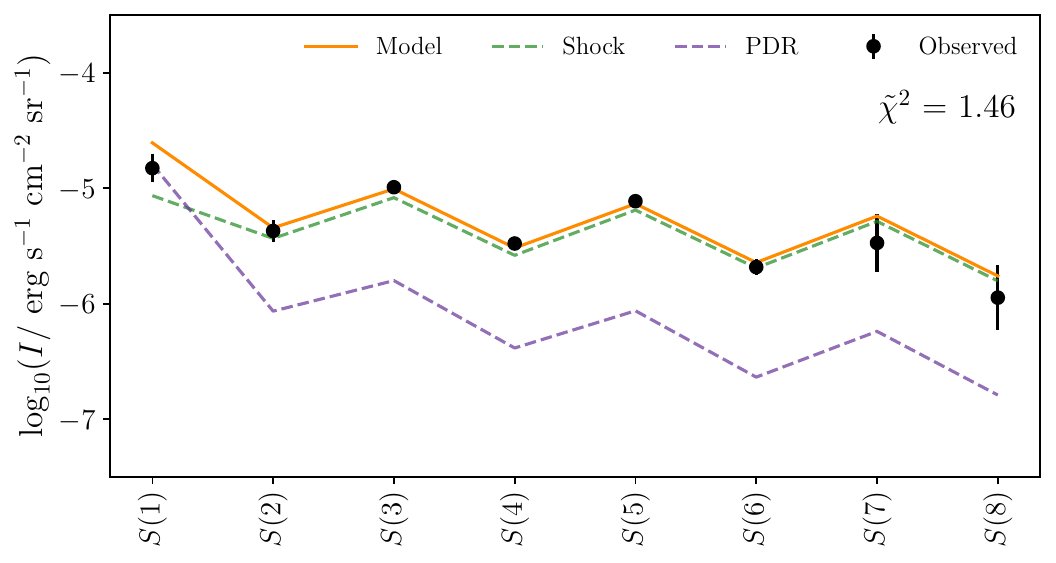}
    \caption{\rev{The H$_2$ line intensities integrated over the whole field of view are shown alongside a toy model that decomposes the emission into 2 components: shocks (green) and PDRs (purple). The total model is shown in orange. The individual rotational H$_2$ lines are labeled along the x-axis.  Shocks appear to be the predominant emission mechanism for most of the H$_2$ lines, though the PDR contribution is still important for producing a model that represents the data well, particularly in the $S(1)$ line.}}
    \label{fig:PDR_Shock_Models}
\end{figure}

\rev{Even with this relatively simple model decomposition, we are nevertheless able to reproduce the observed H$_2$ intensities quite well, within a reduced \rrev{$\chi^2 \sim 1.5$}. Figure \ref{fig:PDR_Shock_Models} shows the final model decomposition, which reveals that the majority of the H$_2$ emission is a result of shocks rather than PDRs (note the logarithmic y-axis).  On average, \rrev{$\sim 88$\%} of each line's emission comes from shocks, with the remaining \rrev{$\sim 12$\%} from PDRs.  The only line that has a significantly higher PDR contribution is the $S(1)$ line, with \rrev{$\sim 35$\%}. \rrev{We caution that the relative PDR/shock contributions are highly uncertain, because they are primarily informed by a single line ($S(1)$), which itself has a high flux uncertainty.  The PDR contributions for the $S(2)$--$S(8)$ lines may lie anywhere from 2--24\%, and the PDR contribution for $S(1)$ is constrained to be $\gtrsim$ 9\%. Despite these limitations, the large shock contributions provide} a potential explanation for the enhanced H$_2$-to-PAH ratios from Figure \ref{fig:L24_H2_PAH}, and \rrev{are} in line with the H$_2$-to-continuum ratios from Figure \ref{fig:L24_H2} if we assume $L_{24}$ is being enhanced by AGN emission. The velocity width for the shock distribution is $\sigma_s \simeq 4$ \kms, indicating a contribution primarily from low-velocity shocks.}

\rev{According to the morphology from Figure \ref{fig:L15_H2_map} and the model decomposition from Figure \ref{fig:PDR_Shock_Models}, it appears that the majority of H$_2$ emission is produced by low-velocity jet-induced shocks in the surrounding medium of the central galaxy. As a sanity check on this interpretation, we now look at the energy budget to confirm the plausibility that the observed level of H$_2$ emission could be powered by the AGN jets. We calculate the total mechanical power dissipated by a continuous distribution of shocks at different velocities following equation 6 in \citet{2024A&A...688A..96V}:}
\rev{\begin{equation}
    L_{\rm K} = \frac{1}{2}\rho_0\bigg(\int f_s(v_s)v_s^3{\rm d}v_s\bigg)D_A^2
\end{equation}}
\rev{where $\rho_0 = \mu m_pn_{\rm H}$ is the pre-shock density, $v_s$ is the shock velocity, $f_s(v_s)$ is the probability distribution over $v_s$, and $D_A$ is the angular diameter distance of the source. A mean molecular weight of $\mu \simeq 1.4$ is typical for these types of regions. Using this formula, we obtain \rrev{$L_{\rm K} \simeq (4$--$7) \times 10^{42}$ \ergs}.  The mechanical jet power, on the other hand, has been estimated from the power needed to inflate the large buoyant bubbles to the north and south of the nucleus by \citet{2019ApJ...885...63M}. They find a jet power of $L_{\rm jet} \simeq 10^{46}$ \ergs. In other words, the mechanical power dissipation from shocks represents only a tiny fraction of the available jet power: \rrev{$L_{\rm K}/L_{\rm jet} \simeq (0.04$--$0.07)$\%}.  This is in line with other analyses, which have found this fraction to not exceed $\sim 1$\% \citep{2022A&A...665L..11P}.
}

\rev{We can similarly check the total reprocessed UV luminosity that would be required to generate a distribution of PDRs in accordance with the modeled PDR contribution to the H$_2$ luminosity.  Following equation 8 in \citet{2024A&A...688A..96V}:}
\rev{\begin{equation}
    L_{\rm UV} = 1.92 \times 10^{-3} (G_0+1) \bigg(\int f_p(G_0){\rm d}G_0\bigg)D_A^2
\end{equation}}
\rev{where $G_0$ is the PDR illumination factor and $f_p(G_0)$ is the probability distribution over $G_0$ (which, in our case, is just a delta function). From this, we obtain \rrev{$L_{\rm UV} \simeq (1$--$6) \times 10^{45}$ \ergs}. We can compare this to the reprocessed UV luminosity observed as infrared emission in the SED.  The integrated 5--1000 $\mu$m luminosity from our CIGALE model, subtracting the AGN component, is $L_{\rm TIR} \simeq 3 \times 10^{46}$ \ergs. The observed value is higher by $\sim$an order of magnitude, which is to be expected when the emission is dominated by shocks.  A large fraction of the observed $L_{\rm TIR}$ likely \textit{does not} originate in reprocessed stellar light, but instead originates from dust that has been heated from shocks and direct radiation from the AGN.}

\subsubsection{\rev{Cooling H$_2$ from the Hot Atmosphere}}
\label{sec:h2_cool}

\begin{figure}
    \centering
    \includegraphics[width=\columnwidth]{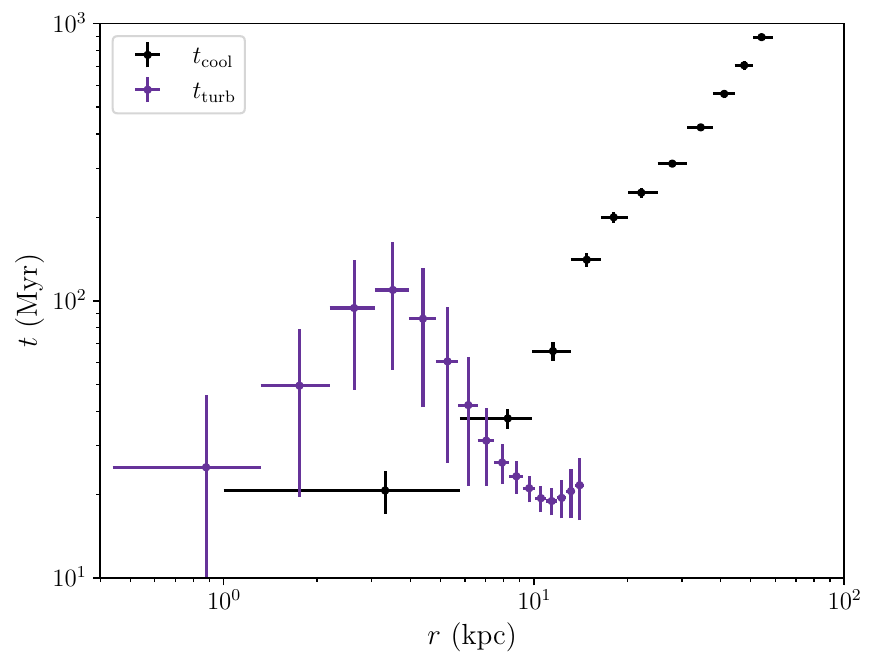}
    \caption{\rev{A radial profile of the turbulent dissipation time of the H$_2$ gas ($t_{\rm turb}$, in purple) and the cooling time of the hot gas ($t_{\rm cool}$, in black) in the core region of the cluster. The cooling time profile is taken from \citet{2019ApJ...885...63M}. $t_{\rm turb}$ rises above $t_{\rm cool}$ at distances $\lesssim 6$--$8$ kpc, indicating that turbulence may have a significant effect on cooling gas in the core of the BCG.}}
    \label{fig:tturb}
\end{figure}

\rev{In general, the cooling signal in the molecular phase, especially in a system with such a vigorous starburst, will be much weaker than the heating signal from stellar and shock ionization. However, we can analyze the cooling H$_2$ gas from an energetic and timescale perspective.}

\rev{The turbulent kinetic energy in the H$_2$ gas can be estimated from the mass $M_{\rm H_2}$ and velocity dispersion $\sigma_{\rm H_2}$ as
\begin{equation}
    E_{\rm H_2} = \frac{3}{2}M_{\rm H_2}\sigma_{\rm H_2}^2
\end{equation}
where the factor of 3 comes from converting the observed 1D velocity dispersion to a 3D velocity dispersion ($\sigma = \sqrt{3}\sigma_{\rm 1D}$). We can use this in conjunction with the observed H$_2$ luminosity, $L_{\rm H_2}$, to calculate the expected timescales for the turbulent energy to dissipate:
\begin{equation}
    t_{\rm turb} \equiv \frac{E_{\rm H_2}}{L_{\rm H_2}} = \frac{3M_{\rm H_2}\sigma_{\rm H_2}^2}{2L_{\rm H_2}}~.
\end{equation}
We calculate $t_{\rm turb}$ in a series of annuli centered on the point source nucleus, each with a width of 1 pixel (\ang[angle-symbol-over-decimal]{;;0.13}), extending out to a radius of $\sim\ang[angle-symbol-over-decimal]{;;2.3}$ or $\sim 15$ kpc.  We compare this to the cooling time of the hot gas, $t_{\rm cool}$, measured in \citet{2019ApJ...885...63M} in Figure \ref{fig:tturb}.}  

\rev{As $\sigma_{\rm H_2}$ is based on the line width of the H$_2$ profile, it represents an upper limit on the amount of turbulence that can be expected. Bulk flows and winds can also contribute to broadening in the observed line profiles through large apertures. Therefore, we interpret the turbulent dissipation timescale as an upper limit. At large radii, $t_{\rm turb} \ll t_{\rm cool}$, such that along the filaments, cooling gas from the hot atmosphere should be unaffected by turbulence. Moving inwards, there is a turnover point at around $\sim 6$--$8$ kpc where the two timescales are comparable. Inwards from this point, $t_{\rm turb} \gtrsim t_{\rm cool}$.  In this core region, turbulence may have a significant effect on the cooling and condensing of gas into molecular clouds.  Primarily, it will prevent the gravitational collapse of cold H$_2$ clouds into stars, lowering the SFR. Effectively, the actual time for the gas to cool and form stars will be dictated by the maximum of $t_{\rm cool}$ and $t_{\rm turb}$.  However, turbulence also likely plays a key role in enhancing cooling at hotter temperatures by fostering mixing between gas phases. 
% A ratio of $t_{\rm turb}/t_{\rm cool} \sim 1$ is suggestive of chaotic top-down condensation, as it indicates a correlation between the gas kinematics in the hot and cold phases. 
We note that this turnover point lies at the same radius as a cloud of rapidly cooling $10^{5.5}$ K [\ion{Ne}{6}] gas reported in R25, which is believed to be seeded by turbulence and mixing.  Taken as a whole, these timescale profiles, combined with the kinematic maps from Figure \ref{fig:h2s3}, support a picture in which the H$_2$ gas forms as a result of cooling along the filaments, flows inwards to the core, and a small fraction of it ($\lesssim 1$\% by mass) is (re)heated to a few 1000 K by turbulent mixing, shock heating, and stellar ionization, in the core region.}

\subsubsection{\rev{PAH Emission}}
\label{sec:pah_correlations}

\begin{figure}
    \centering
    \quad
    \includegraphics[width=\columnwidth]{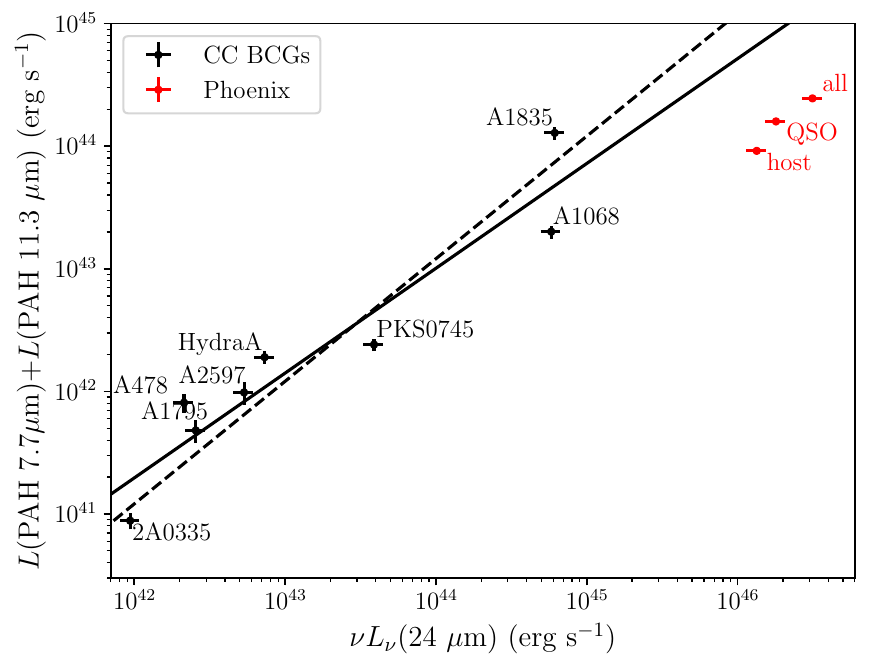}
    \caption{The combined 7.7 $\mu$m and 11.3 $\mu$m PAH luminosity as a function of the 24 $\mu$m continuum luminosity. The ensemble value for Phoenix is plotted against the D11 sample of BCGs. The solid line shows the best-fit power law, and the dashed line shows a slope of 1.}
    \label{fig:L24_pah}
\end{figure}

\begin{figure*}
    \centering
    \includegraphics[width=\textwidth]{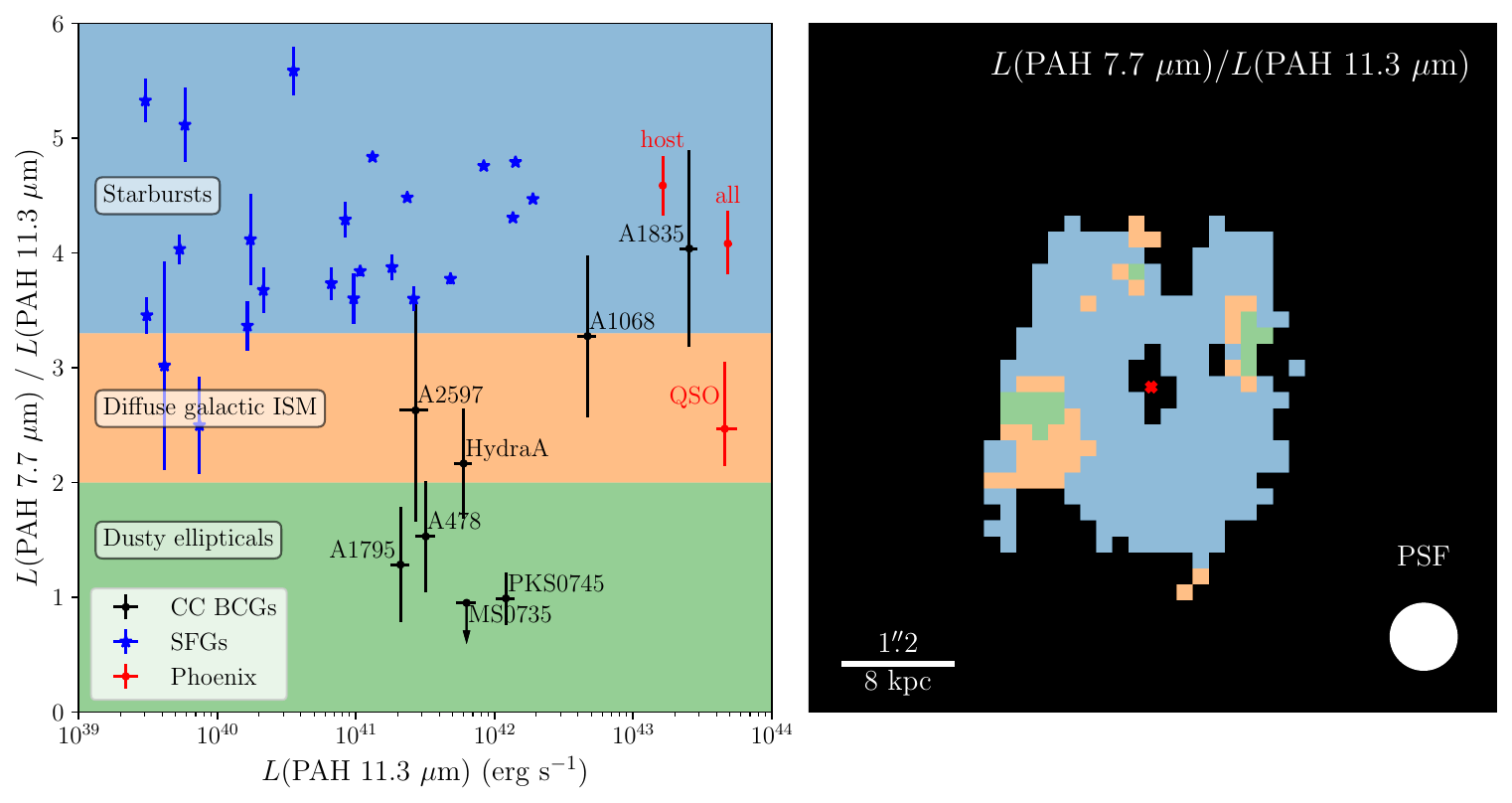}
    \caption{\textit{Left}: The ratio of the 7.7 $\mu$m PAH luminosity over the 11.3 $\mu$m PAH luminosity as a function of the 11.3 $\mu$m PAH luminosity. Different regions are colored according to the dominant ionization mechanisms expected to produce PAH ratios in this regime: starbursts (blue), diffuse galactic ISM (orange), and dusty ellipticals (green). The D11 sample is plotted in black, Phoenix is shown in red, and the SINGS star-forming galaxies are shown in blue. The data points are labeled identically to Figure \ref{fig:L24_H2}. \textit{Right}: A 2D map of the PAH luminosity ratio in each spaxel. Only spaxels where both PAHs are detected with an $S/N \geqslant 1$ are shown. The spaxels are colored according to which region they fall in on the left panel. A scale bar with physical and angular units is shown in the bottom left, and a circle displaying the size of the PSF FWHM is shown in the bottom right.}
    \label{fig:pah_ratios}
\end{figure*}

The PAH features are emitted by bending and stretching modes of the bonds between carbon and hydrogen atoms in very small dust grains \citep{1984A&A...137L...5L, 1998A&A...339..194B, 2000A&A...357.1013V}---these modes can be excited by the absorption of UV photons \citep{1985ApJ...290L..25A, 1999ApJ...513L..65S}, making PAH emission a strong tracer of the stellar ionization field \citep{2004ApJ...613..986P, 2006ApJ...653.1129B, 2004A&A...419..501F}. Thus, in typical star-forming galaxies where the IR continuum is primarily a result of reprocessed stellar light by larger dust grains, the PAH and continuum luminosity are observed to be strongly correlated with each other \citep{2010ApJ...723..895W}.  D11 found that such a correlation also exists in their sample of cool core BCGs, indicating that starlight is still the main contributor to the observed PAH luminosities, even in massive elliptical galaxies. We plot this relationship, recreating D11's Figure 8a, in Figure \ref{fig:L24_pah}. Phoenix noticeably falls below the trend seen in the other BCGs, even after subtracting the QSO PSF, by 2.5--3$\sigma$ (where $\sigma$ here is the scatter in the $L_{24}$-$L_{\rm PAH}$ relation). This may be due to \rev{contamination of the underlying continuum with non-stellar-reprocessed emission, and/or it may be the result of} PAH emission being suppressed by the presence of \rev{a hard radiation field from an} AGN, as has been shown to be the case in numerous studies \citep{1991MNRAS.248..606R, 2022ApJ...925..218X}.  \rev{Such hard radiation can also be emitted from the hot ICM in a cluster environment,} leading D11 to conclude that the IR-emitting PAHs seen in BCGs must somehow be shielded from the hot gas, perhaps through a layer of dusty gas with larger dust grains. Additionally, Phoenix's powerful AGN is heavily dust obscured, leading to the potential for localized pockets where PAHs can exist, shielded from both the AGN and the ICM.

We can constrain the ionization source \rev{of the PAHs} by considering the ratio of the 7.7 $\mu$m to 11.3 $\mu$m PAH features, $L_{7.7}/L_{11.3}$, which is sensitive primarily to the fraction of ionized and neutral PAHs in the system \citep{2007ApJ...657..810D}. Starburst systems with a large population of young, hot stars and a hard radiation field typically having elevated ratios of $\sim$3--7\citep{2007ApJ...656..770S}. The diffuse galactic ISM, with more evolved stars and an intermediate radiation field, has ratios between 2--3.3 \citep{2004ApJ...609..203S}. Finally, dusty elliptical galaxies, with only evolved stars and a soft radiation field, typically have much lower ratios between 1--2 \citep{2008ApJ...684..270K}. We examine $L_{7.7}/L_{11.3}$ in each spaxel and classify them into 3 categories, for simplicity: ``starbursts'' (young stars; $L_{7.7}/L_{11.3} > 3.3$), ``diffuse galactic ISM'' (middle-aged stars; $2 < L_{7.7}/L_{11.3} < 3.3$), and ``dusty ellipticals'' (old stars; $L_{7.7}/L_{11.3} < 2$). The results are shown in Figure \ref{fig:pah_ratios}. Compared with the D11 sample and the SINGS sample, we see that most of Phoenix A's host galaxy has ratios typical of starburst galaxies. However, there are a few pockets to the southeast and northwest of the nucleus where the ratio drops all the way into dusty elliptical territory. This may indicate that the cooling flow is more suppressed in these regions compared to their surroundings, preventing younger stars from forming and causing the primary ionization source to be old stars. Alternatively, these regions may have a higher dust obscuration around the young stars, but we do not see any coincidence between the pockets of low PAH ratios here and regions of enhanced dust obscuration in Figure \ref{fig:opticaldepth}. 

The ratio also drops significantly in the nucleus (as shown by the ``QSO'' data point), following the trends seen in \citet{2007ApJ...656..770S} for AGN-dominated systems.  \rev{The hard ionizing radiation from an AGN is thought to have a profound effect on PAHs, changing both their grain size distribution and ionization fraction. Ionized PAHs are preferentially destroyed relative to neutral PAHs, leading to lowered $L_{7.7}/L_{11.3}$ ratios. Recent \textit{JWST} results from the GATOS collaboration \citep{2024A&A...691A.162G} suggest that AGN radiation can have noticeable effects on the PAH ionization fraction even at kpc scales, particularly in regions showing signs of jet or outflow activity from the AGN.  This may provide an alternate explanation for the pockets of lowered $L_{7.7}/L_{11.3}$ ratios to the east and west of the nucleus, although these do not lie along the jet axis of the AGN in this system.  There is also an indirect effect on the PAH ratios caused by the suppression of star formation within the inner $\sim$1 kpc from the SMBH---}at this scale, the closer to the SMBH, the more effective heating becomes. This has been observed in simulations with substantial rises in the innermost entropy and turbulence \citep[e.g.][]{2020MNRAS.498.4983W}.

% \begin{figure}
%     \centering
%     \includegraphics[width=\columnwidth]{donahue_pah_NeII_spaxels.pdf}
%     \caption{The combined 7.7 $\mu$m and 11.3 $\mu$m PAH luminosity is shown as a function of the \neii\, luminosity. The points show values for individual spaxels.  Only spaxels that detect both PAH features to an $S/N \geqslant 1$ and the \neii\, line to an $S/N \geqslant 3$ are shown.}
%     \label{fig:PAH_NeII}
% \end{figure}

% Additionally, we see a strong link between PAH emission and \neii\, emission (Figure \ref{fig:PAH_NeII}), with an $r = 0.81$.  Since \neii\, is also strongly linked with H$_2$ (Figure \ref{fig:NeII_H2}), this creates a direct correspondence between the excitation of the PAHs, the molecular gas, and the warm ionized gas, with star formation being the primary mechanism (as shown by the PAH ratios and the $L_{{\rm H}_2}$/$L_{15}$ ratio).

\subsubsection{Star Formation Rates}

\begin{figure*}
    \centering
    \quad
    \includegraphics[width=\columnwidth]{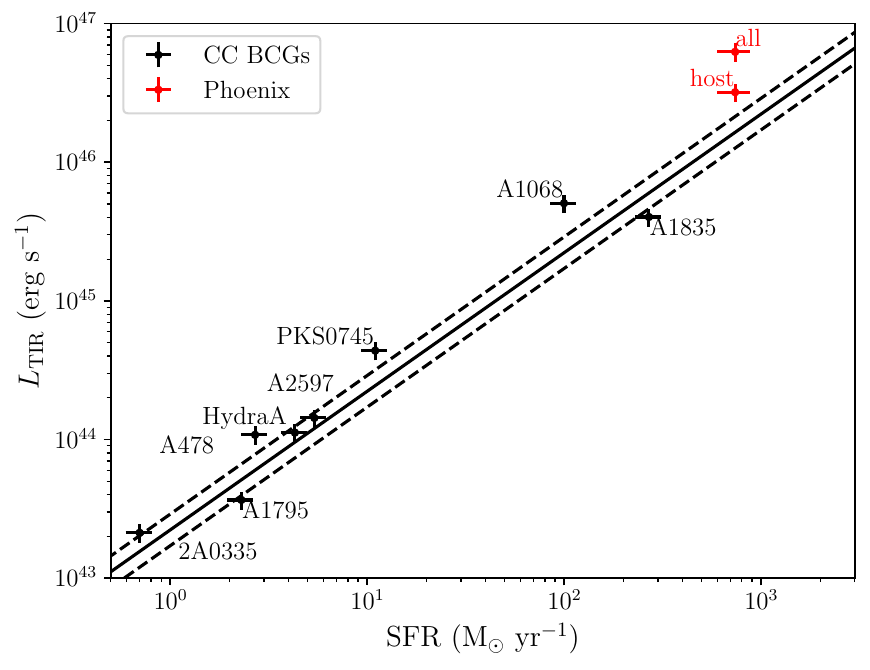}
    \includegraphics[width=\columnwidth]{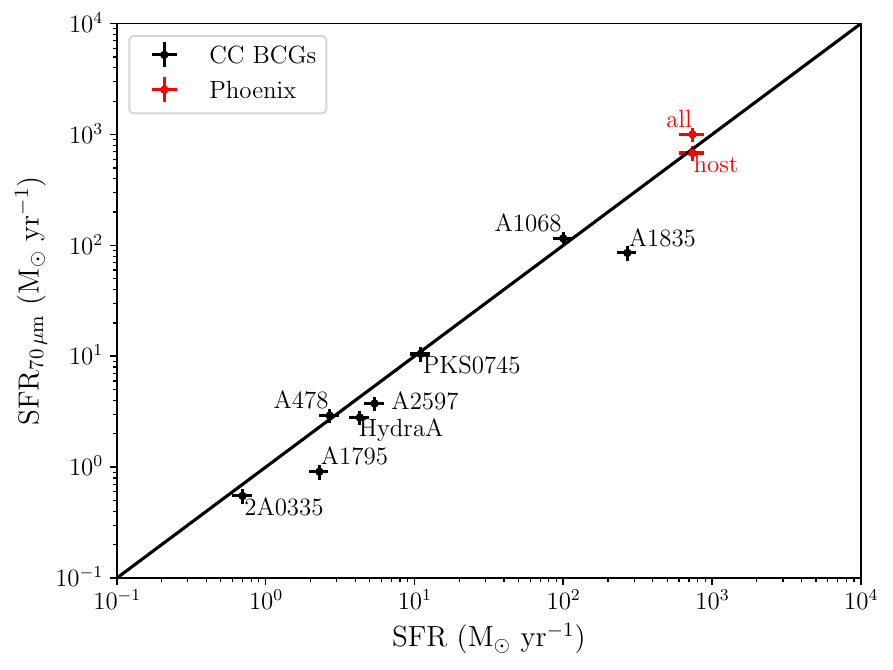}
    \caption{\textit{Left}: The total infrared luminosity $L_{\rm TIR}$ integrated from 5--1000 $\mu$m, as a function of the modeled SFR. The solid line shows the \citet{1998ARA&A..36..189K} relation, and the dashed lines show a 30\% scatter on that relation. \textit{Right}: The SFR estimated from the monochromatic 70 $\mu$m luminosity as a function of the modeled SFR. The solid line shows the \citet{2010ApJ...714.1256C} relation.}
    \label{fig:SFR_LTIR}
\end{figure*}

The literature has many calibrated star formation rate correlations with host galaxy properties, including integrated continuum luminosities, single-band continuum luminosities, and emission line luminosities. These correlations are applicable only in certain circumstances, so they can be compared and taken advantage of to infer whether these conditions hold. Here, we analyze SFRs from a number of different indicators in comparison to our modeled SFR from the SED ($\S$\ref{sec:stellarpop}), which we take to be the ``true'' (read: most accurate) value.

Firstly, the SFR can be estimated from the total infrared luminosity $L_{\rm TIR}$ (integrated from 5--1000 $\mu$m) using \citet{1998ARA&A..36..189K}, and the 70 $\mu$m luminosity $L_{70} \equiv \nu L_\nu(70\,{\rm \mu m})$ using \citet{2010ApJ...714.1256C}. We recreate D11's Figure 4 in Figure \ref{fig:SFR_LTIR}, showing both of these correlations for Phoenix, using $L_{\rm TIR}$ and $L_{70}$ obtained from the SED fitting. We notice that the $L_{\rm TIR}$ correlation in particular would vastly overpredict the SFR at $>2000$ \msunyr\, if no AGN correction is accounted for.  However, it lines up fairly well after subtracting the AGN contribution. These discrepancies confirm, as we saw in the SED modeling, that the AGN, despite being heavily obscured in the optical, accounts for a significant fraction of the underlying continuum luminosity throughout the whole infrared range. This includes contributions directly from the accretion disk itself, those which get absorbed and re-emitted by dust in the torus and the polar regions, and synchrotron emission (at longer wavelengths) from the bipolar radio jets.

\begin{figure*}
    \centering
    \quad
    \includegraphics[width=\columnwidth]{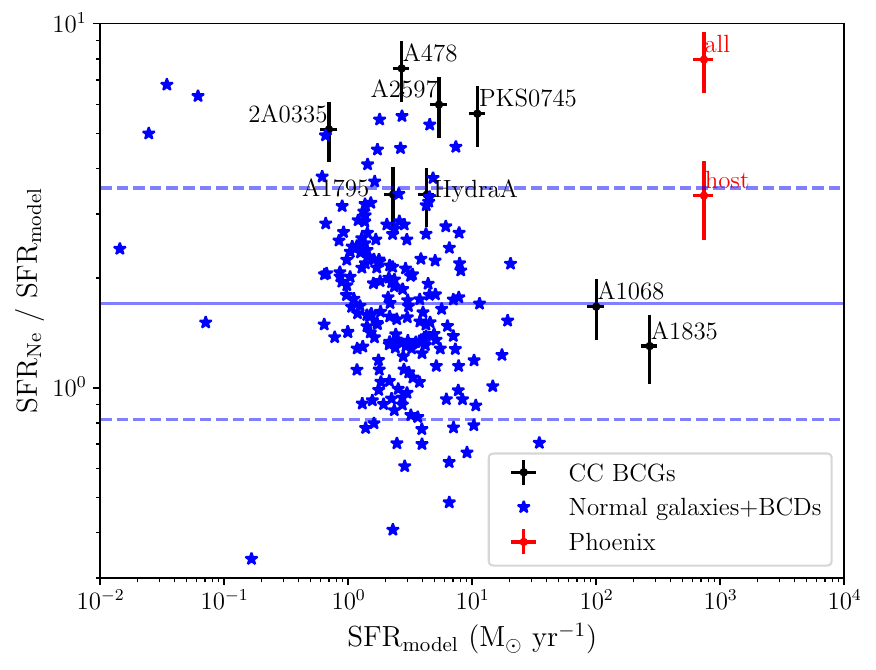}
    \includegraphics[width=\columnwidth]{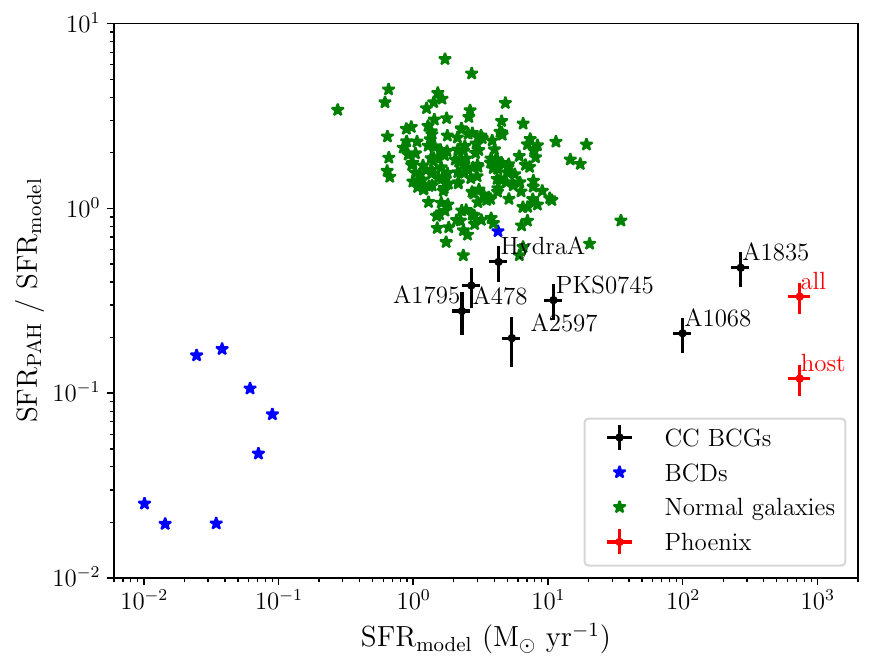}
    \caption{\textit{Left}: A comparison between the SFR estimated from the [\ion{Ne}{2}] and [\ion{Ne}{3}] line luminosities and the SFR modeled from the SED. The data points for the BCGs are in black, Phoenix is in red, and \citet{2019ApJ...884..136X} galaxies are in blue. For \citet{2019ApJ...884..136X} galaxies, ``SFR$_{\rm model}$'' refers to SFRs from the MPA/JHU catalogue. The solid blue line shows the average value of the \citet{2019ApJ...884..136X} galaxies, while the dashed blue line shows the 1-$\sigma$ standard deviation. \textit{Right}: A comparison between the SFR estimated from the 7.7 $\mu$m and 11.3 $\mu$m PAH luminosities and the SFR \rev{estimated from SED modeling}. The colors here are the same, except that the \citet{2019ApJ...884..136X} galaxies are split up into BCDs (in blue) and non-BCDs (in green). The BCDs here do \textit{not} use the different scaling relation parameters for galaxies with $M_* < 10^9$ \msun\, as suggested by \citet{2019ApJ...884..136X}, to highlight the apparent deficit in PAHs in these systems.}
    \label{fig:SFR_Ne}
\end{figure*}

The SFR can also be measured from the [\ion{Ne}{2}] and [\ion{Ne}{3}] lines \citep{2007ApJ...658..314H}. Following D11, we take Ne$^+$/Ne $= 0.75$ and Ne$^{2+}$/Ne $= 0.15$. We recreate their Figure 12 in Figure \ref{fig:SFR_Ne} (left). We include the sample of star-forming galaxies used by \citet{2019ApJ...884..136X} as a reference point---for these galaxies, ``SFR$_{\rm model}$'' refers to SFRs obtained from the MPA/JHU catalogue\footnote{\href{https://wwwmpa.mpa-garching.mpg.de/SDSS/DR7/}{https://wwwmpa.mpa-garching.mpg.de/SDSS/DR7/}}. After subtracting the contribution to the [\ion{Ne}{2}] and [\ion{Ne}{3}] lines from AGN photoionization, we see that the ``host'' point lies close to the expected value for SFGs (solid line), \rev{but is still marginally enhanced, lying closer to the $+1\sigma$ line (dashed blue line). This may be due to enhanced emission from shocks, as we have seen that the H$_2$ emission in this system is primarily emitted by shocks (Figure \ref{fig:PDR_Shock_Models}), and the H$_2$ and [\ion{Ne}{2}] emission are correlated (Figure \ref{fig:NeII_H2}).} Unfortunately, we cannot compare with the H$_2$-calibrated SFR since it relies on the $S(0)$ line, which falls outside of the wavelength coverage of MIRI at Phoenix's redshift. But considering the strong correlation between the H$_2$ and [\ion{Ne}{2}] luminosities, we can infer that we would likely see good agreement with the modeled SFR value here as well.

Additionally, we can look at the SFR inferred from the PAH features, using the relation calibrated by \citet{2019ApJ...884..136X}. For each BCG, we take an average of the SFRs inferred from the 7.7 $\mu$m and 11.3 $\mu$m PAH features. Comparing these SFRs to the modeled values in Figure \ref{fig:SFR_Ne} (right) shows a systematic deficit, with the PAH-based SFRs being \rev{40--80\%} lower. In this panel, we also show the \citet{2019ApJ...884..136X} sample of galaxies, but now split up between blue compact dwarfs (BCDs) in blue and normal galaxies in green. The normal galaxies have a typical PAH-based SFR in agreement with the \rev{modeled} SFR. However, we see a decline in both the BCDs and BCGs.  \citet{2019ApJ...884..136X} address this for the BCD population by adopting a different scaling offset for galaxies with $M_* < 10^9$ \msun---we have not included this correction in \rev{Figure \ref{fig:SFR_Ne}} to highlight the apparent PAH deficit in these galaxies and to show the similarities they have with the BCGs. Phoenix, in particular, has a PAH-based SFR almost an order of magnitude smaller than its modeled value. This may be evidence for a PAH deficit in the high-mass galaxy population similar to the deficit in the BCDs: the \citet{2019ApJ...884..136X} relation we utilize is calibrated for the general galaxy population with stellar masses between $10^9$--$10^{11.4}$ \msun, whereas the BCGs we examine here all have stellar masses $\gtrsim 10^{11}$ \msun. \rev{This is reinforced by the results of our CIGALE SED models, in which the PAH mass fraction parameter ($q_{\rm PAH}$) is driven to the lowest possible value of 0.47\%, an order of magnitude lower than a typical Milky-Way-like galaxy with $\sim$4.6\% \citep{2007ApJ...657..810D}.} The cluster environment may have a significant effect on the survivability of PAHs: collisions with high-energy electrons from the ICM (and an AGN, if active) can sputter and destroy the dust grains on extremely short timescales.  D11 argue, using the analysis of \citet{2010A&A...510A..37M}, that PAHs embedded in a 1 keV, 0.1 cm$^{-3}$ plasma could not survive longer than a few hundred years.  However, dust can also locally enhance cooling, nurturing the formation of cold gas clouds around the dust that act as a protective shield against the energetic electrons and ions in the ICM, allowing the dust to survive longer in these localized cooling regions.  \citet{2010A&A...510A..37M} estimate that for conditions similar to the Orion nebula (7000 K, $10^4$ cm$^{-3}$), the PAH survivability is much higher, at roughly 10 Myr.  The feeding and feedback cycle of the AGN may also play an important role in periodically refreshing the dust content of the ICM by blowing out dust from the nucleus during outbursts and allowing for more dust formation (or, rather, lessened dust sputtering) during periods of inactivity.

\subsection{An Undermassive Black Hole?} \label{sec:discuss}

With the SFRs of order $\sim$1000 \msunyr\, and molecular gas masses of order $\sim$10$^{10}$ \msun\, pointing to massive amounts of cooling gas in the Phoenix cluster, the question then becomes: What makes this cluster unique?  Why is this cluster able to host a massive cooling flow while every other observed cluster cannot?  The simplest explanation is that there is nothing particularly special about the Phoenix cluster, but we just happened to be observing it at a time when it is undergoing a short-lived cooling spike that will not last much longer than $\sim$a few Myr (as we proposed in R25). \rev{This scenario is supported by simulations at both the cluster scale \citep{2017MNRAS.466..677G} and black-hole scale \citep{2019MNRAS.483.3488B}, which predict that feeding and feedback follow an episodic cycle.} If indeed all clusters undergo \rev{a} phase of rapid cooling once throughout their lifetimes, then we should expect to see about 1 in every 10,000 clusters \citep[which is roughly the amount of clusters known---e.g.][]{2024A&A...690A.322K} experiencing such an event. This seems to line up with  expectations, but it becomes problematic when we consider that clusters should still exhibit post-starburst signatures long after they have experienced these rapid cooling spikes.  These features are generally not seen, even in cool core clusters, and are more often associated with other processes such as galaxy interactions, mergers, and ram-pressure stripping \citep{2022ApJ...930...43W}.  Therefore, it seems that we can rule out the presupposition that there is \textit{nothing} special at all about the Phoenix cluster.

Another idea, previously explored in \citet{2018ApJ...858...45M} and \citet{2019ApJ...885...63M}, is that the Phoenix cluster hosts an undermassive SMBH relative to the size of its cool core, which is expected to occur in the most massive galaxy clusters.  In this scenario, rapid cooling drives the accretion rate of the AGN to high Eddington ratios ($\dot{M}/\dot{M}_{\rm Edd} \lesssim 5\%$).  But at these higher Eddington ratios, much of the power output of the AGN comes in the form of radiative feedback as opposed to mechanical feedback, which is less effective at suppressing cooling and may even enhance it through the inverse Compton effect close to the AGN \citep{2019ApJ...885...63M}. Indeed, \citet{2013MNRAS.432..530R} showed that mechanical power output tends to become saturated at $\sim$1\% of the Eddington luminosity, while radiative power continues to grow with the accretion rate.  

While there are currently only weak constraints on the mass of Phoenix's black hole, it is confirmed to show signs of both radiative \citep[bright SEDs; $\S$\ref{sec:stellarpop};][]{2012Natur.488..349M} and mechanical \citep[X-ray cavities, radio jets;][]{2015ApJ...805...35H, 2021A&A...646A..38T} feedback.  Estimates of the mechanical power output place it at 0.5--1$\times 10^{46}$ \ergs\, \citep{2019ApJ...885...63M}, while we can estimate from our SED fitting in $\S$\ref{sec:stellarpop} a radiative power output of $L_{\rm bol} = 3 \times 10^{46}$ \ergs.  The fact that these are similar in magnitude suggests that the AGN has only recently undergone (or perhaps is still undergoing) a transition in feedback modes \citep{2015ApJ...811..111M}.  In addition, the mechanical power is \textit{just} high enough to balance with the cooling luminosity of $\sim$10$^{46}$ \ergs\, \citep{2019ApJ...885...63M}, and it corresponds to the mechanical feedback limit of a $\sim$10$^{10}$ \msun\, black hole.  If Phoenix's black hole has only recently experienced a period of rapid growth up to $10^{10}$ \msun\,, this would explain the puzzlingly high rates of cooling despite the black hole's current power output being seemingly large enough that it should offset this cooling.

As the cluster continues to evolve, in this picture, we should expect the cooling to begin being regulated more efficiently by the recently grown black hole.  Now that the mechanical power output has reached levels sufficient to balance cooling, the cooling rates, star formation rates, and accretion rate of the SMBH should all slow down until they reach roughly the percent level of the classical cooling rate.  In fact, we can estimate from the bolometric AGN luminosity that the black hole's current accretion rate is $\dot{M} = L_{\rm bol}/\eta c^2 \approx 5$ \msunyr\, (assuming an efficiency $\eta = 0.1$), which is about an order of magnitude lower than previously thought \citep[due to uncertain X-ray-to-bolometric luminosity corrections;][]{2012Natur.488..349M}. This may indeed be an indication that the black hole's growth has now slowed after reaching a mass sufficient to balance cooling. This trend will ultimately drive Phoenix towards Abell 1835 and Abell 1068 on many of our figures, decreasing dramatically in $L_{24}$ as the AGN radiative power output diminishes, and decreasing less dramatically in other quantities ($L_{{\rm H}_2}$, $L_{\rm [Ne II]}$, $L_{\rm PAH}$).  The ratio of the 7.7 $\mu$m to 11.3 $\mu$m PAH features will also decrease with the SFR, trending towards values more typical of the diffuse galactic ISM.  The PAH-derived SFRs, however, will likely stay suppressed relative to other SFR indicators, as this seems to be largely due to the influence of the ICM and is not a feature specific to Phoenix.  Whether this scenario holds up in reality would largely be determined with more precise constraints on the SMBH mass.

\section{Conclusion} \label{sec:conc}

Since its discovery, the Phoenix Cluster has stood out as a unique system, even among cool core clusters, due to its hosting of the most massive and efficient cooling flow in the known universe \citep{2012Natur.488..349M}. The results of our analysis of the molecular gas, dust, and star formation from MIRI/MRS data have only further exemplified this uniqueness. We list the important results of our analysis, in sequence, below:

\begin{enumerate}
    \item We have \rev{measured} the optical depth due to silicate absorption in the warm molecular line-emitting gas based on fitting excitation models of the rotational H$_2$ lines\rev{, finding that the silicate dust traces cooling regions and has a similar morphology to other types of dust}.
    
    \item Using this technique in conjunction with the LOKI spectral modeling software, we have measured the total molecular gas mass in the Phoenix cluster to be \rrev{$1.9^{+0.5}_{-0.4} \times 10^{10}$ \msun}, in agreement with previous estimates made using CO emission. We have estimated the CO-to-H$_2$ conversion factor to be \rrev{$\alpha_{\rm CO} = 0.8 \pm 0.2 M_\odot\,{\rm pc}^{-2}\,({\rm K}\,{\rm km}\,{\rm s}^{-1})^{-1}$}.
    
    \item We have filled in the MIR gap in Phoenix's SED and obtained refined measurements of the stellar mass, $M_* = 2.6 \pm 0.5 \times 10^{12}$ \msun, and star formation rates, $\langle{\rm SFR}\rangle_{\rm 10} = 1340 \pm 100$ \msunyr\, and $\langle{\rm SFR}\rangle_{\rm 100} = 740 \pm 80$ \msunyr, in agreement within $\pm$1$\sigma$ of previous estimates based on a number of different scaling relations \citep{2012Natur.488..349M}.

    \item \rev{We have developed a complete picture to explain the observed H$_2$ emission in the Phoenix cluster, motivated by a comprehensive analysis of the kinematics, energetics, and morphology.  The hot atmosphere cools and condenses along the filaments at distances $\gtrsim 8$ kpc, before flowing inwards towards the core.  In the core region $\lesssim 8$ kpc, the molecular gas encounters turbulence from the AGN's radio jets and stellar feedback, which slows down the cooling process by $\sim$an order of magnitude. The molecular gas is (re)heated by stellar radiation, shocks, and potentially by mixing with hotter gas phases, all of which also contribute towards enhancing the H$_2$ luminosity per unit mass, decreasing $\alpha_{\rm CO}$.}
    
    % \item We have found, using correlations between different MIR emission features, that star formation---and in particular young, hot stars formed via the cooling flow---in the Phoenix Cluster's BCG plays a major role in the excitation and heating of the $\sim 100$ K molecular H$_2$ gas, the $\sim 10^4$ K ionized [\ion{Ne}{2}]/[\ion{Ne}{3}] gas, the dust grains responsible for the IR continuum, and the smaller PAH grains responsible for the broad PAH features. This is in contrast to other cool cores, which require an additional source of particle heating.
    
    % \item The other primary heating mechanism is the AGN at the center of the BCG. This also contributes substantially to the heating of the IR continuum, the ionized gas, as well as the hotter phases (see i.e. R25).
    
    % \item Other heating mechanisms, such as cosmic rays, suprathermal ICM particles, and magnetohydrodynamic waves, may also be present, but at suppressed importance relative to the general cool core population. 
    
    \item We have uncovered an apparent PAH deficit in the $\gtrsim 10^{11}\,\msun$ cool core BCG population, causing their PAH-based SFRs to be underpredicted by up to an order of magnitude. This is distinct from the deficit found in galaxies with strong AGN like Phoenix, as the other BCGs in the sample all show little to no signs of AGN activity. This may be a result of suprathermal ICM particle heating, creating a harsh environment unsuitable for the sustained existence of PAH grains.

    % \item \rev{We have revisited the proposed scenario in which Phoenix contains an undermassive central black hole relative to the size of its halo. This would provide a neat explanation for the extreme levels of cooling, despite the AGN having seemingly enough power to balance this cooling, with a mechanical power of $L_{\rm jet} \simeq 10^{46}$ \ergs\, and a bolometric radiative power of $L_{\rm bol} \simeq 3 \times 10^{46}$ \ergs.}

    \item \rev{We have updated the measurement of the bolometric luminosity of Phoenix's AGN, finding $L_{\rm bol} = 3 \times 10^{46}$ \ergs, an order of magnitude lower than previous estimates based on an X-ray-to-bolometric correction \citep{2012Natur.488..349M}.  The fact that this is comparable to the mechanical power output from the jets ($L_{\rm jet} \simeq 10^{46}$ \ergs) suggests a transition in feedback modes is currently underway and lends credence to the theory that Phoenix hosts an undermassive central black hole relative to the size of its halo.}
    
\end{enumerate}

\rev{Taken together, our analyses} provide even more evidence to suggest that the Phoenix cluster is experiencing a highly efficient, but still turbulent and chaotic, top-down cascade of condensation and cooling from the hot ICM down to the cold molecular phase. \rev{We are seeing the effects of this cooling propagate through the young stellar populations, and then feeding back into the gas, dust, and PAHs through the interstellar radiation field, shocks, and turbulence.}

\begin{acknowledgments}
This work is based on observations with the NASA/ESA/CSA James Webb Space Telescope obtained from the Data Archive at the Space Telescope Science Institute, which is operated by the Association of Universities for Research in Astronomy, Incorporated, under NASA contract NAS 5-03127. Support for Program number JWST-GO-02439.001-A was provided through a grant from the STScI under NASA contract NAS 5-03127.
M Reefe acknowledges support from the National Science Foundation Graduate Research Fellowship under Grant No. 2141064.  
M Chatzikos acknowledges support from NASA (19-ATP19-0188, 22-ADAP22-0139) and NSF (1910687).
MG acknowledges support from the ERC Consolidator Grant \textit{BlackHoleWeather} (101086804).
HRR acknowledges an Anne McLaren Fellowship from the University of Nottingham.
\end{acknowledgments}

\vspace{5mm}
\facilities{\textit{JWST}}
\software{\texttt{julia} \citep{doi:10.1137/141000671}, \texttt{LOKI.jl} \citep[][\href{https://github.com/Michael-Reefe/Loki.jl}{https://github.com/Michael-Reefe/Loki.jl}]{reefe_2025_15069243}, \texttt{python} \citep{10.5555/1593511}, \texttt{numpy} \citep{2020Natur.585..357H}, \texttt{scipy} \citep{2020NatMe..17..261V}, \texttt{astropy} \citep{2013A&A...558A..33A}, \texttt{matplotlib} \citep{Hunter:2007}, \texttt{SAOImage DS9} \citep{2000ascl.soft03002S}}

\vspace{5mm}
All \textit{JWST} data used in this paper can be found in MAST: \dataset[10.17909/pc04-3145]{http://dx.doi.org/10.17909/pc04-3145}.  24 $\mu$m fluxes for the \citet{2007ApJ...667..149F} sample were obtained with NED \citep{https://doi.org/10.26132/ned1}.

% \appendix

\bibliography{main}{}

\begin{thebibliography}{}
\expandafter\ifx\csname natexlab\endcsname\relax\def\natexlab#1{#1}\fi
\providecommand{\url}[1]{\href{#1}{#1}}
\providecommand{\dodoi}[1]{doi:~\href{http://doi.org/#1}{\nolinkurl{#1}}}
\providecommand{\doeprint}[1]{\href{http://ascl.net/#1}{\nolinkurl{http://ascl.net/#1}}}
\providecommand{\doarXiv}[1]{\href{https://arxiv.org/abs/#1}{\nolinkurl{https://arxiv.org/abs/#1}}}

\bibitem[{{Allamandola} {et~al.}(1985){Allamandola}, {Tielens}, \& {Barker}}]{1985ApJ...290L..25A}
{Allamandola}, L.~J., {Tielens}, A.~G.~G.~M., \& {Barker}, J.~R. 1985, \apjl, 290, L25, \dodoi{10.1086/184435}

\bibitem[{{Allen}(1995)}]{1995MNRAS.276..947A}
{Allen}, S.~W. 1995, \mnras, 276, 947, \dodoi{10.1093/mnras/276.3.947}

\bibitem[{{Astropy Collaboration} {et~al.}(2013){Astropy Collaboration}, {Robitaille}, {Tollerud}, {Greenfield}, {Droettboom}, {Bray}, {Aldcroft}, {Davis}, {Ginsburg}, {Price-Whelan}, {Kerzendorf}, {Conley}, {Crighton}, {Barbary}, {Muna}, {Ferguson}, {Grollier}, {Parikh}, {Nair}, {Unther}, {Deil}, {Woillez}, {Conseil}, {Kramer}, {Turner}, {Singer}, {Fox}, {Weaver}, {Zabalza}, {Edwards}, {Azalee Bostroem}, {Burke}, {Casey}, {Crawford}, {Dencheva}, {Ely}, {Jenness}, {Labrie}, {Lim}, {Pierfederici}, {Pontzen}, {Ptak}, {Refsdal}, {Servillat}, \& {Streicher}}]{2013A&A...558A..33A}
{Astropy Collaboration}, {Robitaille}, T.~P., {Tollerud}, E.~J., {et~al.} 2013, \aap, 558, A33, \dodoi{10.1051/0004-6361/201322068}

\bibitem[{{Beckmann} {et~al.}(2019){Beckmann}, {Devriendt}, \& {Slyz}}]{2019MNRAS.483.3488B}
{Beckmann}, R.~S., {Devriendt}, J., \& {Slyz}, A. 2019, \mnras, 483, 3488, \dodoi{10.1093/mnras/sty2890}

\bibitem[{Bezanson {et~al.}(2017)Bezanson, Edelman, Karpinski, \& Shah}]{doi:10.1137/141000671}
Bezanson, J., Edelman, A., Karpinski, S., \& Shah, V.~B. 2017, SIAM Review, 59, 65, \dodoi{10.1137/141000671}

\bibitem[{{Bicknell} {et~al.}(2000){Bicknell}, {Sutherland}, {van Breugel}, {Dopita}, {Dey}, \& {Miley}}]{2000ApJ...540..678B}
{Bicknell}, G.~V., {Sutherland}, R.~S., {van Breugel}, W. J.~M., {et~al.} 2000, \apj, 540, 678, \dodoi{10.1086/309343}

\bibitem[{{Bolatto} {et~al.}(2013){Bolatto}, {Wolfire}, \& {Leroy}}]{2013ARA&A..51..207B}
{Bolatto}, A.~D., {Wolfire}, M., \& {Leroy}, A.~K. 2013, \araa, 51, 207, \dodoi{10.1146/annurev-astro-082812-140944}

\bibitem[{{Boquien} {et~al.}(2019){Boquien}, {Burgarella}, {Roehlly}, {Buat}, {Ciesla}, {Corre}, {Inoue}, \& {Salas}}]{2019A&A...622A.103B}
{Boquien}, M., {Burgarella}, D., {Roehlly}, Y., {et~al.} 2019, \aap, 622, A103, \dodoi{10.1051/0004-6361/201834156}

\bibitem[{{Bordoloi} {et~al.}(2017){Bordoloi}, {Wagner}, {Heckman}, \& {Norman}}]{2017ApJ...848..122B}
{Bordoloi}, R., {Wagner}, A.~Y., {Heckman}, T.~M., \& {Norman}, C.~A. 2017, \apj, 848, 122, \dodoi{10.3847/1538-4357/aa8e9c}

\bibitem[{{Boulanger} {et~al.}(1998){Boulanger}, {Boisssel}, {Cesarsky}, \& {Ryter}}]{1998A&A...339..194B}
{Boulanger}, F., {Boisssel}, P., {Cesarsky}, D., \& {Ryter}, C. 1998, \aap, 339, 194

\bibitem[{{Brandl} {et~al.}(2006){Brandl}, {Bernard-Salas}, {Spoon}, {Devost}, {Sloan}, {Guilles}, {Wu}, {Houck}, {Weedman}, {Armus}, {Appleton}, {Soifer}, {Charmandaris}, {Hao}, {Higdon}, {Marshall}, \& {Herter}}]{2006ApJ...653.1129B}
{Brandl}, B.~R., {Bernard-Salas}, J., {Spoon}, H.~W.~W., {et~al.} 2006, \apj, 653, 1129, \dodoi{10.1086/508849}

\bibitem[{{Calura} {et~al.}(2008){Calura}, {Pipino}, \& {Matteucci}}]{2008A&A...479..669C}
{Calura}, F., {Pipino}, A., \& {Matteucci}, F. 2008, \aap, 479, 669, \dodoi{10.1051/0004-6361:20078090}

\bibitem[{{Calzadilla} {et~al.}(2023){Calzadilla}, {Bleem}, {McDonald}, {Gladders}, {Mantz}, {Allen}, {Bayliss}, {Eilers}, {Floyd}, {Hlavacek-Larrondo}, {Khullar}, {Kim}, {Mahler}, {Sharon}, {Somboonpanyakul}, {Stalder}, {Stark}, \& {SPT Collaboration}}]{2023ApJ...947...44C}
{Calzadilla}, M.~S., {Bleem}, L.~E., {McDonald}, M., {et~al.} 2023, \apj, 947, 44, \dodoi{10.3847/1538-4357/acc6c2}

\bibitem[{{Calzetti} {et~al.}(2010){Calzetti}, {Wu}, {Hong}, {Kennicutt}, {Lee}, {Dale}, {Engelbracht}, {van Zee}, {Draine}, {Hao}, {Gordon}, {Moustakas}, {Murphy}, {Regan}, {Begum}, {Block}, {Dalcanton}, {Funes}, {Gil de Paz}, {Johnson}, {Sakai}, {Skillman}, {Walter}, {Weisz}, {Williams}, \& {Wu}}]{2010ApJ...714.1256C}
{Calzetti}, D., {Wu}, S.~Y., {Hong}, S., {et~al.} 2010, \apj, 714, 1256, \dodoi{10.1088/0004-637X/714/2/1256}

\bibitem[{{Canizares} {et~al.}(1988){Canizares}, {Markert}, \& {Donahue}}]{1988ASIC..229...63C}
{Canizares}, C.~R., {Markert}, T.~H., \& {Donahue}, M.~E. 1988, in NATO Advanced Study Institute (ASI) Series C, Vol. 229, Cooling Flows in Clusters and Galaxies, ed. A.~C. {Fabian}, 63, \dodoi{10.1007/978-94-009-2953-1_7}

\bibitem[{{Crawford} {et~al.}(1999){Crawford}, {Allen}, {Ebeling}, {Edge}, \& {Fabian}}]{1999MNRAS.306..857C}
{Crawford}, C.~S., {Allen}, S.~W., {Ebeling}, H., {Edge}, A.~C., \& {Fabian}, A.~C. 1999, \mnras, 306, 857, \dodoi{10.1046/j.1365-8711.1999.02583.x}

\bibitem[{{Dale} {et~al.}(2007){Dale}, {Gil de Paz}, {Gordon}, {Hanson}, {Armus}, {Bendo}, {Bianchi}, {Block}, {Boissier}, {Boselli}, {Buckalew}, {Buat}, {Burgarella}, {Calzetti}, {Cannon}, {Engelbracht}, {Helou}, {Hollenbach}, {Jarrett}, {Kennicutt}, {Leitherer}, {Li}, {Madore}, {Martin}, {Meyer}, {Murphy}, {Regan}, {Roussel}, {Smith}, {Sosey}, {Thilker}, \& {Walter}}]{2007ApJ...655..863D}
{Dale}, D.~A., {Gil de Paz}, A., {Gordon}, K.~D., {et~al.} 2007, \apj, 655, 863, \dodoi{10.1086/510362}

\bibitem[{{David} {et~al.}(2001){David}, {Nulsen}, {McNamara}, {Forman}, {Jones}, {Ponman}, {Robertson}, \& {Wise}}]{2001ApJ...557..546D}
{David}, L.~P., {Nulsen}, P.~E.~J., {McNamara}, B.~R., {et~al.} 2001, \apj, 557, 546, \dodoi{10.1086/322250}

\bibitem[{{Donahue} {et~al.}(2011){Donahue}, {de Messi{\`e}res}, {O'Connell}, {Voit}, {Hoffer}, {McNamara}, \& {Nulsen}}]{2011ApJ...732...40D}
{Donahue}, M., {de Messi{\`e}res}, G.~E., {O'Connell}, R.~W., {et~al.} 2011, \apj, 732, 40, \dodoi{10.1088/0004-637X/732/1/40}

\bibitem[{{Donahue} {et~al.}(2015){Donahue}, {Connor}, {Fogarty}, {Li}, {Voit}, {Postman}, {Koekemoer}, {Moustakas}, {Bradley}, \& {Ford}}]{2015ApJ...805..177D}
{Donahue}, M., {Connor}, T., {Fogarty}, K., {et~al.} 2015, \apj, 805, 177, \dodoi{10.1088/0004-637X/805/2/177}

\bibitem[{{Downes} \& {Solomon}(1998)}]{1998ApJ...507..615D}
{Downes}, D., \& {Solomon}, P.~M. 1998, \apj, 507, 615, \dodoi{10.1086/306339}

\bibitem[{{Downes} {et~al.}(1993){Downes}, {Solomon}, \& {Radford}}]{1993ApJ...414L..13D}
{Downes}, D., {Solomon}, P.~M., \& {Radford}, S.~J.~E. 1993, \apjl, 414, L13, \dodoi{10.1086/186984}

\bibitem[{{Draine} \& {Li}(2007)}]{2007ApJ...657..810D}
{Draine}, B.~T., \& {Li}, A. 2007, \apj, 657, 810, \dodoi{10.1086/511055}

\bibitem[{{Edge}(2001)}]{2001MNRAS.328..762E}
{Edge}, A.~C. 2001, \mnras, 328, 762, \dodoi{10.1046/j.1365-8711.2001.04802.x}

\bibitem[{{Edwards} {et~al.}(2007){Edwards}, {Hudson}, {Balogh}, \& {Smith}}]{2007MNRAS.379..100E}
{Edwards}, L. O.~V., {Hudson}, M.~J., {Balogh}, M.~L., \& {Smith}, R.~J. 2007, \mnras, 379, 100, \dodoi{10.1111/j.1365-2966.2007.11910.x}

\bibitem[{{Fabian}(2012)}]{2012ARA&A..50..455F}
{Fabian}, A.~C. 2012, \araa, 50, 455, \dodoi{10.1146/annurev-astro-081811-125521}

\bibitem[{{Fabian} {et~al.}(2022){Fabian}, {Ferland}, {Sanders}, {McNamara}, {Pinto}, \& {Walker}}]{2022MNRAS.515.3336F}
{Fabian}, A.~C., {Ferland}, G.~J., {Sanders}, J.~S., {et~al.} 2022, \mnras, 515, 3336, \dodoi{10.1093/mnras/stac2003}

\bibitem[{{Fabian} {et~al.}(1984){Fabian}, {Nulsen}, \& {Canizares}}]{1984Natur.310..733F}
{Fabian}, A.~C., {Nulsen}, P.~E.~J., \& {Canizares}, C.~R. 1984, \nat, 310, 733, \dodoi{10.1038/310733a0}

\bibitem[{{Fabian} {et~al.}(2023){Fabian}, {Sanders}, {Ferland}, {McNamara}, {Pinto}, \& {Walker}}]{2023MNRAS.521.1794F}
{Fabian}, A.~C., {Sanders}, J.~S., {Ferland}, G.~J., {et~al.} 2023, \mnras, 521, 1794, \dodoi{10.1093/mnras/stad507}

\bibitem[{{Farrah} {et~al.}(2007){Farrah}, {Bernard-Salas}, {Spoon}, {Soifer}, {Armus}, {Brandl}, {Charmandaris}, {Desai}, {Higdon}, {Devost}, \& {Houck}}]{2007ApJ...667..149F}
{Farrah}, D., {Bernard-Salas}, J., {Spoon}, H.~W.~W., {et~al.} 2007, \apj, 667, 149, \dodoi{10.1086/520834}

\bibitem[{{F{\"o}rster Schreiber} {et~al.}(2004){F{\"o}rster Schreiber}, {Roussel}, {Sauvage}, \& {Charmandaris}}]{2004A&A...419..501F}
{F{\"o}rster Schreiber}, N.~M., {Roussel}, H., {Sauvage}, M., \& {Charmandaris}, V. 2004, \aap, 419, 501, \dodoi{10.1051/0004-6361:20040963}

\bibitem[{{Garc{\'\i}a-Bernete} {et~al.}(2024){Garc{\'\i}a-Bernete}, {Rigopoulou}, {Donnan}, {Alonso-Herrero}, {Pereira-Santaella}, {Shimizu}, {Davies}, {Roche}, {Garc{\'\i}a-Burillo}, {Labiano}, {Hermosa Mu{\~n}oz}, {Zhang}, {Audibert}, {Bellocchi}, {Bunker}, {Combes}, {Delaney}, {Esparza-Arredondo}, {Gandhi}, {Gonz{\'a}lez-Mart{\'\i}n}, {H{\"o}nig}, {Imanishi}, {Hicks}, {Fuller}, {Leist}, {Levenson}, {Lopez-Rodriguez}, {Packham}, {Ramos Almeida}, {Ricci}, {Stalevski}, {Villar Mart{\'\i}n}, \& {Ward}}]{2024A&A...691A.162G}
{Garc{\'\i}a-Bernete}, I., {Rigopoulou}, D., {Donnan}, F.~R., {et~al.} 2024, \aap, 691, A162, \dodoi{10.1051/0004-6361/202450086}

\bibitem[{{Gaspari} {et~al.}(2011){Gaspari}, {Melioli}, {Brighenti}, \& {D'Ercole}}]{2011MNRAS.411..349G}
{Gaspari}, M., {Melioli}, C., {Brighenti}, F., \& {D'Ercole}, A. 2011, \mnras, 411, 349, \dodoi{10.1111/j.1365-2966.2010.17688.x}

\bibitem[{{Gaspari} {et~al.}(2017){Gaspari}, {Temi}, \& {Brighenti}}]{2017MNRAS.466..677G}
{Gaspari}, M., {Temi}, P., \& {Brighenti}, F. 2017, \mnras, 466, 677, \dodoi{10.1093/mnras/stw3108}

\bibitem[{{Gaspari} {et~al.}(2020){Gaspari}, {Tombesi}, \& {Cappi}}]{2020NatAs...4...10G}
{Gaspari}, M., {Tombesi}, F., \& {Cappi}, M. 2020, Nature Astronomy, 4, 10, \dodoi{10.1038/s41550-019-0970-1}

\bibitem[{{Gaspari} {et~al.}(2018){Gaspari}, {McDonald}, {Hamer}, {Brighenti}, {Temi}, {Gendron-Marsolais}, {Hlavacek-Larrondo}, {Edge}, {Werner}, {Tozzi}, {Sun}, {Stone}, {Tremblay}, {Hogan}, {Eckert}, {Ettori}, {Yu}, {Biffi}, \& {Planelles}}]{2018ApJ...854..167G}
{Gaspari}, M., {McDonald}, M., {Hamer}, S.~L., {et~al.} 2018, \apj, 854, 167, \dodoi{10.3847/1538-4357/aaaa1b}

\bibitem[{{Harris} {et~al.}(2020){Harris}, {Millman}, {van der Walt}, {Gommers}, {Virtanen}, {Cournapeau}, {Wieser}, {Taylor}, {Berg}, {Smith}, {Kern}, {Picus}, {Hoyer}, {van Kerkwijk}, {Brett}, {Haldane}, {del R{\'\i}o}, {Wiebe}, {Peterson}, {G{\'e}rard-Marchant}, {Sheppard}, {Reddy}, {Weckesser}, {Abbasi}, {Gohlke}, \& {Oliphant}}]{2020Natur.585..357H}
{Harris}, C.~R., {Millman}, K.~J., {van der Walt}, S.~J., {et~al.} 2020, \nat, 585, 357, \dodoi{10.1038/s41586-020-2649-2}

\bibitem[{{Hatch} {et~al.}(2007){Hatch}, {Crawford}, \& {Fabian}}]{2007MNRAS.380...33H}
{Hatch}, N.~A., {Crawford}, C.~S., \& {Fabian}, A.~C. 2007, \mnras, 380, 33, \dodoi{10.1111/j.1365-2966.2007.12009.x}

\bibitem[{{Hicks} \& {Mushotzky}(2005)}]{2005ApJ...635L...9H}
{Hicks}, A.~K., \& {Mushotzky}, R. 2005, \apjl, 635, L9, \dodoi{10.1086/499123}

\bibitem[{{Hlavacek-Larrondo} {et~al.}(2015){Hlavacek-Larrondo}, {McDonald}, {Benson}, {Forman}, {Allen}, {Bleem}, {Ashby}, {Bocquet}, {Brodwin}, {Dietrich}, {Jones}, {Liu}, {Reichardt}, {Saliwanchik}, {Saro}, {Schrabback}, {Song}, {Stalder}, {Vikhlinin}, \& {Zenteno}}]{2015ApJ...805...35H}
{Hlavacek-Larrondo}, J., {McDonald}, M., {Benson}, B.~A., {et~al.} 2015, \apj, 805, 35, \dodoi{10.1088/0004-637X/805/1/35}

\bibitem[{{Ho} \& {Keto}(2007)}]{2007ApJ...658..314H}
{Ho}, L.~C., \& {Keto}, E. 2007, \apj, 658, 314, \dodoi{10.1086/511260}

\bibitem[{{Hoffer} {et~al.}(2012){Hoffer}, {Donahue}, {Hicks}, \& {Barthelemy}}]{2012ApJS..199...23H}
{Hoffer}, A.~S., {Donahue}, M., {Hicks}, A., \& {Barthelemy}, R.~S. 2012, \apjs, 199, 23, \dodoi{10.1088/0067-0049/199/1/23}

\bibitem[{Hunter(2007)}]{Hunter:2007}
Hunter, J.~D. 2007, Computing in Science \& Engineering, 9, 90, \dodoi{10.1109/MCSE.2007.55}

\bibitem[{{Kaneda} {et~al.}(2008){Kaneda}, {Onaka}, {Sakon}, {Kitayama}, {Okada}, \& {Suzuki}}]{2008ApJ...684..270K}
{Kaneda}, H., {Onaka}, T., {Sakon}, I., {et~al.} 2008, \apj, 684, 270, \dodoi{10.1086/590243}

\bibitem[{{Kaviraj} {et~al.}(2012){Kaviraj}, {Ting}, {Bureau}, {Shabala}, {Crockett}, {Silk}, {Lintott}, {Smith}, {Keel}, {Masters}, {Schawinski}, \& {Bamford}}]{2012MNRAS.423...49K}
{Kaviraj}, S., {Ting}, Y.-S., {Bureau}, M., {et~al.} 2012, \mnras, 423, 49, \dodoi{10.1111/j.1365-2966.2012.20957.x}

\bibitem[{{Kennicutt}(1990)}]{1990ASSL..161..405K}
{Kennicutt}, Robert~C., J. 1990, in Astrophysics and Space Science Library, Vol. 161, The Interstellar Medium in Galaxies, ed. J.~{Thronson}, Harley~A. \& J.~M. {Shull}, 405--435, \dodoi{10.1007/978-94-009-0595-5_17}

\bibitem[{{Kennicutt}(1998)}]{1998ARA&A..36..189K}
{Kennicutt}, Robert~C., J. 1998, \araa, 36, 189, \dodoi{10.1146/annurev.astro.36.1.189}

\bibitem[{{Klein} {et~al.}(2024){Klein}, {Mohr}, \& {Davies}}]{2024A&A...690A.322K}
{Klein}, M., {Mohr}, J.~J., \& {Davies}, C.~T. 2024, \aap, 690, A322, \dodoi{10.1051/0004-6361/202451203}

\bibitem[{{Labiano} {et~al.}(2021){Labiano}, {Argyriou}, {{\'A}lvarez-M{\'a}rquez}, {Glasse}, {Glauser}, {Patapis}, {Law}, {Brandl}, {Justtanont}, {Lahuis}, {Mart{\'\i}nez-Galarza}, {Mueller}, {Noriega-Crespo}, {Royer}, {Shaughnessy}, \& {Vandenbussche}}]{2021A&A...656A..57L}
{Labiano}, A., {Argyriou}, I., {{\'A}lvarez-M{\'a}rquez}, J., {et~al.} 2021, \aap, 656, A57, \dodoi{10.1051/0004-6361/202140614}

\bibitem[{{Le Petit} {et~al.}(2006){Le Petit}, {Nehm{\'e}}, {Le Bourlot}, \& {Roueff}}]{2006ApJS..164..506L}
{Le Petit}, F., {Nehm{\'e}}, C., {Le Bourlot}, J., \& {Roueff}, E. 2006, \apjs, 164, 506, \dodoi{10.1086/503252}

\bibitem[{{Leger} \& {Puget}(1984)}]{1984A&A...137L...5L}
{Leger}, A., \& {Puget}, J.~L. 1984, \aap, 137, L5

\bibitem[{{Li} {et~al.}(2017){Li}, {Ruszkowski}, \& {Bryan}}]{2017ApJ...847..106L}
{Li}, Y., {Ruszkowski}, M., \& {Bryan}, G.~L. 2017, \apj, 847, 106, \dodoi{10.3847/1538-4357/aa88c1}

\bibitem[{{Madau} \& {Dickinson}(2014)}]{2014ARAA..52..415M}
{Madau}, P., \& {Dickinson}, M. 2014, \araa, 52, 415, \dodoi{10.1146/annurev-astro-081811-125615}

\bibitem[{{McDonald} {et~al.}(2013){McDonald}, {Benson}, {Veilleux}, {Bautz}, \& {Reichardt}}]{2013ApJ...765L..37M}
{McDonald}, M., {Benson}, B., {Veilleux}, S., {Bautz}, M.~W., \& {Reichardt}, C.~L. 2013, \apjl, 765, L37, \dodoi{10.1088/2041-8205/765/2/L37}

\bibitem[{{McDonald} {et~al.}(2018){McDonald}, {Gaspari}, {McNamara}, \& {Tremblay}}]{2018ApJ...858...45M}
{McDonald}, M., {Gaspari}, M., {McNamara}, B.~R., \& {Tremblay}, G.~R. 2018, \apj, 858, 45, \dodoi{10.3847/1538-4357/aabace}

\bibitem[{{McDonald} {et~al.}(2010){McDonald}, {Veilleux}, {Rupke}, \& {Mushotzky}}]{2010ApJ...721.1262M}
{McDonald}, M., {Veilleux}, S., {Rupke}, D. S.~N., \& {Mushotzky}, R. 2010, \apj, 721, 1262, \dodoi{10.1088/0004-637X/721/2/1262}

\bibitem[{{McDonald} {et~al.}(2012){McDonald}, {Bayliss}, {Benson}, {Foley}, {Ruel}, {Sullivan}, {Veilleux}, {Aird}, {Ashby}, {Bautz}, {Bazin}, {Bleem}, {Brodwin}, {Carlstrom}, {Chang}, {Cho}, {Clocchiatti}, {Crawford}, {Crites}, {de Haan}, {Desai}, {Dobbs}, {Dudley}, {Egami}, {Forman}, {Garmire}, {George}, {Gladders}, {Gonzalez}, {Halverson}, {Harrington}, {High}, {Holder}, {Holzapfel}, {Hoover}, {Hrubes}, {Jones}, {Joy}, {Keisler}, {Knox}, {Lee}, {Leitch}, {Liu}, {Lueker}, {Luong-van}, {Mantz}, {Marrone}, {McMahon}, {Mehl}, {Meyer}, {Miller}, {Mocanu}, {Mohr}, {Montroy}, {Murray}, {Natoli}, {Padin}, {Plagge}, {Pryke}, {Rawle}, {Reichardt}, {Rest}, {Rex}, {Ruhl}, {Saliwanchik}, {Saro}, {Sayre}, {Schaffer}, {Shaw}, {Shirokoff}, {Simcoe}, {Song}, {Spieler}, {Stalder}, {Staniszewski}, {Stark}, {Story}, {Stubbs}, {{\v{S}}uhada}, {van Engelen}, {Vanderlinde}, {Vieira}, {Vikhlinin}, {Williamson}, {Zahn}, \& {Zenteno}}]{2012Natur.488..349M}
{McDonald}, M., {Bayliss}, M., {Benson}, B.~A., {et~al.} 2012, \nat, 488, 349, \dodoi{10.1038/nature11379}

\bibitem[{{McDonald} {et~al.}(2014){McDonald}, {Swinbank}, {Edge}, {Wilner}, {Veilleux}, {Benson}, {Hogan}, {Marrone}, {McNamara}, {Wei}, {Bayliss}, \& {Bautz}}]{2014ApJ...784...18M}
{McDonald}, M., {Swinbank}, M., {Edge}, A.~C., {et~al.} 2014, \apj, 784, 18, \dodoi{10.1088/0004-637X/784/1/18}

\bibitem[{{McDonald} {et~al.}(2015){McDonald}, {McNamara}, {van Weeren}, {Applegate}, {Bayliss}, {Bautz}, {Benson}, {Carlstrom}, {Bleem}, {Chatzikos}, {Edge}, {Fabian}, {Garmire}, {Hlavacek-Larrondo}, {Jones-Forman}, {Mantz}, {Miller}, {Stalder}, {Veilleux}, \& {ZuHone}}]{2015ApJ...811..111M}
{McDonald}, M., {McNamara}, B.~R., {van Weeren}, R.~J., {et~al.} 2015, \apj, 811, 111, \dodoi{10.1088/0004-637X/811/2/111}

\bibitem[{{McDonald} {et~al.}(2019){McDonald}, {McNamara}, {Voit}, {Bayliss}, {Benson}, {Brodwin}, {Canning}, {Florian}, {Garmire}, {Gaspari}, {Gladders}, {Hlavacek-Larrondo}, {Kara}, {Reichardt}, {Russell}, {Saro}, {Sharon}, {Somboonpanyakul}, {Tremblay}, \& {van Weeren}}]{2019ApJ...885...63M}
{McDonald}, M., {McNamara}, B.~R., {Voit}, G.~M., {et~al.} 2019, \apj, 885, 63, \dodoi{10.3847/1538-4357/ab464c}

\bibitem[{{McNamara} \& {Nulsen}(2007)}]{2007ARA&A..45..117M}
{McNamara}, B.~R., \& {Nulsen}, P.~E.~J. 2007, \araa, 45, 117, \dodoi{10.1146/annurev.astro.45.051806.110625}

\bibitem[{{McNamara} \& {O'Connell}(1989)}]{1989AJ.....98.2018M}
{McNamara}, B.~R., \& {O'Connell}, R.~W. 1989, \aj, 98, 2018, \dodoi{10.1086/115275}

\bibitem[{{Micelotta} {et~al.}(2010){Micelotta}, {Jones}, \& {Tielens}}]{2010A&A...510A..37M}
{Micelotta}, E.~R., {Jones}, A.~P., \& {Tielens}, A.~G.~G.~M. 2010, \aap, 510, A37, \dodoi{10.1051/0004-6361/200911683}

\bibitem[{{Mittal} {et~al.}(2015){Mittal}, {Whelan}, \& {Combes}}]{2015MNRAS.450.2564M}
{Mittal}, R., {Whelan}, J.~T., \& {Combes}, F. 2015, \mnras, 450, 2564, \dodoi{10.1093/mnras/stv754}

\bibitem[{{Mo} {et~al.}(2010){Mo}, {van den Bosch}, \& {White}}]{2010gfe..book.....M}
{Mo}, H., {van den Bosch}, F.~C., \& {White}, S. 2010, {Galaxy Formation and Evolution}

\bibitem[{{Molendi} {et~al.}(2016){Molendi}, {Tozzi}, {Gaspari}, {De Grandi}, {Gastaldello}, {Ghizzardi}, \& {Rossetti}}]{2016A&A...595A.123M}
{Molendi}, S., {Tozzi}, P., {Gaspari}, M., {et~al.} 2016, \aap, 595, A123, \dodoi{10.1051/0004-6361/201628338}

\bibitem[{{NASA/IPAC Extragalactic Database (NED)}(2019)}]{https://doi.org/10.26132/ned1}
{NASA/IPAC Extragalactic Database (NED)}. 2019, NASA/IPAC Extragalactic Database (NED),  IPAC, \dodoi{10.26132/NED1}

\bibitem[{{O'Dea} {et~al.}(2008){O'Dea}, {Baum}, {Privon}, {Noel-Storr}, {Quillen}, {Zufelt}, {Park}, {Edge}, {Russell}, {Fabian}, {Donahue}, {Sarazin}, {McNamara}, {Bregman}, \& {Egami}}]{2008ApJ...681.1035O}
{O'Dea}, C.~P., {Baum}, S.~A., {Privon}, G., {et~al.} 2008, \apj, 681, 1035, \dodoi{10.1086/588212}

\bibitem[{{Ogle} {et~al.}(2010){Ogle}, {Boulanger}, {Guillard}, {Evans}, {Antonucci}, {Appleton}, {Nesvadba}, \& {Leipski}}]{2010ApJ...724.1193O}
{Ogle}, P., {Boulanger}, F., {Guillard}, P., {et~al.} 2010, \apj, 724, 1193, \dodoi{10.1088/0004-637X/724/2/1193}

\bibitem[{{Peeters} {et~al.}(2004){Peeters}, {Spoon}, \& {Tielens}}]{2004ApJ...613..986P}
{Peeters}, E., {Spoon}, H.~W.~W., \& {Tielens}, A.~G.~G.~M. 2004, \apj, 613, 986, \dodoi{10.1086/423237}

\bibitem[{{Pereira-Santaella} {et~al.}(2022){Pereira-Santaella}, {{\'A}lvarez-M{\'a}rquez}, {Garc{\'\i}a-Bernete}, {Labiano}, {Colina}, {Alonso-Herrero}, {Bellocchi}, {Garc{\'\i}a-Burillo}, {H{\"o}nig}, {Ramos Almeida}, \& {Rosario}}]{2022A&A...665L..11P}
{Pereira-Santaella}, M., {{\'A}lvarez-M{\'a}rquez}, J., {Garc{\'\i}a-Bernete}, I., {et~al.} 2022, \aap, 665, L11, \dodoi{10.1051/0004-6361/202244725}

\bibitem[{{Peterson} {et~al.}(2003){Peterson}, {Kahn}, {Paerels}, {Kaastra}, {Tamura}, {Bleeker}, {Ferrigno}, \& {Jernigan}}]{2003ApJ...590..207P}
{Peterson}, J.~R., {Kahn}, S.~M., {Paerels}, F.~B.~S., {et~al.} 2003, \apj, 590, 207, \dodoi{10.1086/374830}

\bibitem[{{Prasad} {et~al.}(2015){Prasad}, {Sharma}, \& {Babul}}]{2015ApJ...811..108P}
{Prasad}, D., {Sharma}, P., \& {Babul}, A. 2015, \apj, 811, 108, \dodoi{10.1088/0004-637X/811/2/108}

\bibitem[{{Rawle} {et~al.}(2012){Rawle}, {Edge}, {Egami}, {Rex}, {Smith}, {Altieri}, {Fiedler}, {Haines}, {Pereira}, {P{\'e}rez-Gonz{\'a}lez}, {Portouw}, {Valtchanov}, {Walth}, {van der Werf}, \& {Zemcov}}]{2012ApJ...747...29R}
{Rawle}, T.~D., {Edge}, A.~C., {Egami}, E., {et~al.} 2012, \apj, 747, 29, \dodoi{10.1088/0004-637X/747/1/29}

\bibitem[{Reefe(2025)}]{reefe_2025_15069243}
Reefe, M. 2025, LOKI: Likelihood Optimization of gas Kinematics in IFUs, v2.0,  Zenodo, \dodoi{10.5281/zenodo.15069243}

\bibitem[{Reefe {et~al.}(2025)Reefe, McDonald, Chatzikos, Seebeck, Mushotzky, Veilleux, Allen, Bayliss, Calzadilla, Canning, Floyd, Gaspari, Hlavacek-Larrondo, McNamara, Russell, Sharon, \& Somboonpanyakul}]{Reefe_2025}
Reefe, M., McDonald, M., Chatzikos, M., {et~al.} 2025, Nature, 638, 360–364, \dodoi{10.1038/s41586-024-08369-x}

\bibitem[{{Roche} {et~al.}(1991){Roche}, {Aitken}, {Smith}, \& {Ward}}]{1991MNRAS.248..606R}
{Roche}, P.~F., {Aitken}, D.~K., {Smith}, C.~H., \& {Ward}, M.~J. 1991, \mnras, 248, 606, \dodoi{10.1093/mnras/248.4.606}

\bibitem[{{Rosenthal} {et~al.}(2000){Rosenthal}, {Bertoldi}, \& {Drapatz}}]{2000A&A...356..705R}
{Rosenthal}, D., {Bertoldi}, F., \& {Drapatz}, S. 2000, \aap, 356, 705, \dodoi{10.48550/arXiv.astro-ph/0002456}

\bibitem[{{Roussel} {et~al.}(2007){Roussel}, {Helou}, {Hollenbach}, {Draine}, {Smith}, {Armus}, {Schinnerer}, {Walter}, {Engelbracht}, {Thornley}, {Kennicutt}, {Calzetti}, {Dale}, {Murphy}, \& {Bot}}]{2007ApJ...669..959R}
{Roussel}, H., {Helou}, G., {Hollenbach}, D.~J., {et~al.} 2007, \apj, 669, 959, \dodoi{10.1086/521667}

\bibitem[{{Russell} {et~al.}(2013){Russell}, {McNamara}, {Edge}, {Hogan}, {Main}, \& {Vantyghem}}]{2013MNRAS.432..530R}
{Russell}, H.~R., {McNamara}, B.~R., {Edge}, A.~C., {et~al.} 2013, \mnras, 432, 530, \dodoi{10.1093/mnras/stt490}

\bibitem[{{Russell} {et~al.}(2017){Russell}, {McDonald}, {McNamara}, {Fabian}, {Nulsen}, {Bayliss}, {Benson}, {Brodwin}, {Carlstrom}, {Edge}, {Hlavacek-Larrondo}, {Marrone}, {Reichardt}, \& {Vieira}}]{2017ApJ...836..130R}
{Russell}, H.~R., {McDonald}, M., {McNamara}, B.~R., {et~al.} 2017, \apj, 836, 130, \dodoi{10.3847/1538-4357/836/1/130}

\bibitem[{{Sakon} {et~al.}(2004){Sakon}, {Onaka}, {Ishihara}, {Ootsubo}, {Yamamura}, {Tanab{\'e}}, \& {Roellig}}]{2004ApJ...609..203S}
{Sakon}, I., {Onaka}, T., {Ishihara}, D., {et~al.} 2004, \apj, 609, 203, \dodoi{10.1086/420919}

\bibitem[{{Salom{\'e}} \& {Combes}(2003)}]{2003A&A...412..657S}
{Salom{\'e}}, P., \& {Combes}, F. 2003, \aap, 412, 657, \dodoi{10.1051/0004-6361:20031438}

\bibitem[{{Salom{\'e}} {et~al.}(2008){Salom{\'e}}, {Revaz}, {Combes}, {Pety}, {Downes}, {Edge}, \& {Fabian}}]{2008A&A...483..793S}
{Salom{\'e}}, P., {Revaz}, Y., {Combes}, F., {et~al.} 2008, \aap, 483, 793, \dodoi{10.1051/0004-6361:200809412}

\bibitem[{{Sloan} {et~al.}(1999){Sloan}, {Hayward}, {Allamandola}, {Bregman}, {DeVito}, \& {Hudgins}}]{1999ApJ...513L..65S}
{Sloan}, G.~C., {Hayward}, T.~L., {Allamandola}, L.~J., {et~al.} 1999, \apjl, 513, L65, \dodoi{10.1086/311906}

\bibitem[{{Smith} {et~al.}(2007){Smith}, {Draine}, {Dale}, {Moustakas}, {Kennicutt}, {Helou}, {Armus}, {Roussel}, {Sheth}, {Bendo}, {Buckalew}, {Calzetti}, {Engelbracht}, {Gordon}, {Hollenbach}, {Li}, {Malhotra}, {Murphy}, \& {Walter}}]{2007ApJ...656..770S}
{Smith}, J.~D.~T., {Draine}, B.~T., {Dale}, D.~A., {et~al.} 2007, \apj, 656, 770, \dodoi{10.1086/510549}

\bibitem[{{Smithsonian Astrophysical Observatory}(2000)}]{2000ascl.soft03002S}
{Smithsonian Astrophysical Observatory}. 2000, {SAOImage DS9: A utility for displaying astronomical images in the X11 window environment}, Astrophysics Source Code Library, record ascl:0003.002

\bibitem[{{Spilker} {et~al.}(2023){Spilker}, {Phadke}, {Aravena}, {Archipley}, {Bayliss}, {Birkin}, {B{\'e}thermin}, {Burgoyne}, {Cathey}, {Chapman}, {Dahle}, {Gonzalez}, {Gururajan}, {Hayward}, {Hezaveh}, {Hill}, {Hutchison}, {Kim}, {Kim}, {Law}, {Legin}, {Malkan}, {Marrone}, {Murphy}, {Narayanan}, {Navarre}, {Olivier}, {Rich}, {Rigby}, {Reuter}, {Rhoads}, {Sharon}, {Smith}, {Solimano}, {Sulzenauer}, {Vieira}, {Vizgan}, {Wei{\ss}}, \& {Whitaker}}]{2023Natur.618..708S}
{Spilker}, J.~S., {Phadke}, K.~A., {Aravena}, M., {et~al.} 2023, \nat, 618, 708, \dodoi{10.1038/s41586-023-05998-6}

\bibitem[{{Sutherland} \& {Dopita}(1993)}]{1993ApJS...88..253S}
{Sutherland}, R.~S., \& {Dopita}, M.~A. 1993, \apjs, 88, 253, \dodoi{10.1086/191823}

\bibitem[{{Timmerman} {et~al.}(2021){Timmerman}, {van Weeren}, {McDonald}, {Ignesti}, {McNamara}, {Hlavacek-Larrondo}, \& {R{\"o}ttgering}}]{2021A&A...646A..38T}
{Timmerman}, R., {van Weeren}, R.~J., {McDonald}, M., {et~al.} 2021, \aap, 646, A38, \dodoi{10.1051/0004-6361/202039075}

\bibitem[{{Togi} \& {Smith}(2016)}]{2016ApJ...830...18T}
{Togi}, A., \& {Smith}, J.~D.~T. 2016, \apj, 830, 18, \dodoi{10.3847/0004-637X/830/1/18}

\bibitem[{{Treyer} {et~al.}(2010){Treyer}, {Schiminovich}, {Johnson}, {O'Dowd}, {Martin}, {Wyder}, {Charlot}, {Heckman}, {Martins}, {Seibert}, \& {van der Hulst}}]{2010ApJ...719.1191T}
{Treyer}, M., {Schiminovich}, D., {Johnson}, B.~D., {et~al.} 2010, \apj, 719, 1191, \dodoi{10.1088/0004-637X/719/2/1191}

\bibitem[{{Tychoniec} {et~al.}(2024){Tychoniec}, {van Gelder}, {van Dishoeck}, {Francis}, {Rocha}, {Caratti o Garatti}, {Beuther}, {Gieser}, {Justtanont}, {Linnartz}, {Le Gouellec}, {Perotti}, {Devaraj}, {Tabone}, {Ray}, {Brunken}, {Chen}, {Kavanagh}, {Klaassen}, {Slavicinska}, {G{\"u}del}, \& {{\"O}stlin}}]{2024A&A...687A..36T}
{Tychoniec}, {\L}., {van Gelder}, M.~L., {van Dishoeck}, E.~F., {et~al.} 2024, \aap, 687, A36, \dodoi{10.1051/0004-6361/202348889}

\bibitem[{{Van De Putte} {et~al.}(2024){Van De Putte}, {Meshaka}, {Trahin}, {Habart}, {Peeters}, {Bern{\'e}}, {Alarc{\'o}n}, {Canin}, {Chown}, {Schroetter}, {Sidhu}, {Boersma}, {Bron}, {Dartois}, {Goicoechea}, {Gordon}, {Onaka}, {Tielens}, {Verstraete}, {Wolfire}, {Abergel}, {Bergin}, {Bernard-Salas}, {Cami}, {Cuadrado}, {Dicken}, {Elyajouri}, {Fuente}, {Joblin}, {Khan}, {Lacinbala}, {Languignon}, {Le Gal}, {Maragkoudakis}, {Okada}, {Pasquini}, {Pound}, {Robberto}, {R{\"o}llig}, {Schefter}, {Schirmer}, {Tabone}, {Vicente}, {Zannese}, {Colgan}, {He}, {Rouill{\'e}}, {Togi}, {Aleman}, {Auchettl}, {Baratta}, {Bejaoui}, {Bera}, {Black}, {Boulanger}, {Bouwman}, {Brandl}, {Brechignac}, {Br{\"u}nken}, {Buragohain}, {Burkhardt}, {Candian}, {Cazaux}, {Cernicharo}, {Chabot}, {Chakraborty}, {Champion}, {Cooke}, {Coutens}, {Cox}, {Demyk}, {Meyer}, {Foschino}, {Garc{\'\i}a-Lario}, {Gerin}, {Gottlieb}, {Guillard}, {Gusdorf}, {Hartigan}, {Herbst}, {Hornekaer}, {Issa}, {J{\"a}ger}, {Janot-Pacheco}, {Kannavou}, {Kaufman},
  {Kemper}, {Kendrew}, {Kirsanova}, {Klaassen}, {Kwok}, {Labiano}, {Lai}, {Le Floch}, {Le Petit}, {Li}, {Linz}, {Mackie}, {Madden}, {Mascetti}, {McGuire}, {Merino}, {Micelotta}, {Morse}, {Mulas}, {Neelamkodan}, {Ohsawa}, {Omont}, {Paladini}, {Palumbo}, {Pathak}, {Pendleton}, {Petrignani}, {Pino}, {Puga}, {Rangwala}, {Rapacioli}, {Rho}, {Ricca}, {Roman-Duval}, {Roser}, {Roueff}, {Salama}, {Sales}, {Sandstrom}, {Sarre}, {Sciamma-O'Brien}, {Sellgren}, {Shenoy}, {Teyssier}, {Thomas}, {Witt}, {Wootten}, {Ysard}, {Zettergren}, {Zhang}, {Zhang}, \& {Zhen}}]{2024A&A...687A..86V}
{Van De Putte}, D., {Meshaka}, R., {Trahin}, B., {et~al.} 2024, \aap, 687, A86, \dodoi{10.1051/0004-6361/202449295}

\bibitem[{{Van Kerckhoven} {et~al.}(2000){Van Kerckhoven}, {Hony}, {Peeters}, {Tielens}, {Allamandola}, {Hudgins}, {Cox}, {Roelfsema}, {Voors}, {Waelkens}, {Waters}, \& {Wesselius}}]{2000A&A...357.1013V}
{Van Kerckhoven}, C., {Hony}, S., {Peeters}, E., {et~al.} 2000, \aap, 357, 1013

\bibitem[{Van~Rossum \& Drake(2009)}]{10.5555/1593511}
Van~Rossum, G., \& Drake, F.~L. 2009, Python 3 Reference Manual (Scotts Valley, CA: CreateSpace)

\bibitem[{{Veilleux} {et~al.}(2009){Veilleux}, {Rupke}, {Kim}, {Genzel}, {Sturm}, {Lutz}, {Contursi}, {Schweitzer}, {Tacconi}, {Netzer}, {Sternberg}, {Mihos}, {Baker}, {Mazzarella}, {Lord}, {Sanders}, {Stockton}, {Joseph}, \& {Barnes}}]{2009ApJS..182..628V}
{Veilleux}, S., {Rupke}, D.~S.~N., {Kim}, D.~C., {et~al.} 2009, \apjs, 182, 628, \dodoi{10.1088/0067-0049/182/2/628}

\bibitem[{{Viaene} {et~al.}(2020){Viaene}, {Nersesian}, {Fritz}, {Verstocken}, {Baes}, {Bianchi}, {Casasola}, {Cassar{\`a}}, {Clark}, {Davies}, {De Looze}, {De Vis}, {Dobbels}, {Galametz}, {Galliano}, {Jones}, {Madden}, {Mosenkov}, {Trcka}, {Xilouris}, \& {Ysard}}]{2020A&A...638A.150V}
{Viaene}, S., {Nersesian}, A., {Fritz}, J., {et~al.} 2020, \aap, 638, A150, \dodoi{10.1051/0004-6361/202037476}

\bibitem[{{Villa-V{\'e}lez} {et~al.}(2024){Villa-V{\'e}lez}, {Godard}, {Guillard}, \& {Pineau des For{\^e}ts}}]{2024A&A...688A..96V}
{Villa-V{\'e}lez}, J.~A., {Godard}, B., {Guillard}, P., \& {Pineau des For{\^e}ts}, G. 2024, \aap, 688, A96, \dodoi{10.1051/0004-6361/202449212}

\bibitem[{{Virtanen} {et~al.}(2020){Virtanen}, {Gommers}, {Oliphant}, {Haberland}, {Reddy}, {Cournapeau}, {Burovski}, {Peterson}, {Weckesser}, {Bright}, {van der Walt}, {Brett}, {Wilson}, {Millman}, {Mayorov}, {Nelson}, {Jones}, {Kern}, {Larson}, {Carey}, {Polat}, {Feng}, {Moore}, {VanderPlas}, {Laxalde}, {Perktold}, {Cimrman}, {Henriksen}, {Quintero}, {Harris}, {Archibald}, {Ribeiro}, {Pedregosa}, {van Mulbregt}, \& {SciPy 1. 0 Contributors}}]{2020NatMe..17..261V}
{Virtanen}, P., {Gommers}, R., {Oliphant}, T.~E., {et~al.} 2020, Nature Methods, 17, 261, \dodoi{10.1038/s41592-019-0686-2}

\bibitem[{{Werle} {et~al.}(2022){Werle}, {Poggianti}, {Moretti}, {Bellhouse}, {Vulcani}, {Gullieuszik}, {Radovich}, {Fritz}, {Ignesti}, {Richard}, {Soucail}, {Bruzual}, {Charlot}, {Mingozzi}, {Bacchini}, {Tomicic}, {Smith}, {Kulier}, {Peluso}, \& {Franchetto}}]{2022ApJ...930...43W}
{Werle}, A., {Poggianti}, B., {Moretti}, A., {et~al.} 2022, \apj, 930, 43, \dodoi{10.3847/1538-4357/ac5f06}

\bibitem[{{Williamson} {et~al.}(2011){Williamson}, {Benson}, {High}, {Vanderlinde}, {Ade}, {Aird}, {Andersson}, {Armstrong}, {Ashby}, {Bautz}, {Bazin}, {Bertin}, {Bleem}, {Bonamente}, {Brodwin}, {Carlstrom}, {Chang}, {Chapman}, {Clocchiatti}, {Crawford}, {Crites}, {de Haan}, {Desai}, {Dobbs}, {Dudley}, {Fazio}, {Foley}, {Forman}, {Garmire}, {George}, {Gladders}, {Gonzalez}, {Halverson}, {Holder}, {Holzapfel}, {Hoover}, {Hrubes}, {Jones}, {Joy}, {Keisler}, {Knox}, {Lee}, {Leitch}, {Lueker}, {Luong-Van}, {Marrone}, {McMahon}, {Mehl}, {Meyer}, {Mohr}, {Montroy}, {Murray}, {Padin}, {Plagge}, {Pryke}, {Reichardt}, {Rest}, {Ruel}, {Ruhl}, {Saliwanchik}, {Saro}, {Schaffer}, {Shaw}, {Shirokoff}, {Song}, {Spieler}, {Stalder}, {Stanford}, {Staniszewski}, {Stark}, {Story}, {Stubbs}, {Vieira}, {Vikhlinin}, \& {Zenteno}}]{2011ApJ...738..139W}
{Williamson}, R., {Benson}, B.~A., {High}, F.~W., {et~al.} 2011, \apj, 738, 139, \dodoi{10.1088/0004-637X/738/2/139}

\bibitem[{{Wittor} \& {Gaspari}(2020)}]{2020MNRAS.498.4983W}
{Wittor}, D., \& {Gaspari}, M. 2020, \mnras, 498, 4983, \dodoi{10.1093/mnras/staa2747}

\bibitem[{{Wu} {et~al.}(2010){Wu}, {Helou}, {Armus}, {Cormier}, {Shi}, {Dale}, {Dasyra}, {Smith}, {Papovich}, {Draine}, {Rahman}, {Stierwalt}, {Fadda}, {Lagache}, \& {Wright}}]{2010ApJ...723..895W}
{Wu}, Y., {Helou}, G., {Armus}, L., {et~al.} 2010, \apj, 723, 895, \dodoi{10.1088/0004-637X/723/1/895}

\bibitem[{{Xie} \& {Ho}(2019)}]{2019ApJ...884..136X}
{Xie}, Y., \& {Ho}, L.~C. 2019, \apj, 884, 136, \dodoi{10.3847/1538-4357/ab4200}

\bibitem[{{Xie} \& {Ho}(2022)}]{2022ApJ...925..218X}
---. 2022, \apj, 925, 218, \dodoi{10.3847/1538-4357/ac32e2}

\bibitem[{{Yang} {et~al.}(2020){Yang}, {Boquien}, {Buat}, {Burgarella}, {Ciesla}, {Duras}, {Stalevski}, {Brandt}, \& {Papovich}}]{2020MNRAS.491..740Y}
{Yang}, G., {Boquien}, M., {Buat}, V., {et~al.} 2020, \mnras, 491, 740, \dodoi{10.1093/mnras/stz3001}

\end{thebibliography}
\bibliographystyle{aasjournal}

\end{document}